%% file: ANA-MUON-2022-01-PAPER.tex
\documentclass[cernpreprint, atlasdraft=false, texlive=2023, UKenglish, coverpage=false, texmf, orcidlogo]{atlasdoc}
 
\usepackage{atlaspackage}
\usepackage{atlasbiblatex}
 
\usepackage[hepprocess=true,hepparticle=true]{atlasphysics}

\graphicspath{{logos/}{figures/}}
 
\usepackage{ANA-MUON-2022-01-PAPER-defs}
 
\DeclareOldFontCommand{\bf}{\normalfont\bfseries}{\mathbf}
\newcommand{\pb}{\text{pb}^{-1}}

\def\Zmm{{Z\to\mu\mu}}

\AtlasTitle{Studies of the muon momentum calibration and performance of the ATLAS detector with $pp$ collisions at $\sqrt{s}$=13~TeV}
 
\AtlasAbstract{
This paper presents the muon momentum calibration and performance studies for the ATLAS detector based on the $pp$ collisions data sample produced at $\sqrt{s}$~=~13~TeV at the LHC during Run~2 and corresponding to an integrated luminosity of 139~$\mathrm{fb^{-1}}$.
An innovative approach is used to correct for potential charge-dependent momentum biases related to the knowledge of the detector geometry, using the $Z\rightarrow\mu^{+}\mu^{-}$ resonance. The muon momentum scale and resolution are measured using samples of $J/\psi\rightarrow\mu^{+}\mu^{-}$ and $Z\rightarrow\mu^{+}\mu^{-}$ events. A calibration procedure is defined and applied to simulated data to match the performance measured in real data. The calibration is validated using an independent sample of $\Upsilon\rightarrow\mu^{+}\mu^{-}$ events.
At the $Z$ ($J/\psi$) peak, the momentum scale is measured with an uncertainty at the 0.05\% (0.1\%) level, and the resolution is measured with an uncertainty at the 1.5\% (2\%) level. 
The charge-dependent bias is removed with a dedicated {\em in situ} correction for momenta up to $450$~GeV with a precision better than 0.03~$\mathrm{TeV^{-1}}$.
}

\AtlasRefCode{MUON-2022-01}
 
\PreprintIdNumber{CERN-EP-2022-274}

\AtlasJournal{EPJC}
\AtlasJournalRef{Eur. Phys. J. C 83 (2023) 686}
\AtlasDOI{10.1140/epjc/s10052-023-11584-x}

\hypersetup{pdftitle={ATLAS document},pdfauthor={The ATLAS Collaboration}}
 
\begin{document}
 
\maketitle
 
\tableofcontents

\section{Introduction}

Muons are a crucial component 
of the physics programme of the ATLAS experiment~\cite{PERF-2007-01} at the Large Hadron Collider (LHC)~\cite{Evans:2008zzb}.
The discovery of the Higgs boson~\cite{HIGG-2012-27,CMS-HIG-12-028}
and the measurement of its properties~\cite{HIGG-2021-23},
precision tests of the Standard Model~\cite{STDM-2014-18},
and searches for physics processes beyond the Standard Model~\cite{EXOT-2018-08,EXOT-2018-30}
all strongly rely on the performance of muon identification and measurement with the ATLAS detector.
 
During its second data-taking campaign (Run~2; 2015--2018), ATLAS collected 139~\ifb{} of $pp$ collisions at a centre-of-mass energy of $\sqrt{s}=13$~\TeV{}, a data sample of unprecedented size. A recent publication presented the performance of the new and optimised muon reconstruction and identification techniques developed for the analysis of the full Run~2 data sample~\cite{MUON-2018-03}. The muon momentum measurement was previously published for early Run~2 data, corresponding to 3.2~\ifb{} collected in 2015~\cite{PERF-2015-10}.
 
This paper describes the methods used to calibrate the momentum measurement of muons reconstructed by the ATLAS detector for the full Run~2 data sample.
First, the potential charge-dependent bias on the scale of the muon momentum measurement, introduced by the imperfect knowledge of the real detector geometry, is measured in reconstructed collision data with a sample of $\Zmm$ events.\footnote{Charges of decay products are omitted in the following for simplicity.} A dedicated correction is derived and applied to the data. Then, $\Jmm$ and $\Zmm$ decays are selected in data and used to measure the resolution and scale of the muon momentum, which are compared with those predicted by simulation. A calibration procedure is applied to simulated events, to improve the agreement between the simulation and the data. Finally, a validation procedure of this calibration is performed using $\Jmm$ and $\Zmm$ events, in addition to an independent sample of $\Upsmm$ decays.
 
In contrast to previous publications, this paper presents a completely new methodology to determine the charge-dependent bias, sensitive to global shifts in the detectors' positions, which could not be measured with previous techniques.
Furthermore, the inclusion of new $\Jmm$ data collected with dedicated trigger strategies, the development of new fitting techniques with better convergence properties, and the significantly larger data sample result in an unprecedented precision of the momentum calibration procedure.
In addition to the calibrations developed separately for the tracks measured in the inner detector and in the muon spectrometer of the ATLAS detector, for the first time a dedicated calibration for tracks obtained by combining information from both sub-detectors is derived. The validation of the calibration procedure is performed using $\Upsmm$ events as well as $\Jmm$ and $\Zmm$ events in regions with finer granularity than those used in the calibration procedure. 
 
This paper is structured as follows: the ATLAS detector is described in \refS{sec:detector};
the simulated and real data samples used for the measurements are presented in \refS{sec:Datasets};
the identification of muon candidates is discussed in \refS{sec:MuonReco};
the methodologies for measuring the muon scale and momentum corrections are presented in \refS{sec:Resolution};
the results of the measurement and the validation of the data-driven corrections derived for simulated samples are presented in \refS{sec::validation};
final conclusions are provided in \refS{sec:conclusion}.


\FloatBarrier
 
\section{ATLAS detector}
\label{sec:detector}

\newcommand{\AtlasCoordFootnote}{
ATLAS uses a right-handed coordinate system with its origin at the nominal interaction point (IP)
in the centre of the detector and the \(z\)-axis along the beam pipe.
The \(x\)-axis points from the IP to the centre of the LHC ring,
and the \(y\)-axis points upwards.
Cylindrical coordinates \((r,\phi)\) are used in the transverse plane,
\(\phi\) being the azimuthal angle around the \(z\)-axis.
The pseudorapidity is defined in terms of the polar angle \(\theta\) as \(\eta = -\ln \tan(\theta/2)\).
Angular distance is measured in units of \(\Delta R \equiv \sqrt{(\Delta\eta)^{2} + (\Delta\phi)^{2}}\).}
 
The ATLAS detector~\cite{PERF-2007-01} at the LHC covers nearly the entire solid angle around the collision point.\footnote{\AtlasCoordFootnote}
It consists of an inner tracking detector surrounded by a thin superconducting solenoid, electromagnetic and hadron calorimeters,
and a muon spectrometer incorporating three large superconducting air-core toroidal magnets.
 
The inner detector system (ID) is immersed in a \qty{2}{\tesla} axial magnetic field (bending the tracks of charged particles in the transverse plane)
and provides charged-particle tracking in the range of $|\eta| < 2.5$.
The high-granularity silicon pixel detector covers the vertex region and typically provides four measurements per track.
The first hit is normally found in the insertable B-layer (IBL) installed before Run~2~\cite{ATLAS-TDR-19,PIX-2018-001}.
It is followed by the silicon microstrip tracker (SCT), which typically provides eight measurements per track.
These silicon detectors are complemented by the transition radiation tracker (TRT),
which enables radially extended track reconstruction up to $|\eta| = 2.0$, typically providing 30 measurements per track.
The TRT also provides electron identification information
based on the fraction of hits above a higher energy-deposit threshold corresponding to transition radiation.
 
The calorimeter system covers the pseudorapidity range $|\eta| < 4.9$.
In the region $|\eta| < 3.2$, electromagnetic calorimetry is provided by barrel and
endcap high-granularity lead/liquid-argon (LAr) calorimeters,
with an additional thin LAr presampler covering $|\eta| < 1.8$
to correct for energy loss in material upstream of the calorimeters.
Hadron calorimetry is provided by the steel/scintillator-tile calorimeter,
segmented into three barrel structures with $|\eta| < 1.7$, and two copper/LAr hadron endcap calorimeters.
The solid angle coverage is completed with forward copper/LAr and tungsten/LAr calorimeter modules
optimised for electromagnetic and hadronic energy measurements respectively.
 
The muon spectrometer (MS) comprises separate trigger and
high-precision tracking chambers measuring the deflection of muons in a magnetic field generated by the superconducting air-core toroidal magnets, which bend charged particles tracks in the $r$--$z$ plane, i.e. in the plane formed by the muon momentum vector and the beam axis.
The field integral of the toroids ranges between \num{2.0} and \SI{6.0}{\tesla\metre}
across most of the detector.
Three layers of precision chambers, each consisting of layers of monitored drift tubes, cover the region $|\eta| < 2.7$,
complemented by cathode-strip chambers in the forward region, where the background is highest.
The muon trigger system covers the range $|\eta| < 2.4$ with resistive-plate chambers in the barrel, and thin-gap chambers in the endcap regions.
 
Events are selected by the first-level trigger system implemented in custom hardware,
followed by selections made by algorithms implemented in software in the high-level trigger~\cite{TRIG-2016-01}.
The first-level trigger accepts events from the \SI{40}{\MHz} bunch crossings at a rate below \SI{100}{\kHz},
which the high-level trigger further reduces to record events to disk at about \SI{1}{\kHz}.
 
An extensive software suite~\cite{ATL-SOFT-PUB-2021-001} is used in data simulation, in the reconstruction and analysis of real
and simulated data, in detector operations, and in the trigger and data acquisition systems of the experiment.


\section{Data and Monte Carlo samples}
\label{sec:Datasets}

\subsection{Data sample}
 
The analysis is performed using the $\sqrt{s}=$13~\TeV{} $pp$ collision data sample recorded by the ATLAS detector between 2015 and 2018. Only events collected in stable beam conditions and with all relevant ATLAS magnets and detector subsystems fully operational are used in the analysis. A combination of single-muon trigger algorithms (for $\Zmm$ events) and trigger algorithms dedicated to $\Jmm$ and $\Upsmm$ topologies~\cite{TRIG-2018-01} were used.
Starting from the middle of 2016, due to the increasing instantaneous luminosity of the LHC, events associated with $\Jmm$ and $\Upsmm$ triggers were collected with a dedicated data acquisition stream for $B$-physics and low-mass final states, 
separated from the main physics stream so that their reconstruction could be delayed if the availability of processing resources was limited~\cite{TRIG-2019-04}.
This additional stream is analysed for a muon momentum calibration for the first time in this paper, significantly increasing the number of $\Jmm$ and $\Upsmm$ events available.
For the data sample used in this analysis a procedure to calibrate the muon detectors was applied, as described in Ref.~\cite{MDET-2018-01}.
The analysed data corresponds to an integrated luminosity of 139~\ifb{} after trigger and data-quality requirements, with an average number of $pp$ collisions per bunch crossing $<\mu>=33.7$. The calibration derived in the analysis does not show any significant dependence on $<\mu>$.

\subsection{Simulated samples}
\label{sec:simulation}
 
Samples of Monte Carlo (MC) simulated inclusive prompt production of $\Jmm$ and $\Psi(2S)\to\mu\mu$ events were generated at leading-order (LO) accuracy using \PYTHIA[8.186]~\cite{Sjostrand:2007gs}, for matrix element (ME) calculations, and for the modelling of the parton shower, hadronisation, and the underlying event.
The \CTEQ[6L1]~\cite{Pumplin:2002vw} parton density function (PDF) set and the A14~\cite{ATL-PHYS-PUB-2014-021} set of tuned generator parameters were used. The effect of QED final-state radiation was simulated with \PHOTOSpp{}~\cite{Golonka:2005pn,Davidson:2010ew}. To increase the effective number of events for the analysis, the generated events were filtered, requiring at least one muon in the event to satisfy $\pt>6$~\GeV{} and $|\eta|<2.5$.
 
Samples of $\Ups{1}$, $\Ups{2}$, and $\Ups{3}$ simulated events with subsequent decays into two muons were generated with a similar configuration to that of the $\Jmm$ samples, though the filter required at least two muons with $\pt>4$~\GeV{} and $|\eta|<2.7$ in the event.
 
The samples of $\Zmm$, $\Ztt$, $\Wmn$, and \mbox{$Z^{*}/\gamma^{*}(\rightarrow\mu\mu)$}
events were simulated using the \POWHEGBOX[v1]~\cite{Nason:2004rx,Frixione:2007vw,Alioli:2010xd,Alioli:2008gx} generator at next-to-leading-order (NLO) accuracy in perturbative QCD, interfaced to \PYTHIA[8.186] with the \AZNLO \cite{STDM-2012-23} set of tuned generator parameters. The \CT[10nlo] PDF set~\cite{Lai:2010vv} was used for the hard-scattering processes, whereas the \CTEQ[6L1] PDF set was used for the modelling of the parton showering, hadronisation, and underlying event. The effect of QED final-state radiation was simulated with \PHOTOSpp{}. The \EVTGEN[1.2.0] program~\cite{Lange:2001uf} was used to decay bottom and charm hadrons.
To improve the agreement with the data, the simulated $Z$ samples are reweighted to reproduce the muon-pair tranverse momentum and rapidity distributions measured in data.
 
A configuration similar to the one used for the $Z$ samples described above was used to simulate diboson processes decaying into final states containing muons, \mbox{$(ZZ, WZ, WW$)}, using \POWHEGBOX[v2] and \PYTHIA[8.210]~\cite{Sjostrand:2014zea} in this case.
 
The production of \ttbar events was modelled using the \POWHEGBOX[v2]~\cite{Frixione:2007nw,Nason:2004rx,Frixione:2007vw,Alioli:2010xd} generator at NLO with the \NNPDF[3.0nlo]~\cite{Ball:2014uwa} PDF set and the \hdamp parameter\footnote{The \hdamp parameter is a re-summation damping factor and one of the  parameters that controls the matching of \POWHEG MEs to the parton shower and thus effectively regulates the high-\pT radiation against which the \ttbar system recoils.} set to 1.5\,\mtop~\cite{ATL-PHYS-PUB-2016-020}. The events were interfaced to \PYTHIA[8.230] with the A14 set of tuned generator parameters and using the \NNPDF[2.3lo] PDF set~\cite{Ball:2012cx}. The decays of bottom and charm hadrons were performed by \EVTGEN[1.6.0].
 
An alternative sample of $\Zmm$ events was simulated with the \SHERPA[2.2.1]~\cite{Bothmann:2019yzt} generator using NLO ME for up to two partons, and LO ME
for up to four partons calculated with Comix~\cite{Gleisberg:2008fv} and \OPENLOOPS~\cite{Buccioni:2019sur,Cascioli:2011va,Denner:2016kdg}. They were matched to the \SHERPA parton shower~\cite{Schumann:2007mg} using the \MEPSatNLO prescription~\cite{Hoeche:2011fd,Hoeche:2012yf,Catani:2001cc,Hoeche:2009rj} using the set of tuned parameters developed by the \SHERPA authors. The \NNPDF[3.0nnlo] PDF set~\cite{Ball:2014uwa} was used and the samples were normalised to a next-to-next-to-leading-order (NNLO) prediction~\cite{Anastasiou:2003ds}.
This sample was used to check the impact of radiative effects on the muon momentum corrections, as \SHERPA features an alternative model than the nominal sample.
The re-weighting procedure applied to the nominal  $\Zmm$ events sample is also applied to this alternative sample.
 
The effect of multiple interactions in the same and neighbouring bunch crossings (\pileup) was modelled by overlaying the simulated hard-scattering event with inelastic proton--proton (\(pp\)) events generated with \PYTHIA[8.186]~\cite{Sjostrand:2007gs} using the \NNPDF[2.3lo] PDF set~\cite{Ball:2012cx} and the A3 set of tuned parameters~\cite{ATL-PHYS-PUB-2016-017}. The MC events were weighted to reproduce the distribution of the average number of interactions per bunch crossing  observed in the data.
 
The detector response was simulated using the \GEANT{} toolkit and reconstructed with the same algorithms as used for real data~\cite{Agostinelli:2002hh,SOFT-2010-01}. The detector simulation assumes perfect alignment for the ID and MS sub-detectors. Dedicated simulated samples with misaligned detector geometry are used for the validation of the calibration procedure. These include samples with a rotation of the ID layers with a linear dependence on the radius~\cite{IDTR-2019-05} and a perfect geometry for the MS\@. They also include samples with the positions of the middle layer of the MS chambers shifted by amounts corresponding to the residual alignment precision~\cite{ATL-MUON-PUB-2021-002}.


\section{Muon reconstruction, identification and event selection}
\label{sec:MuonReco}

\subsection{Muon reconstruction and identification}
 
The muon reconstruction is described in detail in Ref.~\cite{MUON-2018-03}; only a brief summary is provided here. Muons are reconstructed using information from the ID and/or MS sub-detectors, which provide two independent measurements of muons crossing the ATLAS detector. In this analysis, the following reconstruction algorithms are used.
\begin{itemize}
\item ID tracks: tracks reconstructed using ID hits only.
\item Standalone tracks (SA): muon candidates obtained using MS hits only; to define kinematic quantities at the interaction point, SA tracks are extrapolated to the beam-spot taking into account energy loss and multiple scattering in the material upstream of the muon spectrometer.
\item Combined tracks (CB): muon candidates obtained by starting from a MS track and matching it to an ID track. A combined fit to the hits, taking into account energy loss in the calorimeters and multiple scattering in the spectrometer, is performed.
\end{itemize}
 
Several working points (WPs) are introduced in Ref.~\cite{MUON-2018-03}, defining sets of quality cuts applied to reconstructed muons.
The \mediumwp{} WP is the baseline for this analysis; this algorithm only selects CB muons in $|\eta|<2.5$ and applies a set of requirements to reject poorly reconstructed tracks.
The \highptwp{} WP applies tighter cuts on CB muons to ensure an optimal muon reconstruction for analyses interested in the very high \pt{} regime, with a better momentum resolution.
Other working points target either lower background rates, at the expense of a lower efficiency, or higher efficiency but with a lower purity for muon identification; these WPs have a momentum resolution performance similar to the \mediumwp{} one.
The individual ID and SA tracks associated with the CB muons are still accessible and used in the analysis, as discussed in the following. For simplicity,  muon candidates reconstructed using the ID, MS or combined ID+MS information are referred to as ID, MS or CB muons, respectively.

\subsection{Event selection}
 
Proton--proton collision vertices are constructed from reconstructed trajectories of charged particles in the ID with transverse momentum $\pt> 500~\MeV$. Events are required to have at least one collision vertex. The vertex with the highest $\sum{\pt^2}$ of reconstructed tracks is selected as the primary vertex of the hard interaction. The data are subjected to quality requirements to reject events in which detector components were not operating correctly~\cite{DAPR-2018-01}.
 
Events are required to be selected by a combination of the following trigger chains:
\begin{itemize}
\item For $\Zmm$ candidate events, unprescaled single-muon trigger chains with the lowest kinematic threshold available in each data sample are used.
\begin{itemize}
\item Data collected in 2015: at least one muon with $\pt>20~\GeV$ and passing a loose isolation requirement based on the scalar sum of the \pt{} of tracks in a cone around the muon candidate track. Events are also retained if selected by a second chain requiring at least one muon with $\pt>40~\GeV$ without any isolation requirement.
\item Data collected in 2016, 2017, 2018: a similar strategy as the one used in 2015 was employed, but with the first chain requiring a muon with $\pt>26~\GeV$ passing a tighter isolation requirement, and the second chain requiring at least one muon with $\pt>50~\GeV$.
\end{itemize}
\item For $\Jmm$ and $\Upsmm$ candidate events, trigger chains requiring at least two muons in the event were used. The trigger algorithms also performed common vertex fits to pairs of oppositely charged muon candidates, requiring at least one of the fitted vertices to satisfy fit quality criteria and have an  invariant mass consistent with that of a \jpsi or $\Upsi$ resonance~\cite{TRIG-2018-01}.  The last requirement is significantly looser than that applied by the offline analysis, to avoid introducing a bias. The trigger algorithms applied the following kinematic requirements on the muons associated with the candidate resonance:
\begin{itemize}
\item Data collected in 2015: both muons must satisfy $\pt>4~\GeV$.
\item Data collected in 2016: both muons must satisfy $\pt>6~\GeV$.
\item Data collected in 2017 and 2018: both muons must satisfy $\pt>6~\GeV$. Additional requirements were also applied on the angular distance between the two muons associated with the vertex and on the displacement of the fitted vertex in the transverse plane. Events were also selected using auxiliary trigger algorithms requiring the leading (sub-leading) muon to have $\pt>11~\GeV$ ($\pt>6~\GeV$) but without any additional requirement.
\end{itemize}
\end{itemize}

Events are required to have two oppositely charged CB muons satisfying the \mediumwp{} WP requirements and spatially matched with the muon candidates reconstructed by the respective trigger chains. Further kinematic requirements are applied to ensure that selected muons are in the regions where the triggers used for the analysis were fully efficient:
\begin{itemize}
\item For $\Zmm$ candidate events, the leading muon must satisfy $\pt>27~\GeV$.
\item For $\Jmm$ and $\Upsmm$ candidate events, both muons must satisfy $\pt>6.25~\GeV$. For events selected only by the auxiliary trigger chain in the 2017 and 2018 data samples, the requirement on the leading muon is changed to $\pt>11.25~\GeV$.
\end{itemize}
 
For each muon, requirements are also applied on the transverse ($d_0$) and longitudinal ($z_0$) impact parameters.
The $d_0$ of a charged-particle track is defined in the plane transverse to the beam as the distance from the primary vertex to the track's point of closest approach. The $z_0$ is the distance in the $z$ direction between this track point and the primary vertex; this distance is represented by $z_0 \sin{\theta}$ in the longitudinal projection. The candidate muons are required to satisfy:
\begin{itemize}
\item $\frac{|d_{0}|}{\sigma(d_{0})}<3$, with $\sigma(d_{0})$ representing the uncertainty in the measured $d_{0}$ value;
\item $|z_0 \sin{\theta}|<$0.5~mm.
\end{itemize}
After application of these selections, the non-prompt component of $\Jmm$ production is reduced to less than 0.5\%.
 
Finally the invariant mass of the muon pair is required to satisfy:
 
\begin{itemize}
\item $2.6<m_{\mu\mu}<3.5~\GeV$ for $\Jmm$ candidate events;
\item $7<m_{\mu\mu}<14~\GeV$ for $\Upsmm$ candidate events;
\item $70<m_{\mu\mu}<130~\GeV$ for $\Zmm$ candidate events.
\end{itemize}


\section{Momentum scale corrections}
\label{sec:Resolution}

The muon momentum scale and resolution are studied using $\Jmm$ and $\Zmm$ decays in data and simulated samples. While dedicated procedures are applied to correct the detector alignment~\cite{PERF-2007-01,IDTR-2019-05,ATL-MUON-PUB-2021-002}, residual misalignments introduce a charge-dependent bias in the momentum measurement. Studies of this effect are discussed in \refS{Sec:SagittaBias}, together with the determination of an appropriate set of corrections for the data. Simulated samples assume ideal detector alignment and thus do not require these corrections.
 
Although the simulation contains an accurate description of the ATLAS detector,
the level of detail is not sufficient to reach the accuracy of 0.1\% on the muon momentum scale and the percent level precision on the resolution measured in data.
The analysis of this sample improves upon the excellent simulated description of the ATLAS detector and its interaction with muons, encompassing the best knowledge of the detector geometry, material distribution, and physics modelling at the time of event simulation.
\refS{Sec:CorrectionExtraction} details the measurement of momentum scale and resolution in data and simulation, and the determination of corresponding corrections for the simulated samples.
Validation studies for the corrections are presented in \refS{sec::validation}.

\subsection{Charge-dependent momentum scale calibration in data}\label{Sec:SagittaBias}

After dedicated detector alignment procedures are applied, residual effects can still introduce a bias in the momentum measurement of the muon. These effects arise from both the ID and the MS sub-detectors.
 
The ID alignment is derived by a global $\chi^2$ minimisation of the track-hit residuals~\cite{IDTR-2019-05}. Residual detector displacements relative to the nominal detector geometry may however still be present, the  {\em weak modes}, that are not fully corrected for by the current alignment procedures. ID alignment studies of these modes are periodically performed and data is reconstructed with an improved description of the detector geometry. Nevertheless, residual effects after this correction procedure still prevent the best possible precision being attained, motivating the study discussed in the following.
 
For the MS, the alignment procedures are less sensitive to charge-dependent biases, though residual effects may remain due to the limited precision of the procedure. An optical alignment system~\cite{PERF-2007-01} monitors the position of the muon chambers relative to each other and relative to fiducial marks in the detector.
The system compensates for chamber position variations as the data is collected. However, several aspects limit the accuracy of this system, including access shafts for detector operations and construction precision in some detector regions.
Additionally, the analysis of dedicated data runs, either collecting cosmic-ray data or in proton--proton collisions with the magnetic field generated by the toroid systems switched off, allows the precision of the relative position of the muon chambers to be further constrained. This leads to a precision at the level of tens of micrometers on the sagitta of the muon track, and up to 120--130~$\mathrm{\mu m}$ in specific detector regions~\cite{ATL-MUON-PUB-2021-002}. This residual uncertainty in the alignment system manifests as charge-dependent effects in the muon spectrometer reconstruction procedure.
 
The precision of the muon momentum reconstruction can be further improved with a dedicated correction procedure accounting for the ID weak modes and MS alignment system uncertainties, which would otherwise degrade the momentum resolution. The charge-dependent bias can be approximated as:
\begin{equation}
\frac{q}{\hat{p}}=\frac{q}{p} + q \cdot \delta_s,
\label{eq:bias_definition}
\end{equation}
 
where $q=\pm 1$ is the charge of the muon, $p$ and $\hat{p}$ are the corrected and uncorrected momentum of the muon, respectively, and $\delta_s$ is the strength of the bias. This parameterisation forms the basis for developing a correction to the data, 
recovering the residual bias and improving the momentum resolution.
 
To estimate (and later correct) the bias, the large sample of $Z\rightarrow\mu\mu$ decays is used. The mass of the dimuon system, $m_{\mu\mu}$, can be expressed as a function of the positively and negatively charged muons' transverse momenta $\pt^+$ and $\pt^-$, respectively:
\begin{equation} 
m_{\mu\mu}^{2} = 2 \cdot \pt^+\pt^- \cdot (\text{cosh}(\Delta\eta) - \text{cos}(\Delta\phi)),
\end{equation}
\noindent where $\Delta\eta$ and $\Delta\phi$ are the differences between the pseudorapidity and azimuthal angles of the two muons.
This expression is valid in the approximation of the muon mass being negligible with respect to the energies of the muons, and has been verified to hold in the context of these studies to sub-\MeV precision.
The bias is parameterised as a function of a grid of 48$\times$48 equal-sized $\eta$--$\phi$ detector regions,
a granularity chosen to ensure that the measurement is not affected by large statistical fluctuations while still correctly reflecting local biases or deformations.
This high granularity allows the transverse momentum \pt{} of the measurement to be used instead of $p$, which is necessary when comparing ID and MS results, since the two sub-detectors have different bending planes due to their respective magnet systems.
Combining the previous two equations, the biased invariant mass $\hat{m}_{\mu\mu}^{2}$ can be expressed as:
\begin{equation}\label{eq:masSagitta2}
\hat{m}_{\mu\mu}^{2} = \frac{m_{\mu\mu}^{2}}{(1+\delta_{s} \pt^+)(1-\delta_{s} \pt^-)}.
\end{equation}
 
\noindent Assuming the bias is small, \refE{eq:masSagitta2} can be approximated as:
\begin{equation}
m_{\mu\mu}^{2} = \hat{m}_{\mu\mu}^{2}(1+\delta_{s}(\eta,\phi) \pt^+ - \delta_{s}(\eta,\phi) \pt^-).
\label{eq:masSagitta3}
\end{equation}
 
Since $\pt^{+}\sim\pt^{-}$ in $Z$ boson decays, charge-dependent biases largely cancel in the expression of $m_{\mu\mu}^{2}$, hence do not impact the average
dimuon mass over all detector regions, but only broaden the resolution of the $Z\rightarrow\mu\mu$ peak.
Therefore, the reconstructed invariant mass $m_{\mu\mu}$ distribution defined in \refE{eq:masSagitta3} is sensitive to the bias through its impact on the variance of the distribution. The values of the sagitta biases $\hat{\delta}_{s}(\eta,\phi)$ are evaluated by minimising the variance of the invariant mass distributions; then,
the biased momentum of the muon $\hat{\pt}$ is corrected using the following equation:
\begin{equation}
\label{eq:sgcorr}
\pt = \frac{\hat{\pt}}{ 1 - q \hat{\delta}_{s}(\eta,\phi)\hat{\pt}}.
\end{equation}
Finally, the updated \pt{} values are used to recalculate the invariant mass distribution and a new iteration is started. Given the large number of degrees of freedom induced by the dependence of the biases on $\eta$ and $\phi$, a direct solution of the equation becomes impractical. Instead, an iterative approach is chosen. At each iteration the values of $\hat{\delta}_{s}(\eta,\phi)$ obtained from the previous iteration are used as an input to $\hat{\pt}$ of the next.
 
A complementary method to estimate this bias is also defined by approximating the average $\pt$ of the muon to half the invariant mass of the dimuon pair.
In this case, the bias can be approximated by:
\begin{equation}
\label{eq:detaqM}
\delta_s (\eta,\phi) = 4\cdot \frac{q (\hat{m}_{\mu\mu}(\eta,\phi) - \langle m_{\mu\mu} \rangle)}{\langle m_{\mu\mu} \rangle^2},
\end{equation}
where $\langle m_{\mu\mu} \rangle$ is the average of the invariant mass of the dimuon pairs used to derive the correction, while $\hat{m}_{\mu\mu}(\eta,\phi)$ is the average invariant mass of the dimuon pairs when the muon with the highest transverse momentum is in the given $(\eta,\phi)$ region.

Both methods additionally assume that the kinematic properties of positively and negatively charged muons produced by $Z$ decays are symmetrical. In addition to the small asymmetry introduced by the weak mixing angle, this assumption breaks down due to the detector acceptance and selection requirements restricting the phase space of the selected muon pairs. The impact of these acceptance and kinematic restrictions can be estimated by applying the same procedure to events simulated with ideal detector alignment, obtaining a residual bias term $\delta_{s}^{\text{MC}}$. The $\delta_{s}^{\text{data}}$ measured in data can be corrected by subtracting the values measured in MC simulation, $\delta_{s}^{\text{MC}}$, to define the final corrections:
\begin{equation}
\delta_{s}(\eta,\phi)  = \delta_{s}^{\text{data}}(\eta,\phi)-\delta_{s}^{\text{MC}}(\eta,\phi).
\end{equation}
 
Typical values for $\delta_{s}^{\text{MC}}(\eta,\phi)$ range from up to $0.01~\TeV^{-1}$ in the central pseudorapidity region to up to about $0.05~\TeV^{-1}$ in the forward pseudorapidity region, after event selection. The procedure is applied separately for the muon momentum measured using the CB, ID, and MS information, and dedicated corrections for each measurement are derived. Both procedures target a residual $\delta_{s}$ of less than $10^{-3}~\TeV{}^{-1}$ through simultaneous iterative updates on the ID, MS, and CB momenta. The convergence is achieved once the target is reached for all momenta. For the first method this happens after five iterations on average, while the second typically requires 25 iterative updates.
In both methods, the angular correlations between the two muons have a small impact on the estimate of $\delta_s$, given the fine granularity chosen in $\eta\times\phi$. \refF{fig:sgBefore} shows the measured sagitta corrections evaluated on CB momenta, with biases up to 0.4~$\TeV{}^{-1}$.
To avoid introducing a bias in the measurement due to the value assumed for $m_{Z}$, the unbiased value of $m_{\mu\mu}$ is taken from the data when integrating over all $\eta$ and $\phi$ bins for both methods.
Results from the two methods, which rely on different assumptions, are compared.
For the first method, a comparable bias for the leading and subleading $\pt$ is assumed, while for the second method the approximation of $\pt^{+} \simeq \pt^{-}$ is less precise at high $\eta$ values.
The $\delta_s (\eta,\phi)$ corrections obtained with the two methods have a correlation close to 100\%, with an agreement between the two methods at the level of $0.01~\TeV{}^{-1}$ or better,
with this difference taken as an additional systematic uncertainty.
Results for the second method are given in \refA{app:sagitta}.
 
\begin{figure*}[!htb]
\centering
\subfloat[ ]{\includegraphics[width=0.45\linewidth]{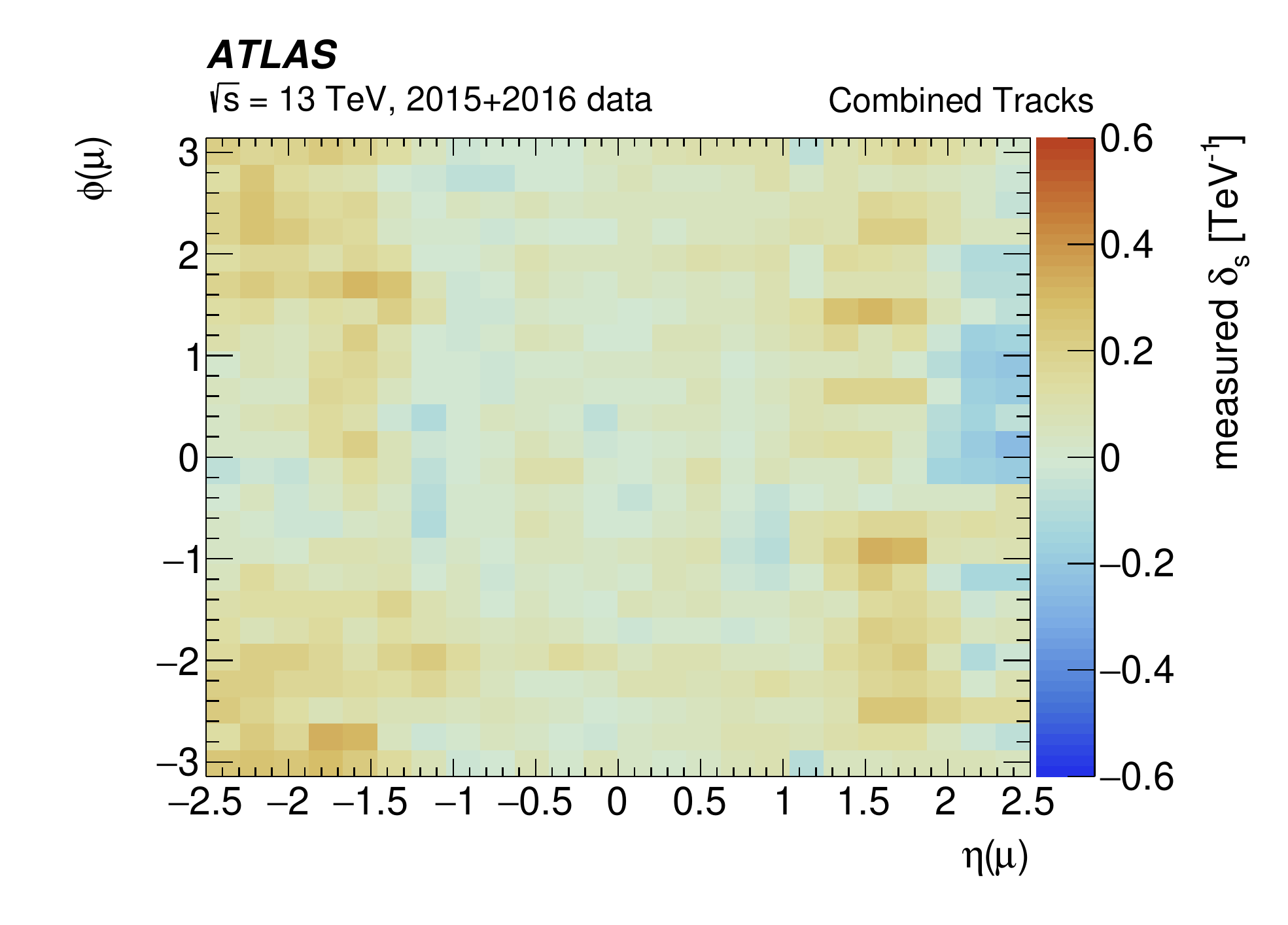}}
\subfloat[ ]{\includegraphics[width=0.45\linewidth]{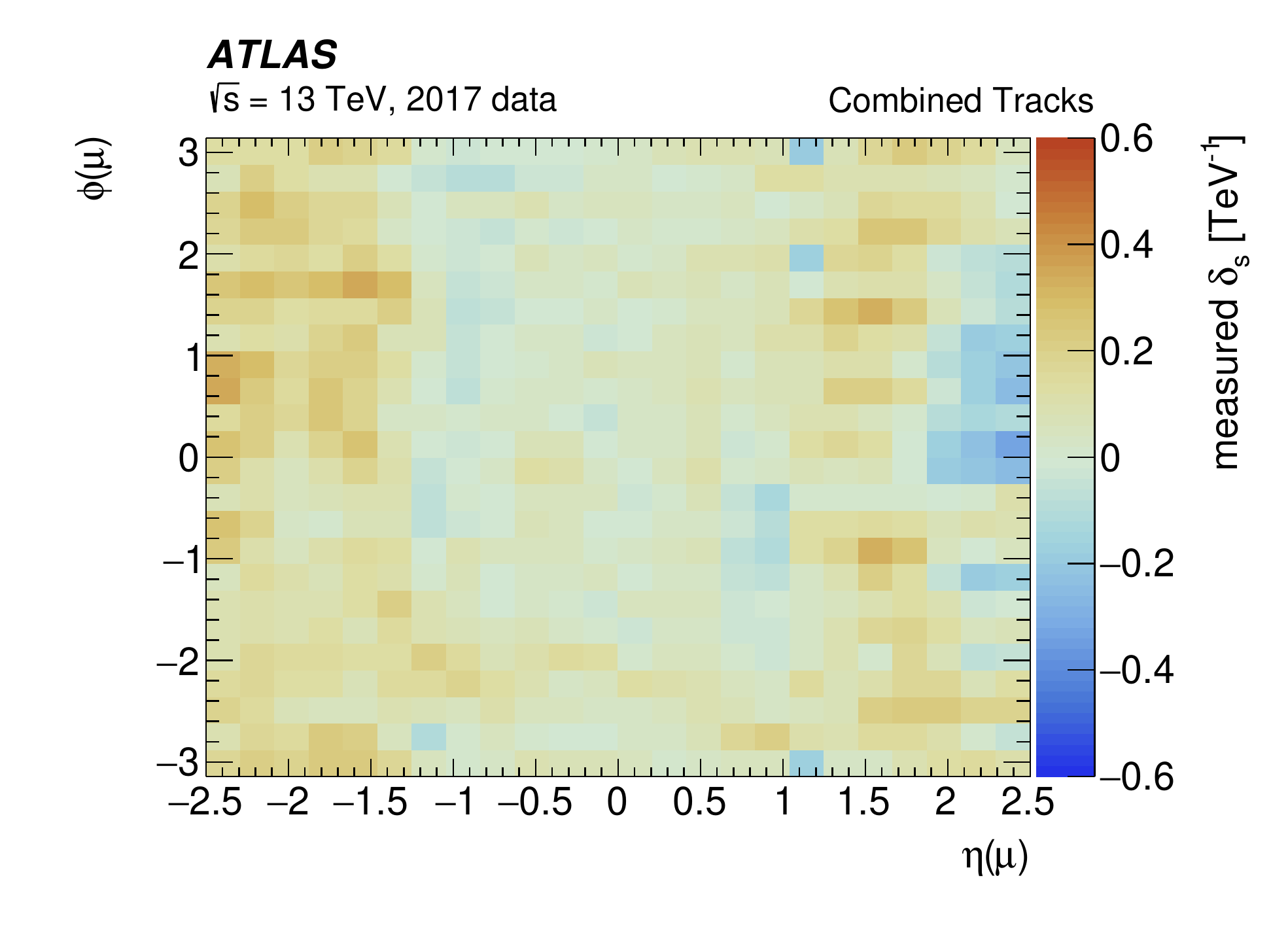}} \\
\subfloat[ ]{\includegraphics[width=0.45\linewidth]{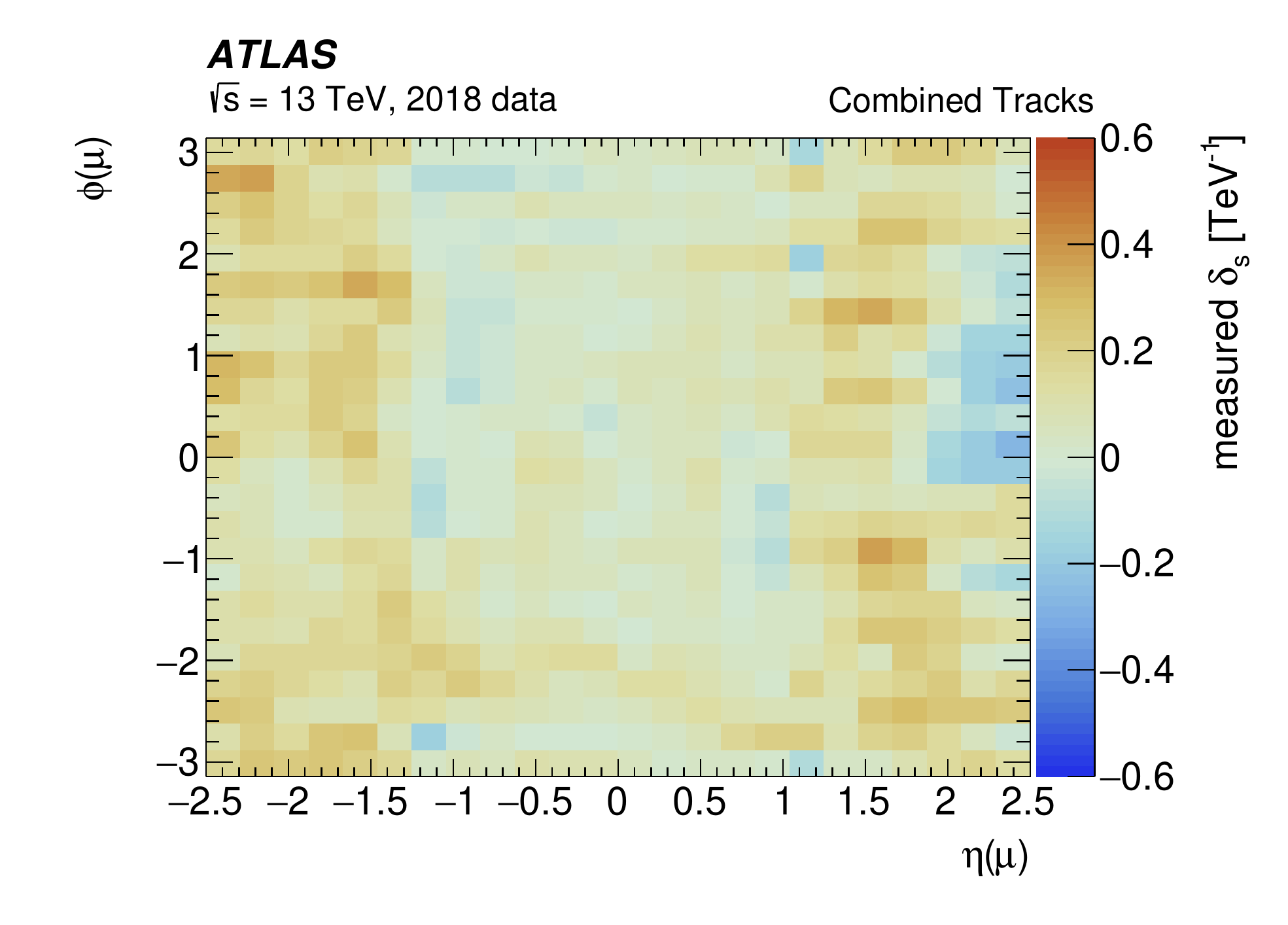}}
\caption{Charge-dependent biases on the muon \pt\ as evaluated on data for CB momenta after alignment and before applying the dedicated corrections. The biases are shown separately for the three data-taking periods,  (a) years 2015 and 2016, (b) 2017, and (c) 2018. The bias is defined in Eq.~(\ref{eq:bias_definition}).}
\label{fig:sgBefore}
\end{figure*}

Using the derived bias results, the transverse momenta of the muon are corrected based on \refE{eq:sgcorr} using the $\hat{\delta}_{s}$ derived after the last iteration. The biases are re-evaluated after the application of the corrections and the results are shown in \refF{fig:sgAfterCorr} for CB momenta. Both methods address relative differences between nearby $\eta, \phi$ regions, as well as biases that are constant across full sub-detector regions.
The biases are reduced to less than $2\cdot10^{-4}~\TeV{}^{-1}$ in all regions of the detector, validating the method to several orders of magnitude better than the typical size of the correction.

To validate the correction procedure, simulated samples with misaligned detector configurations, as described in \refS{sec:simulation}, were used. The injected biases were produced by a rotation of the ID detector layers with a linear dependence on the radial distance from the interaction point. The sizes of these biases in the simulation are the same as those observed in data from $\Zmm$ decays before applying the correction procedure. \refF{fig:truth} uses these biased detector geometry samples to compare the reconstructed and generated momentum distributions, 
as evaluated before and after the correction, plotted separately for positive and negative muons. The application of the correction reduces the differences between the positively and negatively charged muons, improving the reconstructed \pt\ resolution.
Further closure tests are performed using simulated samples with distorted geometry on $H\rightarrow ZZ\rightarrow \mu\mu\mu\mu$ and hypothetical narrow resonances decaying into two muons with masses greater than $500~\gev$, to estimate the impact of the correction in muon momentum regimes far from that of muons originating from $Z$ decays. Closure is observed at the same level of precision.
 
\begin{figure*}[!htb]
\centering
\subfloat[ ]{\includegraphics[width=0.45\linewidth]{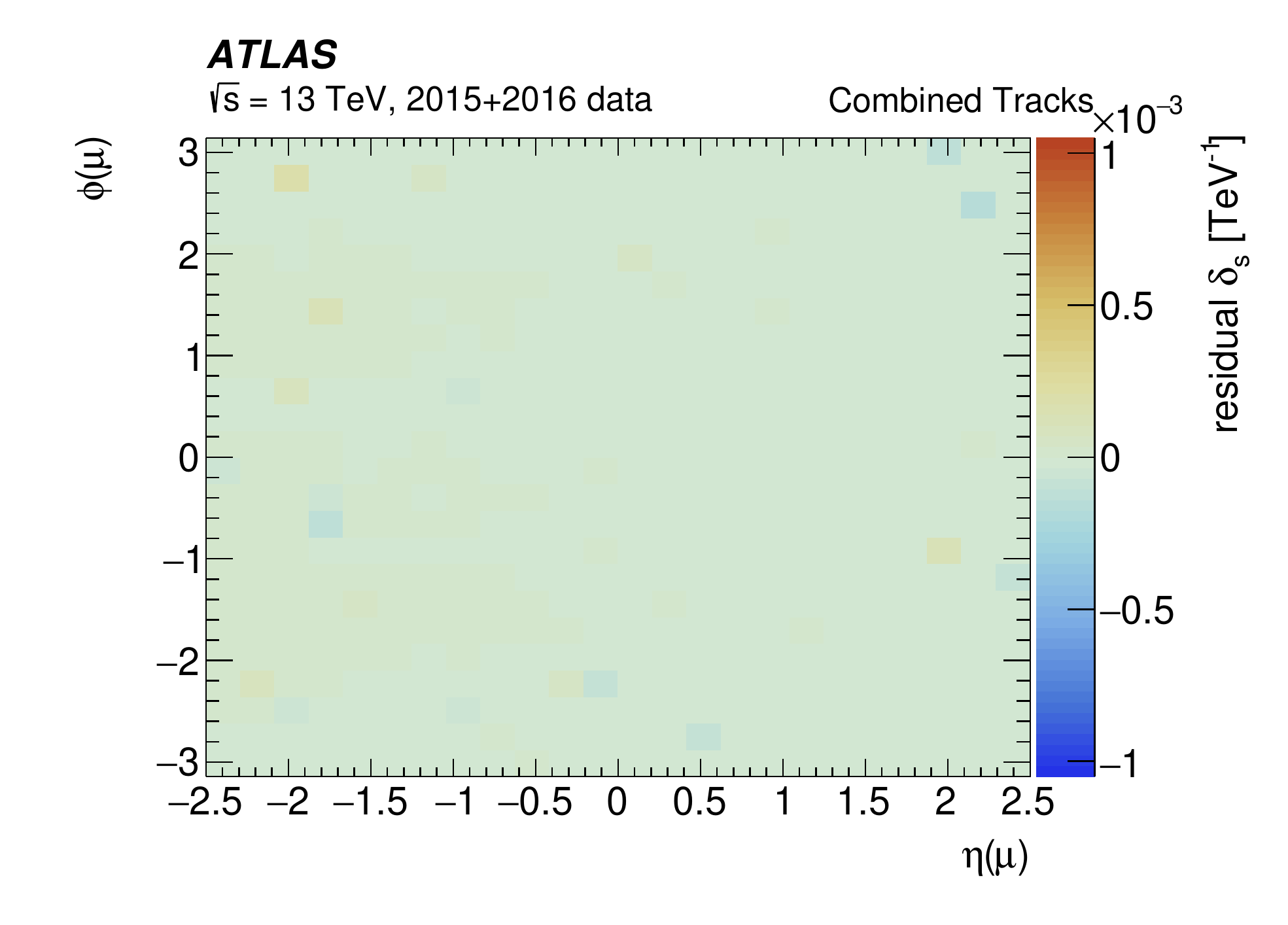}}
\subfloat[ ]{\includegraphics[width=0.45\linewidth]{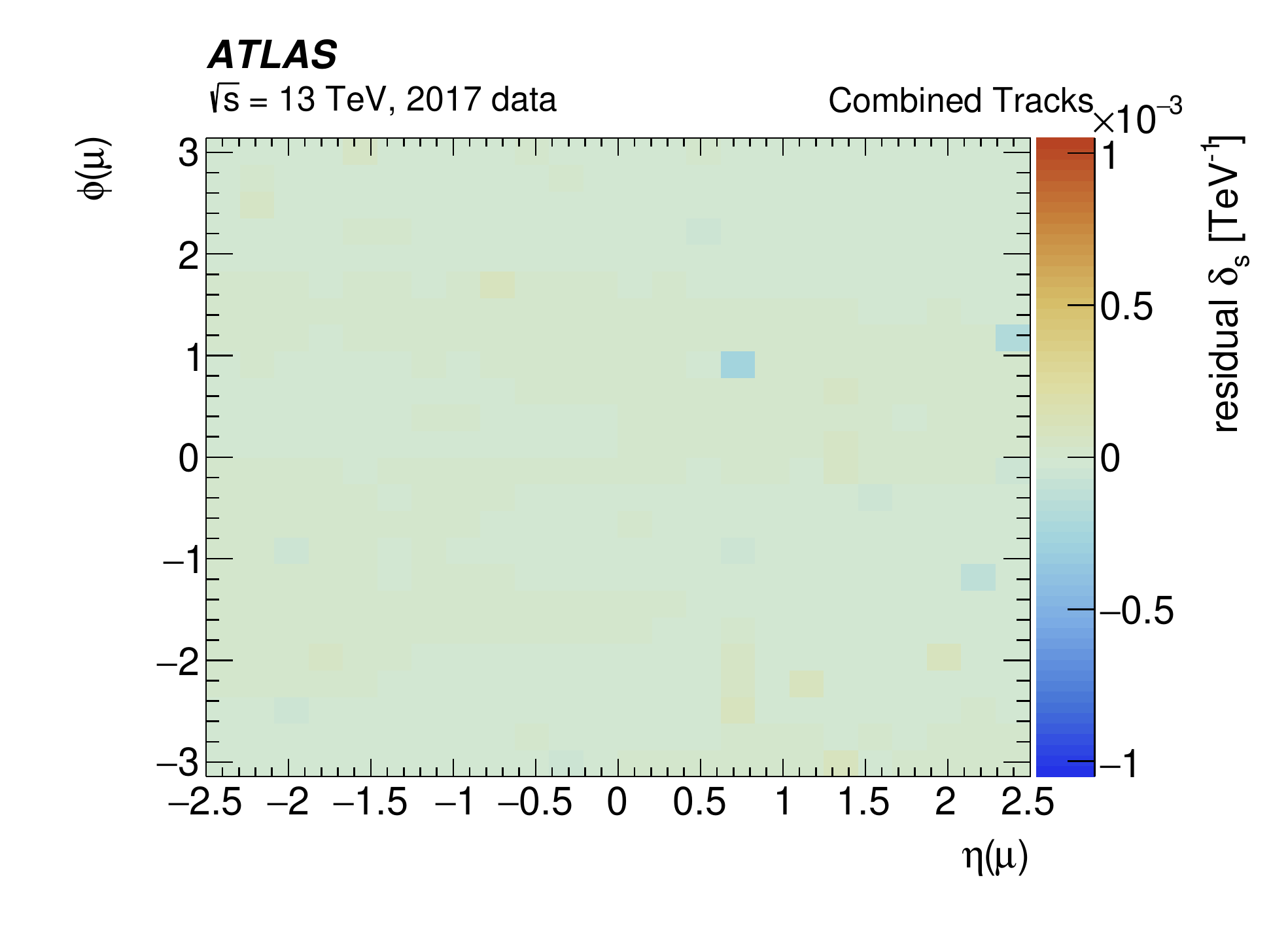}} \\
\subfloat[ ]{\includegraphics[width=0.45\linewidth]{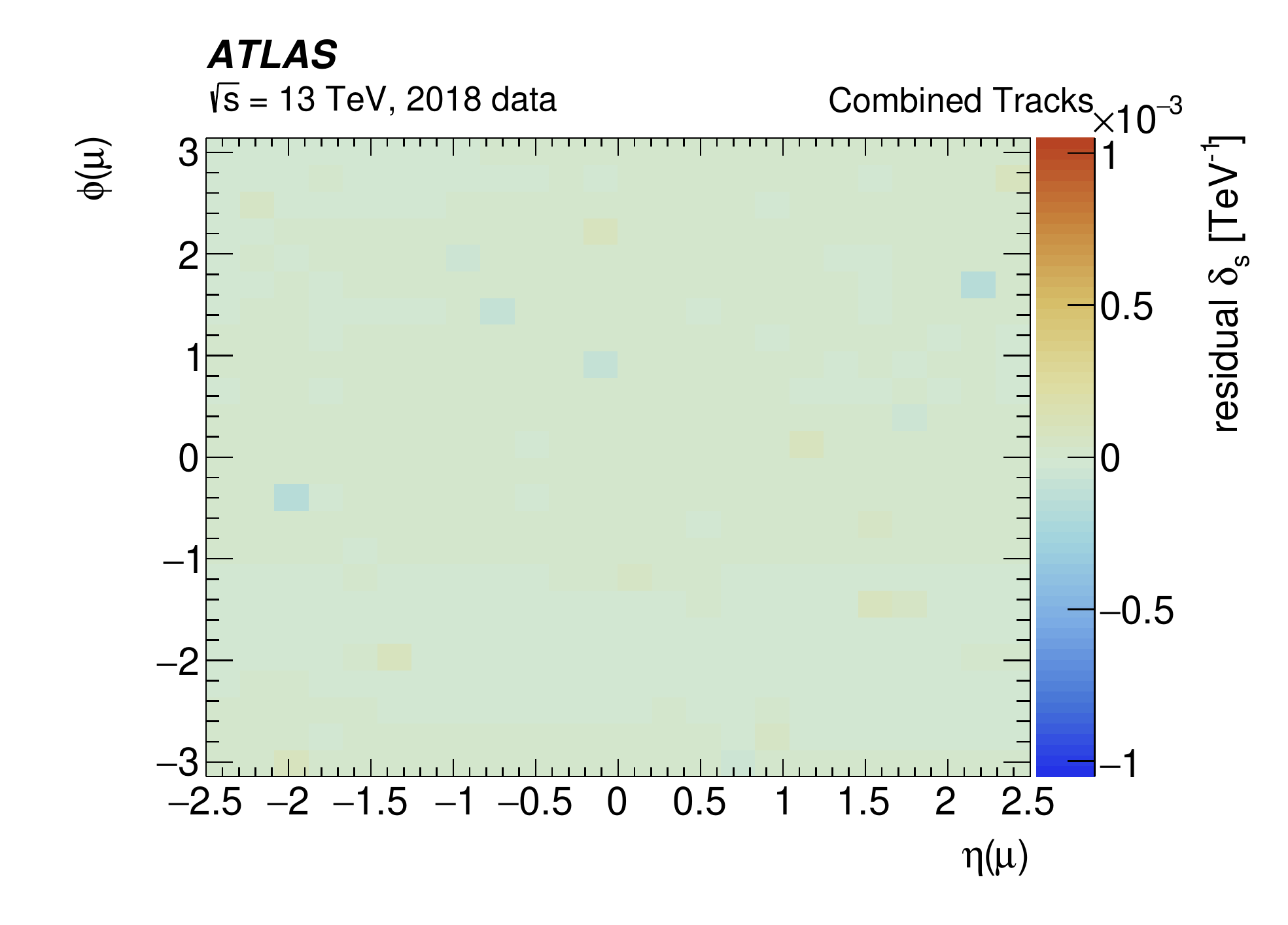}}
\caption{Residual charge-dependent biases on the muon \pt\ as evaluated on data for CB momenta after alignment and after application of the correction procedure. The residual biases are shown separately for the three data-taking periods, (a) years 2015 and 2016, (b) 2017, and (c) 2018. The bias is defined in Eq.~(\ref{eq:bias_definition}).}
\label{fig:sgAfterCorr}
\end{figure*}

\begin{figure*}[!htb]
\centering
\subfloat[ ]{\includegraphics[width=0.45\linewidth]{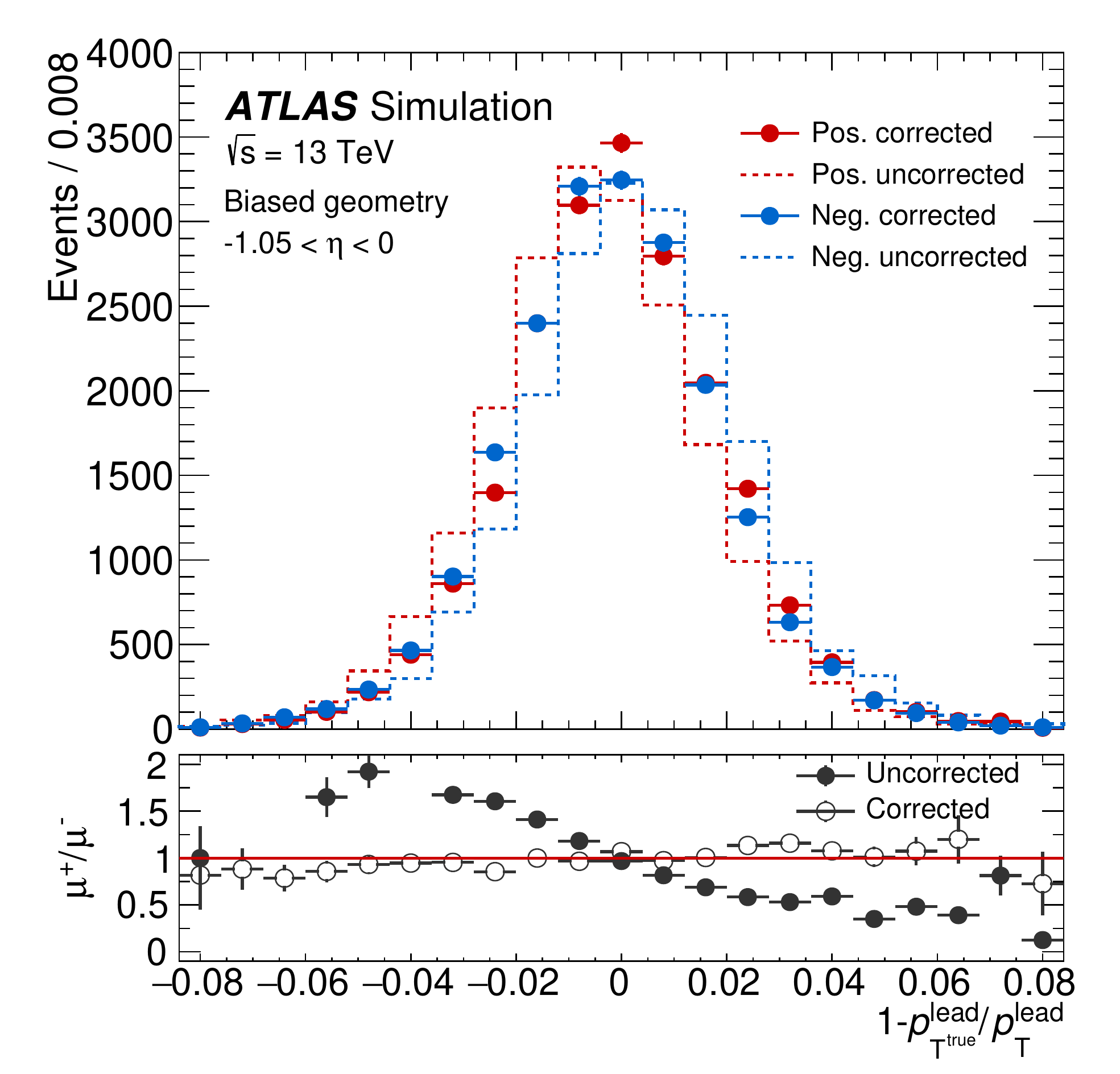}}
\subfloat[ ]{\includegraphics[width=0.45\linewidth]{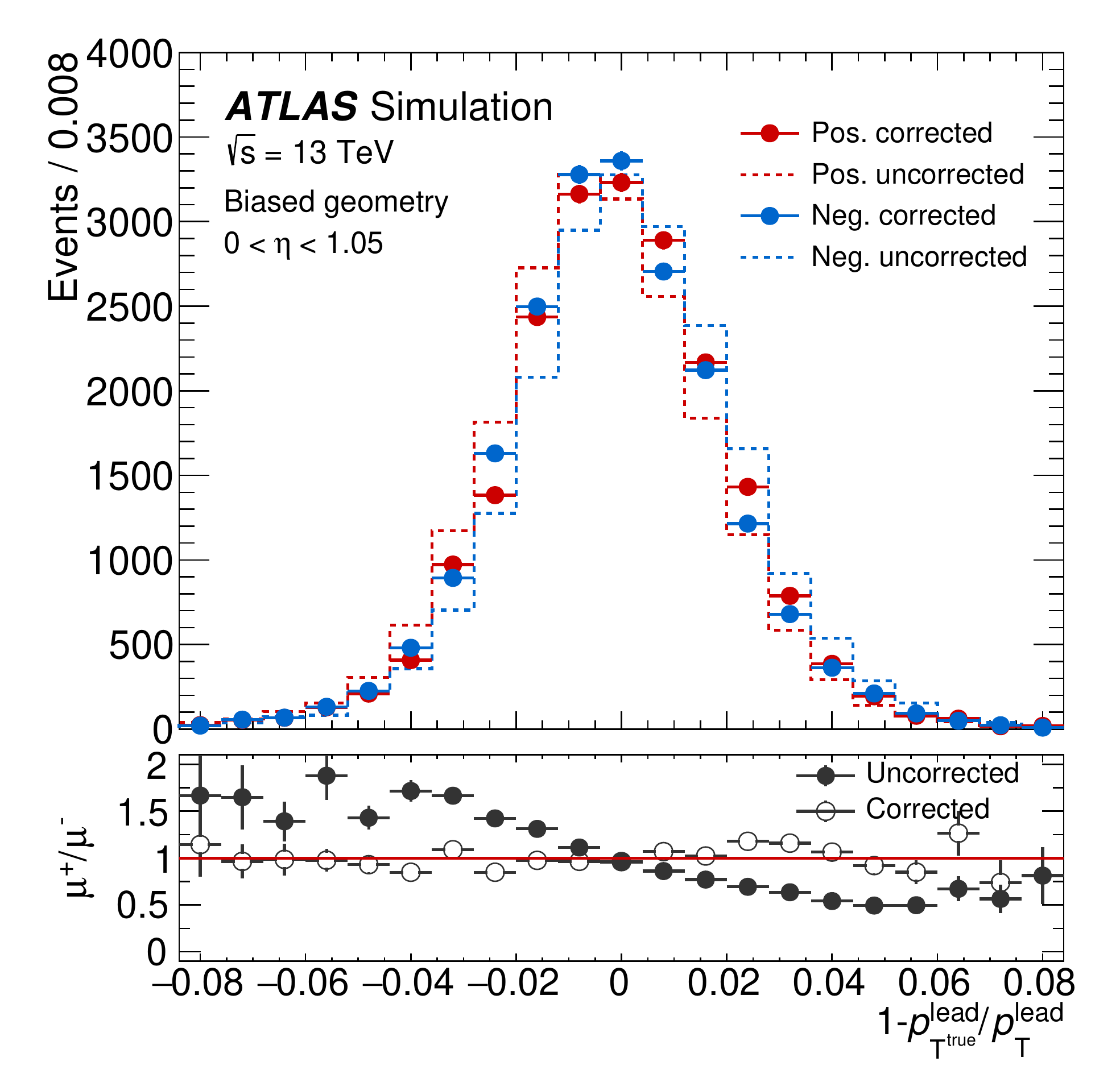}}
\caption{The bias of the transverse momentum measured in simulated data. The upper panels show the leading muon in the event ($\pt^\textrm{lead}$) compared with its true value separately for positive muons and negative muons, (a) for $-1.05<\eta<0$ and (b) for $0<\eta<1.05$. The dotted lines correspond to the momentum bias as evaluated on a simulation with a biased geometry, for positive and negative muons. The markers correspond to the bias evaluated with the same simulation after the application of the correction procedures. The lower panels show the ratios of positive to negative muons before and after correction.}
\label{fig:truth}
\end{figure*}
 
Several sources of systematic uncertainty associated with the correction procedure are studied. They relate to both the accuracy of the measurement of $\delta_{s}$ and the assumptions used in the correction procedure.
 
The systematic uncertainty originating from the residual non-closure of the correction compared with a perfectly aligned detector simulation is evaluated as follows.
First, the $\delta_{s}$ value obtained after the last iteration on data is used to bias the $\pt$ in a perfectly aligned simulation. This is done using \refE{eq:sgcorr}, by flipping the sign of $\delta_{s}$, and using the same granularity in $\eta$ and $\phi$ as used for deriving the correction. The resulting $\pt$ is compared with the unbiased momentum and the difference is taken as a systematic uncertainty. A second component of the residual non-closure is estimated in simulation by injecting a set of constant known biases and by taking the difference of $\delta_{s}$ between the injected and measured $\delta_{s}$. The injected biases range from percent fractions up to twice those observed in the data at the first iteration of the correction procedure; the largest resulting $\delta_{s}$ difference is taken as the uncertainty for all bins.
Both methods show very good closure for the full range of injected biases.
The uncertainties originating from the non-closures amount to about 0.02~$\TeV{}^{-1}$, depending on the pseudorapidity region.
 
A second contribution to the systematic uncertainties arises from extrapolations to momenta outside the phase-space region covered by $Z$ decays. In fact, the relative contribution of the ID and MS detectors on the CB momentum measurement may vary in a non-trivial way. This may lead to biases on the estimate of $\delta_{s}$ and of the correction. In this case, the residual bias is evaluated as the difference between simulation and data and as a function of $\eta$ and $\pt$ using the alternative method of \refE{eq:detaqM}. The difference compared with the simulation is used to account for the kinematic dependencies of the $\delta_{s}$ estimate.  The resulting uncertainty is small, compared with the non-closure described previously, in the central regions of pseudorapidity. However, in the forward pseudorapidity regions, it becomes sizeable and evolves as $2~\times10^{-4}~\TeV{}^{-2}~\times (\pt\ - 0.045~\TeV)$.
 
The statistical uncertainties due to the observed and simulated number of $Z$ decays are also taken into account. These are approximately an order of magnitude smaller than the first two components, but in the forward region they reach up to 0.02~$\TeV^{-1}$.
 
Given the lack of sufficient $\Zmm$ decays above $450~\GeV$ in data to validate the methodology, the data are not corrected for charge-dependent biases beyond this value. Instead, an uncertainty   accounting for the charge-dependent biases before any correction is assigned to the simulated momenta. This is estimated by biasing the simulated muon momenta in a perfectly aligned simulation, according to \refE{eq:sgcorr}, using the values of $\delta_{s}$ as measured before any correction. 
As done for the non-closure uncertainty, the uncertainty is derived by injecting bias into simulated events and taking the difference between the biased and the non-biased muon momentum values, resulting in an uncertainty of up to 0.4~$\TeV^{-1}$ as shown in \refF{fig:sgBefore}.


\subsection{Muon momentum calibration procedure in simulation}\label{Sec:CorrectionExtraction}

In the following, the \textit{muon momentum calibration} is defined as the procedure used to identify
the corrections to the reconstructed muon transverse momenta in simulation in order to match the measurement of the same quantities in data. This procedure is performed after correcting for the charge-dependent bias discussed in the previous section.
The transverse momenta of the ID and MS tracks associated with a CB muon, referred to as $\pt^{\textrm{ID}}$ and $\pt^{\textrm{MS}}$, respectively, are used in addition to the transverse momentum of the CB track, referred to as $\pt^{\textrm{CB}}$.
 
The corrected transverse momentum after applying the calibration procedure, $\pt^{\textrm{Cor,Det}}$  (Det$=$  CB, ID, MS), is described as:
\begin{eqnarray}\label{eq:pt_cor}
\pt^{\textrm{Cor,Det}} &=&  \frac{\pt^{\textrm{MC,Det}} + \sum\limits^{1}_{n=0} s^\mathrm{Det}_{n}(\eta,\phi)\left(\pt^{\textrm{MC,Det}}\right)^{n}}{1+\sum\limits^{2}_{m=0} \Delta r^\mathrm{Det}_{m}(\eta,\phi)\left(\pt^{\textrm{MC,Det}}\right)^{m-1}g_{m}},
\end{eqnarray}
where
$\pt^{\textrm{MC,Det}}$ is the uncorrected transverse momentum in simulation,
$g_{m}$ are normally distributed random variables
with zero mean and unit width,
and the terms  $\Delta r_{m}^{\textrm{Det}}(\eta,\phi)$ and $s_n^{\textrm{Det}}(\eta, \phi)$  describe the momentum resolution smearing and scale corrections respectively, applied in a specific ($\eta$, $\phi$) detector region. A possible $s_2^{\textrm{Det}}(\eta, \phi)$ term in the numerator is neglected because it would model effects already corrected in data with the procedure described in \refS{Sec:SagittaBias}.
 
The corrections described in \refE{eq:pt_cor} are defined in $\eta$--$\phi$ detector regions homogeneous in detector technology and performance.
All corrections are  divided into 18 pseudorapidity regions.
In addition, the CB and MS corrections are divided into two $\phi$ bins separating the two types of MS sectors: those that include the magnet coils ({\em small sectors}) and those between two coils ({\em large sectors}). The small and large MS sectors are subjected to independent alignment techniques and cover detector areas with different material distribution, leading to scale and resolution differences.
 
The numerator of \refE{eq:pt_cor} describes the momentum scales.
The $s_1^\mathrm{Det}(\eta, \phi)$ term corrects for inaccuracy in the description of the magnetic field integral and the dimension of the detector in the direction perpendicular to the magnetic field.
The $s^\mathrm{Det}_0(\eta, \phi)$ term models the effect on the CB and MS momentum from the inaccuracy in the simulation of the energy loss
in the calorimeter and other material between the interaction point and the exit of the MS.
As the energy loss between the interaction point and the ID is negligible,  $s^\mathrm{ID}_0(\eta)$ is set to zero~\cite{PERF-2015-10}.
 
The denominator of \refE{eq:pt_cor} describes the momentum smearing that broadens the relative \pt\ resolution in simulation,
$\sigma(\pt)/ \pt$,
to properly describe the data.
The corrections to the resolution assume that the relative \pt\ resolution can be parameterized as follows:
\begin{equation}\label{eq:qoverpt_reso}
\frac{ \sigma(\pt)}{\pt} = r_0/\pt \oplus r_1 \oplus r_2\cdot \pt\textrm{,}
\end{equation}
 
with $\oplus$ denoting a sum in quadrature. In \refE{eq:qoverpt_reso}, the first term ($r_0$)
mainly accounts for fluctuations of the energy loss in the traversed material;
the second term ($r_1$) mainly accounts for multiple scattering, uncertainties related to, and inhomogeneities in, the modelling of the local magnetic field, and length-scale radial expansions of the detector layers; and the third term ($r_2$) mainly describes intrinsic resolution effects caused by the spatial resolution of the hit measurements and residual misalignments between the different detector elements.
The energy loss term has a negligible impact on the muon resolution in the momentum range considered, and therefore $\Delta r^\mathrm{Det}_{0}$ is set to zero.
 
In a second step, to cross-check the validity of the corrections obtained directly for the CB tracks, the corrected combined momenta from ID and MS measurements $\pt^{\textrm{Corr ID+MS}}$ is also obtained by combining the ID and MS corrected momenta with a weighted average:
\begin{equation} 
\pt^{\textrm{Cor,ID+MS}} = f\cdot \pt^{\textrm{Cor,ID}} + (1 - f)\cdot \pt^{\textrm{Cor, MS}}\textrm{.}
\end{equation}
The weight $f$ is calculated by solving the following linear equation:
\begin{equation} 
\pt^{\textrm{MC,CB}} = f\cdot \pt^{\textrm{MC,ID}} + (1 - f)\cdot \pt^{\textrm{MC, MS}}\textrm{,}
\end{equation}
which assumes that the relative contribution of the two sub-detectors to the combined track remains unchanged before and after momentum corrections.
 
\subsubsection{Determination of the \pt\ calibration constants}
 
The CB, ID, and MS correction parameters contained in \refE{eq:pt_cor}
are extracted from data
using a fitting procedure that compares the invariant mass distributions for $\Jmm$ and $\Zmm$ candidates in data and simulation, selected as discussed in \refS{sec:MuonReco}.
 
When extracting correction parameters, muons are assigned to $\eta$--$\phi$ regions of fit (ROFs), which are defined separately for the ID and the MS.
The values of $\Delta r^\mathrm{ID}_{0}$, $ \Delta r^\mathrm{MS}_{0}$, $ \Delta r^\mathrm{CB}_{0}$, and $s^\mathrm{ID}_0$,
are set to zero as previously discussed, while $ \Delta r^\mathrm{MS}_{2}$ is determined from alignment studies using cosmic-ray data and
special runs with the toroidal magnetic field turned off~\cite{ATL-MUON-PUB-2021-002}, and from detector calibration procedures~\cite{MDET-2018-01}.
 
The corrections are extracted using the distributions of the dimuon invariant mass, $m_{\mu\mu}^\mathrm{Det}$.
When fitting a specific ROF, one muon is required to belong to the ROF and the other can be anywhere in the detector.
To enhance the sensitivity to  $\pt$-dependent correction effects, the dimuon pair is classified according to the $\pt$ of the muons.
For $\Jmm$ decays, the fit is performed in two exclusive categories defined by the subleading muon transverse momentum: $6.25 < \pt^{\textrm{Det,sublead}} < 9~\GeV$ and $\pt^{\textrm{Det,sublead}} > 9~\GeV$.
For $\Zmm$ decays the two categories are defined by the leading muon transverse momentum: $ 20< \pt^{\textrm{Det,lead}} < 50~\GeV$ and $\pt^{\textrm{Det,lead}}> 50~\GeV$.  
 
Templates for the  $m_{\mu\mu}^{\textrm{Det}}$ variables are built using $\Jmm$ and  $\Zmm$ simulated signal samples.
The first step in the fitting procedure consists of estimating the contribution of the background processes.
For the $\Zmm$ selection, the small background component (approximately 0.1\%) is extracted from simulation and added to the templates.
A much larger (up to 15\%) non-resonant background from decays of light and heavy hadrons and from continuum Drell--Yan production is present for events selected in the $\Jmm$ channel, estimated with a data-driven approach as this background is difficult to simulate.
The dimuon invariant mass distribution in data is fitted in each ROF using
a Crystal Ball function~\cite{Crystal} with an exponential background distribution. This background model is then combined with the simulated signal templates used in the fit.
 
To estimate the scale and smearing parameters, a multi-stage procedure is used. A two-step random walk minimisation is first performed where the parameters are sampled within a specified range. The parameter configuration that leads to the best agreement for the invariant mass distribution between simulation and data is kept. In the first step, the best agreement is defined by a simplified metric where the compatibility between the means and standard deviations of the distributions is compared. In the second step, a binned $\chi^2$ is computed that compares the dimuon mass spectrum in simulation to that of data. Around this best-agreement parameter configuration, signal templates are then generated at discrete intervals and interpolated using moment morphing~\cite{Baak:2014fta}. Using this signal model, a full minimisation using the binned $\chi^2$ is performed. This procedure is iteratively repeated, where the second muon outside the ROF definition is calibrated using results from the previous iteration, until the change of each parameter stays within the root mean squared of its values as estimated from the last five iterations. This happens after 16 iterations. Each calibration parameter's values from the last five iterations are averaged to produce their final value.
 
The calibration parameters obtained from the fits to the data are summarised in Tables~\ref{tab:id_cor}, ~\ref{tab:ms_cor}, and ~\ref{tab:cb_cor}, averaged over three $\eta$ regions. The sums in quadrature of the statistical and systematic uncertainties are shown, with the latter dominating. All sources of uncertainties are evaluated by varying the parameters of the template fit.
The higher uncertainty for $\Delta r_1^\mathrm{ID}$ in the second bin is associated with the region of the ID with the largest amount of material, corresponding to the transition between the barrel and endcap region of its subdetectors.
The increase in uncertainty as a function of $|\eta|$ for $s_1^\mathrm{ID}$ is due to the decrease of the magnetic field integral of the solenoid as a function of $|\eta|$.
The larger uncertainties for some of the MS terms in the $1.05\le |\eta|<2.0$ region are associated with the low bending power of the magnet system of the MS in part of that region.
 
The main contributions to the final systematic uncertainty are:
 
\begin{itemize}
\item $J/\psi$ only and $Z$ only fit: The fit is repeated using only the $\Jmm$ or $\Zmm$ decays with only $s_1^\mathrm{Det}$ left as a free parameter and the other parameters fixed to their nominal fitted value. The resulting uncertainty is defined by taking the difference relative to the nominal fitted $s_1^\mathrm{Det}$ and accounts for the extrapolation from the regions dominated by either the $\Jmm$ or the $\Zmm$ in their respective \pt\ spectra.
\item $Z$ kinematic reweighting: As discussed in \refS{sec:simulation}, the reweighting maps are derived separately for each simulated sample. To evaluate the impact of the reweighting on the analysis, new templates of $m_{\mu\mu}^\mathrm{ID}$, $m_{\mu\mu}^\mathrm{CB}$, and $m_{\mu\mu}^\mathrm{MS}$ are produced. The results obtained when using the different $\Zmm$ samples with their respective reweighting scheme and the non-reweighted results for the nominal $\Zmm$ samples are used to derive the uncertainty. This systematic uncertainty impacts the overall $Z$ lineshape by changing the relative weight of various momentum regimes and, by extension, the contributions of their dedicated corrections in the final fit.
\item Decay modelling and final state radiation modelling: The nominal simulated sample of $\Zmm$ events, of which the QED final-state radiation is handled by \PHOTOSpp{}, is compared with $\Zmm$ events simulated with the \SHERPA[2.2.1] sample which uses an alternative radiative modelling. The impact of the uncertainties in the decay modelling are significantly smaller than those originating from the components discussed above, partially due to the already discussed kinematic reweighting. 
\item $J/\psi$ and $Z$ \pt template range: The \pt ranges of the $J/\psi$ and $Z$ templates are varied. For $J/\psi$, the ranges of the two nominal \pt\ templates are $\pt^{J/\psi, \textrm{nom}} \in [6.25, 9]~\GeV$ and $\pt^{J/\psi, \textrm{nom}} \in [9, 20]~\GeV$. For $Z$, the ranges of two templates are $\pt^{Z, \textrm{nom}} \in~[20, 50]~\GeV$ and $\pt^{Z, \textrm{nom}} \in [50, 300]~\GeV$. The boundary of the two templates for $J/\psi$ is varied from 9 to 8 and 12~\GeV, and for $Z$ is varied from 50 to 40 and 80~\GeV. This preserves the numbers of events, and covers any  additional \pt dependencies.
\item Mass binning of $J/\psi$ and $Z$: The number of bins for the $m_{\mu \mu}$ templates are reduced from 200 to 150 for $Z$, and from 90 to 60 for $J/\psi$. This systematic uncertainty covers any binning effect in generating templates.
\item $Z$ mass window: New templates of $m_{\mu \mu}$ are generated for the $\Zmm$ sample changing the $m_{Z}$ window selection to $75 < m_{\mu\mu} < 115~\GeV$ for all track types. The high and low mass regions away from the resonant pole of the $Z$ are more sensitive to initial- and final-state radiation contributions, the running of $\alpha^\text{EM}_{Z}$, and other specific choices of the MC generators used in the $\Zmm$ simulation.
\item $J/\psi$ mass window: The same approach as in the case of the $Z$ is applied, changing the invariant mass window selection to $2.75 < m_{\mu\mu} < 3.4~\GeV$ for all track types. The region furthest from the $m_{J/\psi}$ peak is more sensitive to the shape variation of the combinatorial background, and the specific choices of the MC generators used in the $\Jmm$ simulation.
\item $J/\psi$ background: New $m_{\mu \mu}$ templates are generated for the $\Jmm$ samples using a background parameterisation different from the exponential function. For this systematic uncertainty, background parameterisations including a combination of two Crystal ball functions and a Chebyshev function are used.
\item Closure test and statistical uncertainty: The root mean squared of the last five iterations of each calibration parameter is used as the statistical uncertainty of the method. This accounts for any instabilities in the fit, and the correlations between different parameters. It was observed that small changes in the starting parameters of the fit do not affect its stability.
\end{itemize}
 
The largest component of the total uncertainty in the momentum scale originates from the comparison of the $\Jpsi$ and $Z$ only fits. The remaining components are smaller by about a factor of two.
For the resolution smearing terms, in the same kinematic regime, the largest uncertainty comes from the $\Jpsi$ and $Z$ \pt\ template fit range.
At significantly higher momenta the uncertainty in the $\Jpsi$ and $Z$ \pt\ template fit range still dominates the resolution uncertainty in $\pt^{\text{Cor,CB}}$.

\begin{table*}[t]
\caption{Summary of ID muon momentum resolution and scale corrections used in \refE{eq:pt_cor},
averaged over the three main detector regions.
The corrections are derived in 18 pseudorapidity regions, as described in \refS{sec:Resolution},
and averaged, with each region assigned a weight proportional to its  $\eta$ width.
The uncertainties represent the sum in quadrature of the statistical and systematic uncertainties.}
\label{tab:id_cor}
\centering
\begin{tabular}{c r@{}l r@{}l r@{}l}
\toprule
Region & \multicolumn{2}{c}{$\Delta r_1^{\textrm{ID}}(\times 10^{-3})$} & \multicolumn{2}{c}{$\Delta r_2^{\textrm{ID}}$ [\TeV$^{-1}$]} & \multicolumn{2}{c}{$s_1^{\textrm{ID}}(\times 10^{-3})$}\\
\midrule
 
$|\eta|<1.05$         & $4.4$ & $\phantom{}^{+3.0}_{-2.7}$  & $0.12$ & $\phantom{}^{+0.03}_{-0.03}$ & $-0.82$ & $\phantom{}^{+0.19}_{-0.06}$  \\[2mm]
$1.05\le |\eta|<2.0$  & $6.7$ & $\phantom{}^{+19.5}_{-3.1}$ & $0.31$ & $\phantom{}^{+0.03}_{-0.03}$ & $-0.86$ & $\phantom{}^{+0.36}_{-0.21}$  \\[2mm]
$|\eta|\ge 2.0$       & $9.4$ & $\phantom{}^{+3.6}_{-5.3}$  & $0.08$ & $\phantom{}^{+0.02}_{-0.02}$ & $-0.45$ & $\phantom{}^{+1.19}_{-0.55}$  \\
 
\bottomrule
\end{tabular}
\end{table*}
 
\begin{table*}[t]
\caption{Summary of MS momentum resolution and scale corrections for small and large MS sectors, averaged over three main detector regions.
The corrections for large and small MS sectors are derived in 18 pseudorapidity  regions, as described in \refS{sec:Resolution},
and averaged, with each region assigned a weight proportional to its  $\eta$ width.
The uncertainties represent the sum in quadrature of the statistical and systematic uncertainties. The larger uncertainties for some terms in the $1.05\le |\eta|<2.0$ region are associated with the low bending power of the magnet system of the MS in part of that region.}
\label{tab:ms_cor}
\centering
\begin{tabular}{cr@{}l r@{}l r@{}l r@{}l r@{}l}
\toprule
Region  & \multicolumn{2}{c}{$\Delta r_0^{\textrm{MS}}$ [\MeV]} & \multicolumn{2}{c}{$\Delta r_1^{\textrm{MS}} (\times 10^{-3})$} & \multicolumn{2}{c}{$\Delta r_2^{\textrm{MS}}$ [\TeV$^{-1}$]}  & \multicolumn{2}{c}{$s_0^{\textrm{MS}}$ [\MeV]} & \multicolumn{2}{c}{$s_1^{\textrm{MS}}  (\times 10^{-3})$}\\
\midrule
 
$|\eta|<1.05$ (large)           & $71$ & $\phantom{}^{+41}_{-10}$  & $13.5$ & $\phantom{}^{+3.1}_{-2.6}$  & $0.11$ & $\phantom{}^{+0.02}_{-0.02}$ & $-27$ & $\phantom{}^{+18}_{-14}$ & $2.33$ & $\phantom{}^{+0.61}_{-0.61}$  \\[2mm]
$|\eta|<1.05$ (small)           & $63$ & $\phantom{}^{+35}_{-23}$  & $17.2$ & $\phantom{}^{+2.3}_{-3.6}$  & $0.12$ & $\phantom{}^{+0.02}_{-0.02}$ & $-18$ & $\phantom{}^{+14}_{-24}$ & $-1.51$ & $\phantom{}^{+1.15}_{-0.45}$  \\[2mm]
$1.05\le |\eta|<2.0$  (large)   & $42$ & $\phantom{}^{+401}_{-53}$ & $26.7$ & $\phantom{}^{+19.5}_{-2.8}$ & $0.14$ & $\phantom{}^{+0.02}_{-0.02}$ & $-29$ & $\phantom{}^{+43}_{-57}$ & $-0.9$ & $\phantom{}^{+2.8}_{-1.9}$  \\[2mm]
$1.05\le |\eta|<2.0$  (small)   & $44$ & $\phantom{}^{+437}_{-59}$ & $29.2$ & $\phantom{}^{+8.2}_{-1.5}$  & $0.18$ & $\phantom{}^{+0.02}_{-0.02}$ & $-28$ & $\phantom{}^{+39}_{-43}$ & $-1.3$ & $\phantom{}^{+2.0}_{-1.6}$  \\[2mm]
$|\eta|\ge 2.0$ (large)         & $61$ & $\phantom{}^{+48}_{-27}$  & $16.8$ & $\phantom{}^{+3.2}_{-4.3}$  & $0.11$ & $\phantom{}^{+0.02}_{-0.02}$ & $-1$  & $\phantom{}^{+43}_{-29}$ & $1.7$ & $\phantom{}^{+1.3}_{-1.4}$  \\[2mm]
$|\eta|\ge 2.0$ (small)         & $47$ & $\phantom{}^{+32}_{-40}$  & $17.8$ & $\phantom{}^{+1}_{-1.9}$    & $0.16$ & $\phantom{}^{+0.02}_{-0.02}$ & $-8$  & $\phantom{}^{+14}_{-12}$ & $1.94$ & $\phantom{}^{+0.76}_{-0.54}$  \\
 
\bottomrule
\end{tabular}
\end{table*}

\begin{table*}[t]
\caption{Summary of CB momentum resolution and scale corrections for small and large MS sectors, averaged over three main detector regions.
The corrections for large and small MS sectors are derived in 18 pseudorapidity  regions, as described in \refS{sec:Resolution},
and averaged, with each region assigned a weight proportional to its  $\eta$ width.
The energy loss term $\Delta r_0^{\textrm{CB}}$ is negligible and therefore fixed to zero in the fit for all $\eta$.
The uncertainties represent the sum in quadrature of the statistical and systematic uncertainties.
}
\label{tab:cb_cor}
\centering
\begin{tabular}{cr@{}l r@{}l r@{}l r@{}l}
\toprule
Region  & \multicolumn{2}{c}{$\Delta r_1^{\textrm{CB}} (\times 10^{-3})$} & \multicolumn{2}{c}{$\Delta r_2^{\textrm{CB}}$ [\TeV$^{-1}$]}  & \multicolumn{2}{c}{$s_0^{\textrm{CB}}$ [\MeV]} & \multicolumn{2}{c}{$s_1^{\textrm{CB}}  (\times 10^{-3})$}\\
\midrule
$|\eta|<1.05$ (large)         & $6.7$  & $\phantom{}^{+1.4}_{-0.9}$ & $0.08$ & $\phantom{}^{+0.04}_{-0.05}$ & $-5.0$ & $\phantom{}^{+2.9}_{-4.0}$ & $0.35$ & $\phantom{}^{+0.24}_{-0.22}$  \\[2mm]
$|\eta|<1.05$ (small)         & $6.5$  & $\phantom{}^{+1.3}_{-1.0}$ & $0.11$ & $\phantom{}^{+0.05}_{-0.05}$ & $-0.9$ & $\phantom{}^{+2.5}_{-3.6}$ & $-0.83$ & $\phantom{}^{+0.25}_{-0.14}$  \\[2mm]
$1.05\le |\eta|<2.0$  (large) & $10.3$ & $\phantom{}^{+2.6}_{-2.7}$ & $0.24$ & $\phantom{}^{+0.10}_{-0.07}$ & $-2.0$ & $\phantom{}^{+5.7}_{-6.7}$ & $-0.83$ & $\phantom{}^{+0.39}_{-0.30}$  \\[2mm]
$1.05\le |\eta|<2.0$  (small) & $8.9$  & $\phantom{}^{+1.7}_{-2.7}$ & $0.29$ & $\phantom{}^{+0.08}_{-0.03}$ & $-3.0$ & $\phantom{}^{+3.3}_{-4.0}$ & $-0.80$ & $\phantom{}^{+0.26}_{-0.21}$  \\[2mm]
$|\eta|\ge 2.0$ (large)       & $10.6$ & $\phantom{}^{+2.2}_{-2.7}$ & $0.21$ & $\phantom{}^{+0.10}_{-0.07}$ &  $2.3$ & $\phantom{}^{+13}_{-9.3}$   & $0.80$ & $\phantom{}^{+1.09}_{-0.42}$  \\[2mm]
$|\eta|\ge 2.0$ (small)       & $11.5$ & $\phantom{}^{+2.2}_{-2.1}$ & $0.26$ & $\phantom{}^{+0.08}_{-0.06}$ & $-12.6$ & $\phantom{}^{+8.2}_{-9.7}$ & $1.59$ & $\phantom{}^{+0.47}_{-0.43}$  \\
 
\bottomrule
\end{tabular}
\end{table*}
 
As an additional check, the value of  $\pt^{\textrm{Cor,ID+MS}}$ is compared with $\pt^{\textrm{Cor,CB}}$. Results from both methods are in agreement within their respective systematic uncertainties. The comparison is performed on simulated samples with a biased alignment of the MS on muons with momenta ranging from few tens of \GeV\ to about 1 \TeV. The uncertainties in the momentum scale of $\pt^{\textrm{Cor,CB}}$ are found to be smaller than those of $\pt^{\textrm{Cor, ID+MS}}$ with a slightly larger uncertainty increase in the resolution as a function of the transverse momentum, as shown in \refA{app:results}. At high momenta the dominant uncertainty in $\pt^{\text{Cor,ID+MS}}$ originates from the residual muon spectrometer misalignment.
 
\begin{figure*}[!t]
\centering
\subfloat[ ]{\includegraphics[width=0.45\linewidth]{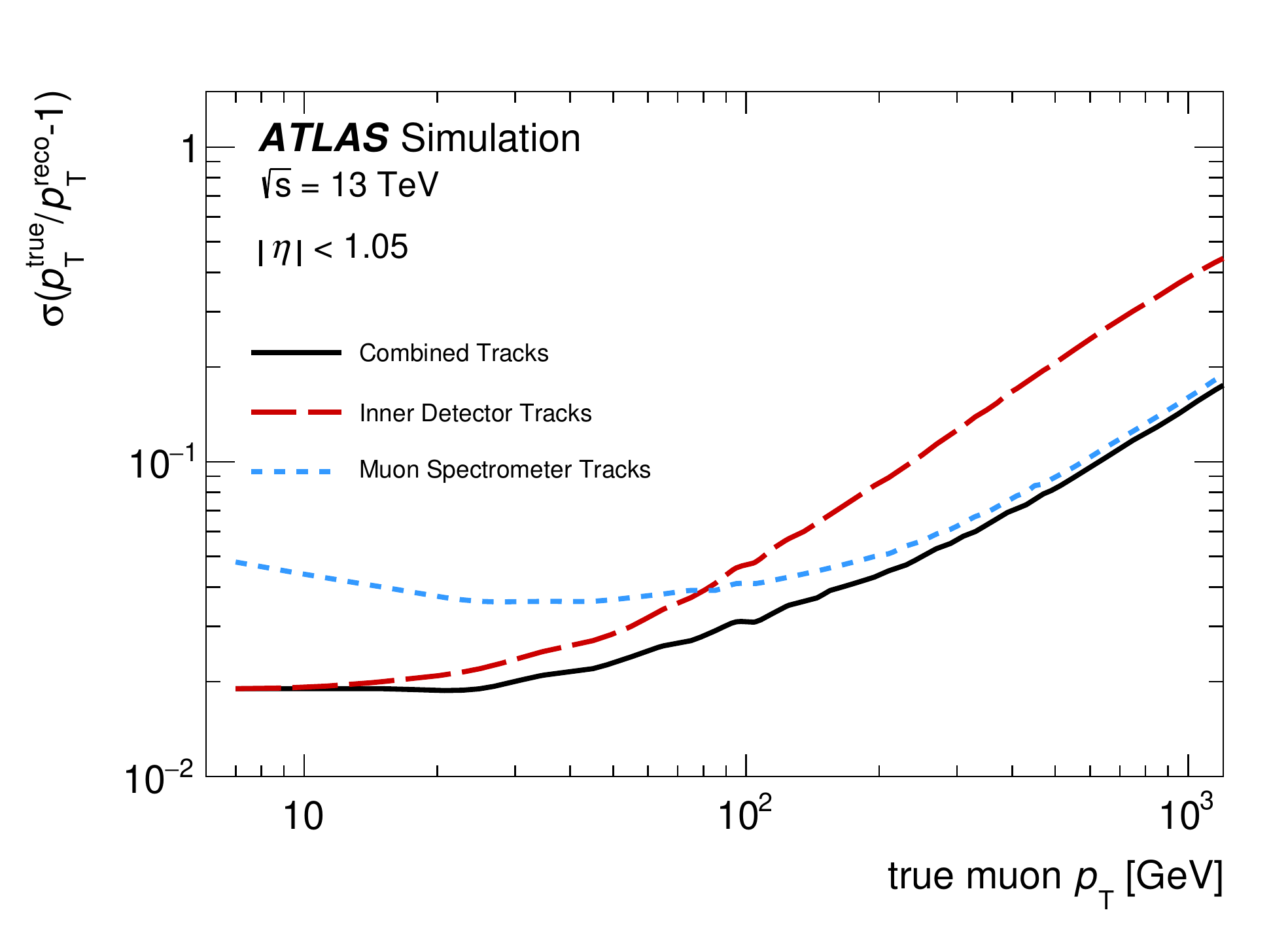}}
\subfloat[ ]{\includegraphics[width=0.45\linewidth]{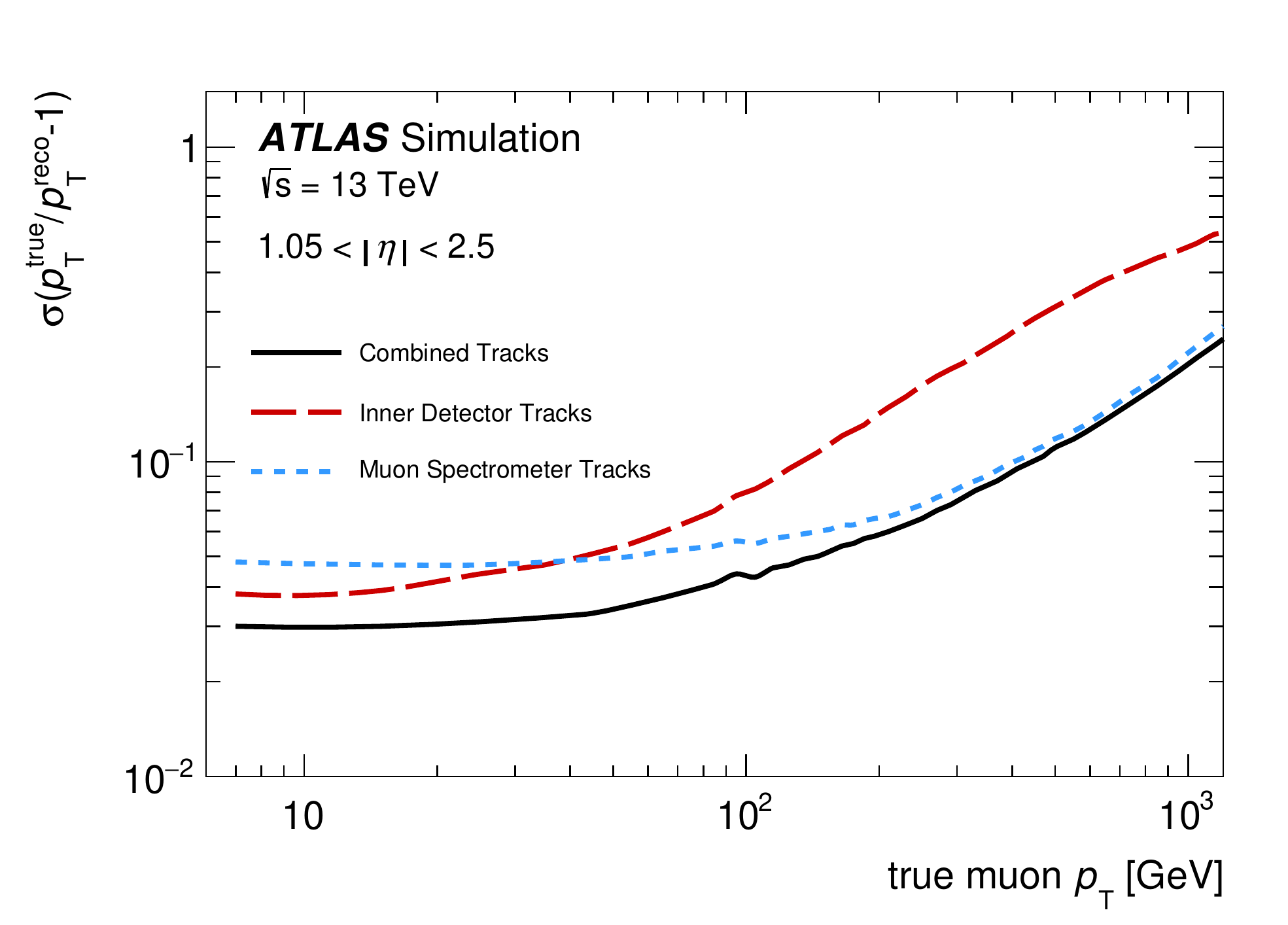}}
\caption{Resolution of the muon \pt\ as obtained from simulation after derivation and application of all correction constants. Muons are selected using the \highptwp{} WP\@. The resolution is shown as a function of the true \pt\ of the muon for a range from 1 \gev\ to 2.5 \tev, (a) for muons with $|\eta|<1.05$ and (b) for muons with $|\eta|>1.05$. The resolution lines are obtained by interpolating between points sampled in steps of $\pt$. The continuous lines correspond to the CB momentum, while the dashed lines to that from the ID, and the dotted lines to that of the MS\@.  Statistical fluctuations, due to the simulated sample size result in small fluctuations on the interpolation between measured values.}
\label{fig:MCresol}
\end{figure*}

The resolution as a function of the muon momentum is shown in \refF{fig:MCresol}, (a) for muons in the barrel region ($|\eta|<1.05$) and (b) for all other muons ($|\eta|>1.05$). The resolution is estimated by taking half of the smallest interval containing the 68\% of the distribution of the relative difference between the corrected and the generated single muon momentum in simulation for muons satisfying the \highptwp{} WP\@. The resolution is approximately 2\% (3\%) at \pt\ values of 45~\GeV{} in the $|\eta|<1.05$ ($|\eta|>1.05$) region and increases with \pt.
The resulting CB track momentum resolution is always better than that of the individual measurements in the ID or the MS\@.
At low \pt\, the ID dominates while in the intermediate regime of few tens of \gev\ to a few hundreds of \gev\ both detector systems contribute equally to the measurement. At \pt\ values higher than few hundred \gev\, the resolution is dominated by the MS\@. The resolution is approximately 14\% (19\%) at 1~\TeV{} in the $|\eta|<1.05$ ($|\eta|>1.05$) region.



\section{Performance studies}
\label{sec::validation}

The samples of $\Jmm$, $\Upsmm$, and $\Zmm$ decays are used to validate the momentum corrections obtained with the template fit method described in the previous sections and measure the muon momentum reconstruction performance. The detector segmentation applied  to $\Jmm$ and $\Zmm$ decays is chosen to be at least twice as fine as that used for deriving the simulation corrections. Furthermore, the momentum requirements used in both cases are looser than those used for deriving the corrections. The combination of the finer segmentation and looser requirements allows the data-driven validation of the correction methods both within the bins assumed in the template fits and extrapolated in \pt\ beyond the range of the calibration procedure. To complete the data-driven validation and performance measurements, an additional study is performed using $\Upsmm$ decays, which are statistically fully independent of the samples used to derived the calibration corrections.

The invariant mass distributions for the $\Jmm$ and $\Zmm$ candidates are shown in \refF{fig:Calib_InvMass} and compared with corrected simulation. The lineshapes of the two resonances in simulation agree with the data within the systematic uncertainties, demonstrating the overall effectiveness of the \pt\ calibration.

\begin{figure*}[th]
\centering
\subfloat[ ]{\includegraphics[width=0.49\linewidth]{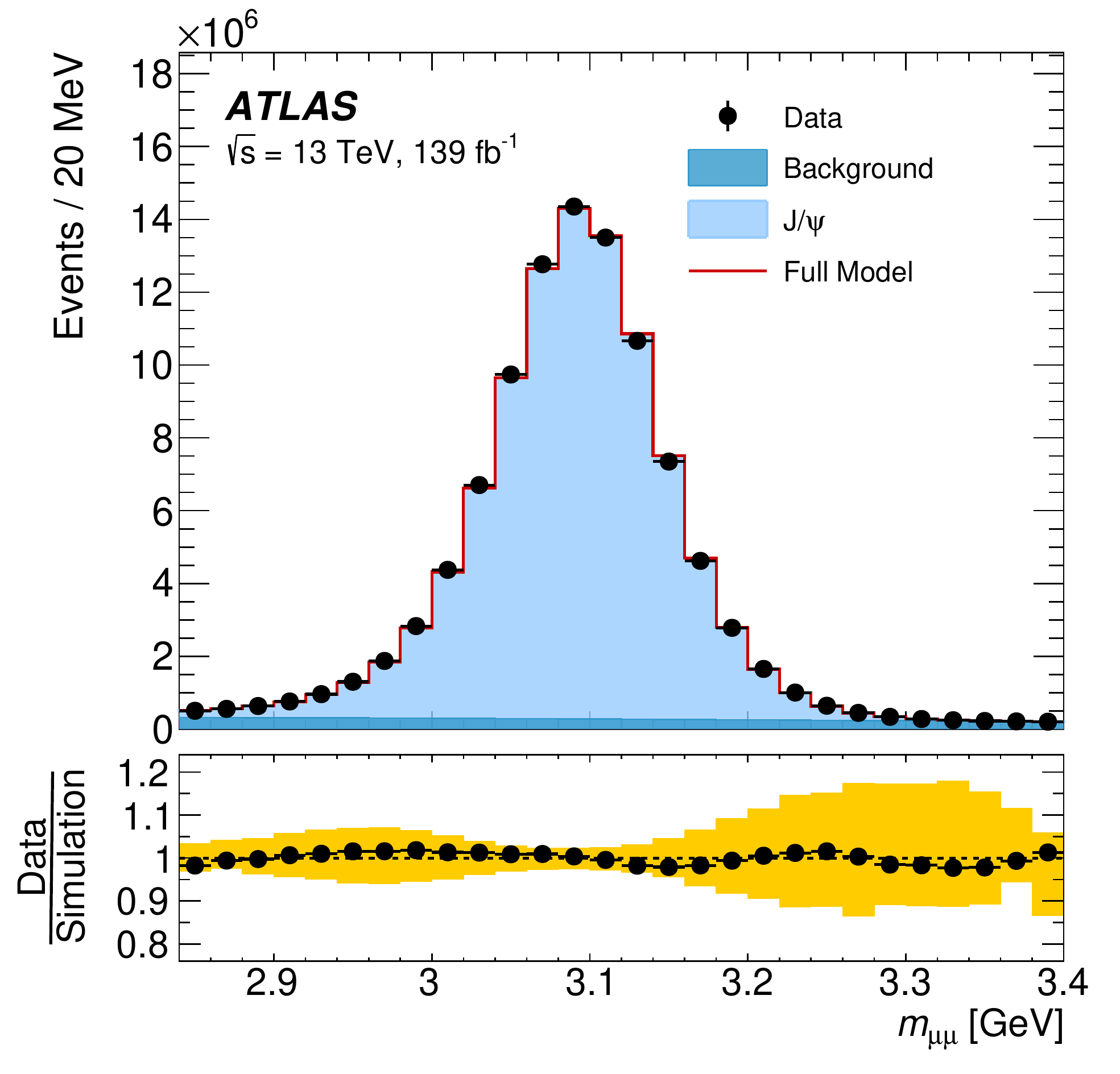}}
\subfloat[ ]{\includegraphics[width=0.49\linewidth]{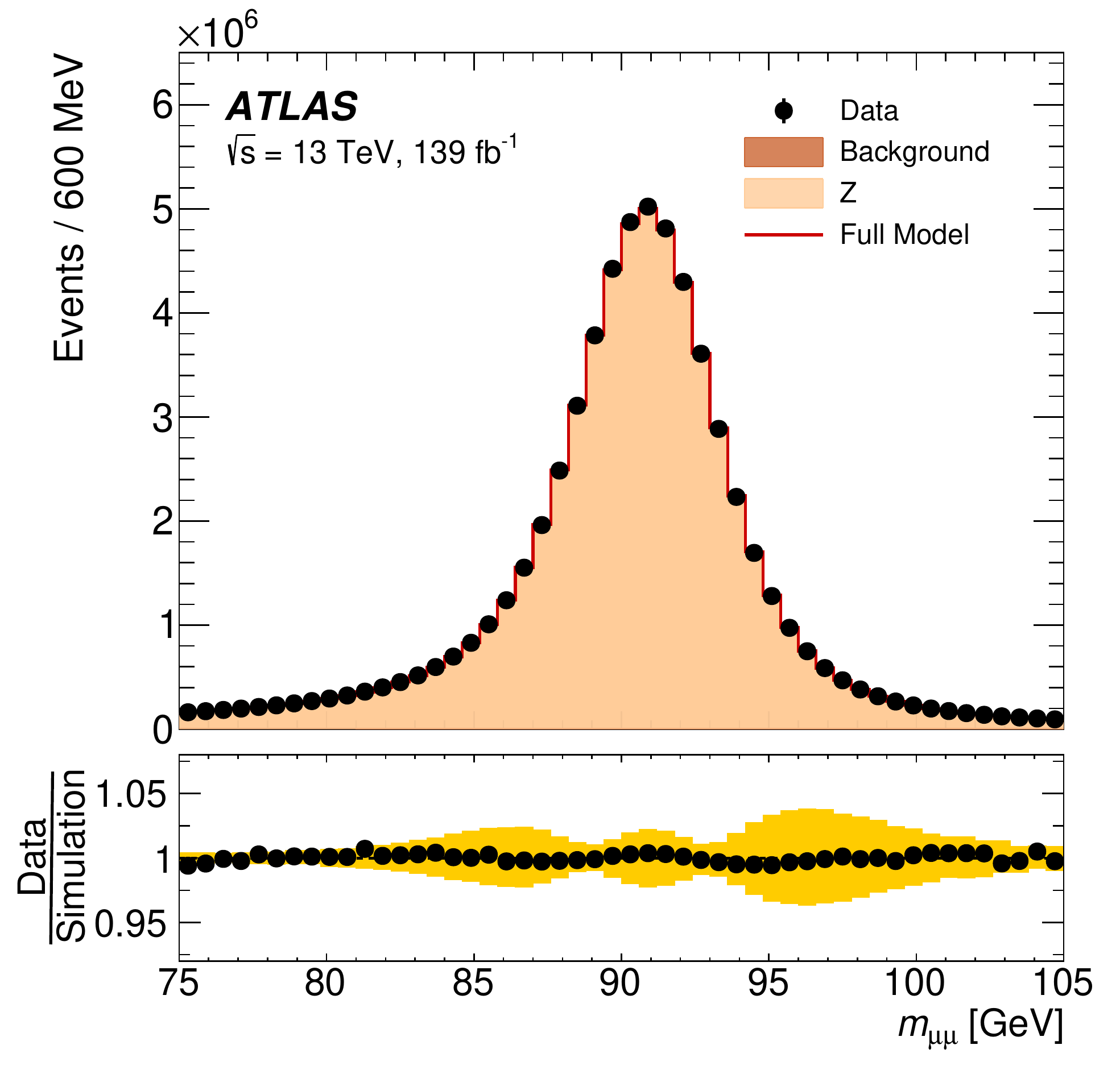}}
\caption{Dimuon invariant mass distribution of
(a) $\Jmm$ and  (b) $\Zmm$ candidate events reconstructed with CB muons.
The upper panels show the invariant mass distribution for data and for the signal simulation plus the background estimate.
The points show the data. The continuous line corresponds to the simulation with the MC momentum corrections applied.
Background estimates are added to the signal simulation. The bands represent the effect of the systematic uncertainties in the MC momentum corrections.  The lower panels show the ratios of the data to the MC simulation.
In the $Z$ sample, the MC background samples are added to the signal sample according to their expected cross-sections.
In the \jpsi sample,  the background is estimated from a fit to the data as described in the text.
The sum of background and signal MC distributions is normalised to the data.}
\label{fig:Calib_InvMass}
\end{figure*}

When the two muons have similar momentum resolution and angular effects are neglected,
the relative mass resolution, $\sigma_{\mu\mu}/m_{\mu\mu}$, is directly proportional to the relative muon momentum resolution,
$\sigma_{p_{\mu}}/p_{\mu}$:
 
\begin{equation}
\frac{\sigma_{\mu\mu}}{m_{\mu\mu}} \, = \,  \frac{1}{\sqrt{2}} \frac{\sigma_{p_{\mu}}}{p_{\mu}}.
\label{eq:dmom}
\end{equation}
Similarly, the total muon momentum scale, defined as $s =  \langle (p^\mathrm{meas} - p^\mathrm{true})/  p^\mathrm{true} \rangle$,
is directly related to the dimuon mass scale, defined as  $ s_{\mu\mu} = \langle (m_{\mu\mu}^\mathrm{meas} - m_{\mu\mu}^\mathrm{true})/  m_{\mu\mu}^\mathrm{true} \rangle$:
\begin{equation}
s_{\mu\mu} = \sqrt{s_{\mu_1}s_{\mu_2}},
\end{equation}
where $s_{\mu_1}$ and $s_{\mu_2}$ are the momentum scales of the two muons. The effectiveness of the momentum calibration is also measured by comparing the mean value $m_{\mu\mu}$, and resolution $\sigma(m_{\mu\mu})$ of the dimuon mass resonances. To measure such quantities, fully analytical fit functions are created for each resonance modelling both the signal and the background. Using the same definition of the resolution function for all resonances, this methodology allows a direct comparison of the resolution quantities as extracted separately for each resonance. In turn, this allows a precise measurement of the momentum reconstruction performance across a wide kinematic regime.  The fitting procedures used to measure the mean value and resolution quantities are optimised, subsequently, for each resonance.
 
In $\Jmm$ decays, the intrinsic width of the resonance is negligible compared with the experimental resolution. In contrast to \refS{Sec:CorrectionExtraction}, the mass window for the $\Jmm$ validation fits is modified to $2.8<m_{\mu\mu}<3.9~\GeV$. The lower bound is raised to remove the abrupt spectrum sculpting due to the trigger criteria, which is difficult to model well analytically. The upper bound is extended to include the $\Psi(2S)$ resonance and add a lever arm to constrain the estimation of the resolution.
 
The resonant peak of the $\jpsi$ is modelled by a double-sided Crystal Ball function; the four parameters modelling the tails are extracted from simulation, and subsequently fixed in the fits to data. A secondary double-sided Crystal Ball function models the $\Psi(2S)$ resonance that follows the $1s$ resonance in the $\jpsi$ validation mass window. Similarly to the $1S$ resonance, the parameters modelling the tails are extracted from simulation and kept fixed when fitting to data. The mean value parameter of the resolution function is kept free floating for the $\Psi(2S)$ resonance for fits on simulated and real data, while the resolution parameter is constrained to scale linearly with the resolution parameter of the $1S$ resonance using the ratio of the mean parameters.
The non-resonant background or the background from mis-identified muons is described by an exponential function. \\
 
In $\Zmm$ decays, the fits use a convolution of the true lineshape (modelled by a Breit--Wigner function)  with an experimental resolution function (a double-sided Crystal Ball). A fit range of $75<m_{\mu\mu}<105~\GeV$ is used. Similarly to the \jpsi, the non-resonant background is described by an exponential function.  The peak position and width of the Crystal Ball function are used as estimators for the  $m_{\mu\mu}$ and $\sigma(m_{\mu\mu})$ variables in each of the $\eta$ and \pt\ bins.

\refF{fig:mass_eta} shows the position of the mean value of the invariant mass distribution, $m_{\mu\mu}$, obtained from the fits to the $Z$ boson and \jpsi  samples
as a function of the pseudorapidity of the highest-$\pt$ muon for each decay.
The distributions are shown for data and corrected simulation, with the ratios of the two in the lower panels.
The simulation is in  good  agreement with the data.
Minor deviations are within the momentum scale systematic uncertainties of
$0.05\%$ in the barrel region increasing with $|\eta|$ to $0.15\%$ for $\Jmm$ decays, and to $0.05\%$ in the barrel region increasing with $|\eta|$ to $0.1\%$ for $\Zmm$ decays.
The systematic uncertainties shown in the plots include the effects of the uncertainties in the calibration constants described in \refS{Sec:CorrectionExtraction}.
The observed  level of agreement  demonstrates that the \pt\ calibration for combined muon tracks described above provides an accurate
description of the momentum scale in all $\eta$ regions, over a wide  \pt\ range. Similar levels of data/MC agreement are observed for the ID and MS components of the combined tracks.

\begin{figure*}[!hbt]
\centering
\subfloat[ ]{\includegraphics[width=0.49\linewidth]{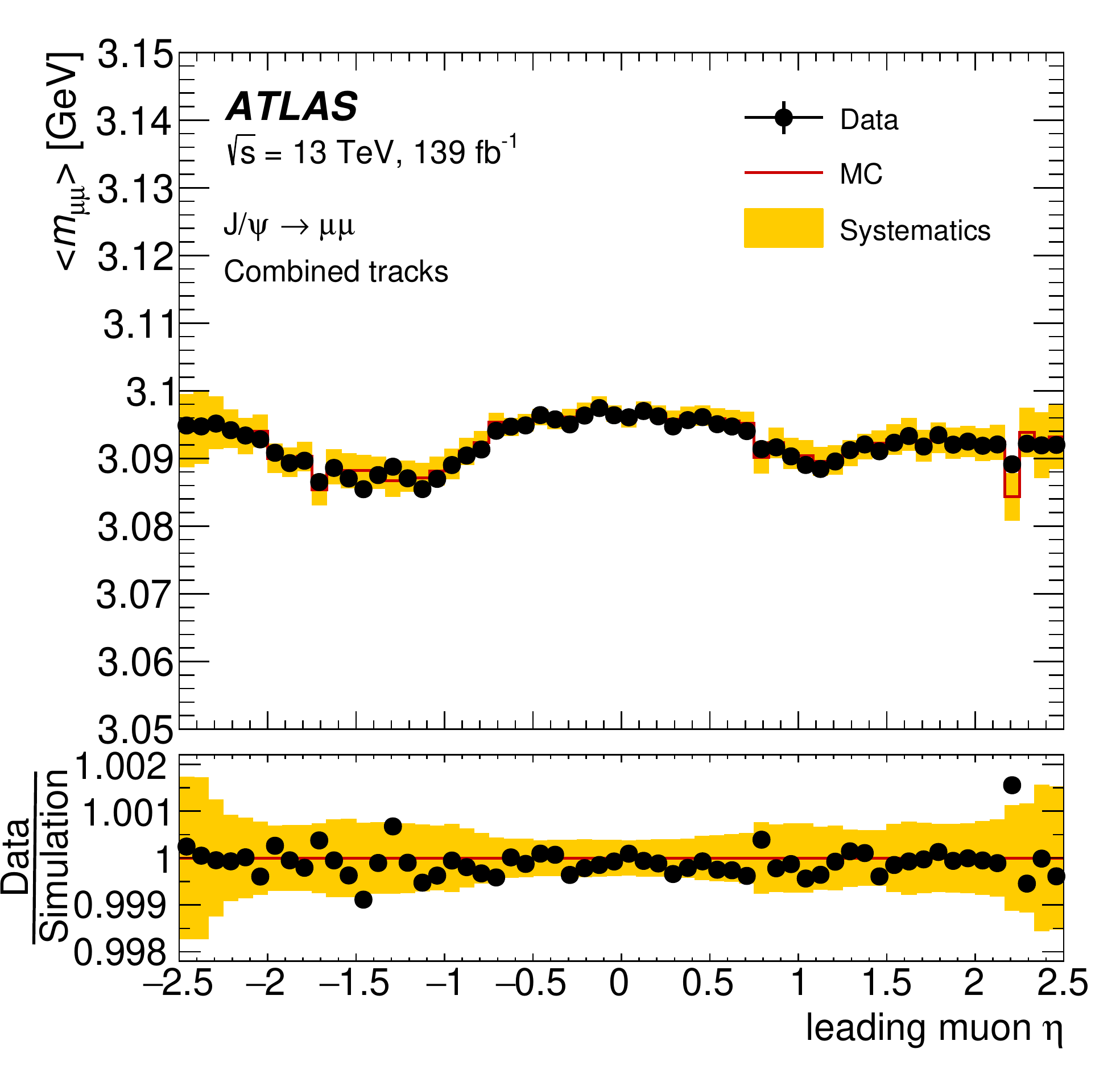}}
\subfloat[ ]{\includegraphics[width=0.49\linewidth]{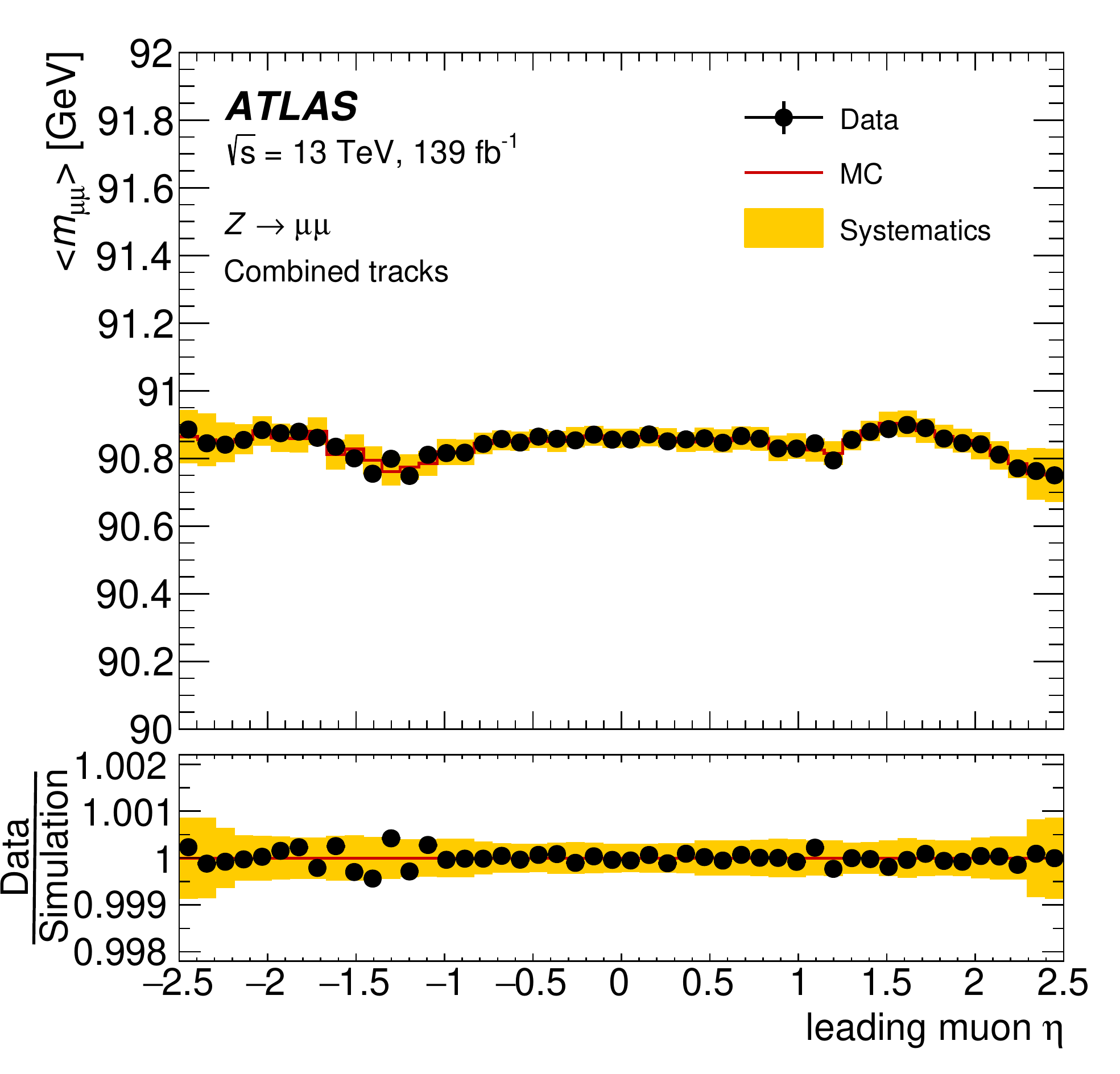}}
\caption{Fitted mean mass of the dimuon system for CB muons for (a) $\Jmm$  and  (b) $\Zmm$ events for
data and corrected simulation as a function of the pseudorapidity of the highest-$\pt$ muon.
The upper panels show the fitted mean mass value for data
and corrected simulation.
The small variations of the invariant mass estimator as a function of pseudorapidity are
due to imperfect energy loss corrections and magnetic field description in the muon reconstruction.
Both effects are  well reproduced in the simulation.
The lower panels show the ratios of the data to the MC simulations.
The error bars represent the statistical uncertainty;  the shaded bands represent the systematic uncertainty in the correction and the systematic uncertainty in the extraction method added in quadrature.}
\label{fig:mass_eta}
\end{figure*}
 
\begin{figure*}[!hbt]
\centering
\subfloat[ ]{\includegraphics[width=0.49\linewidth]{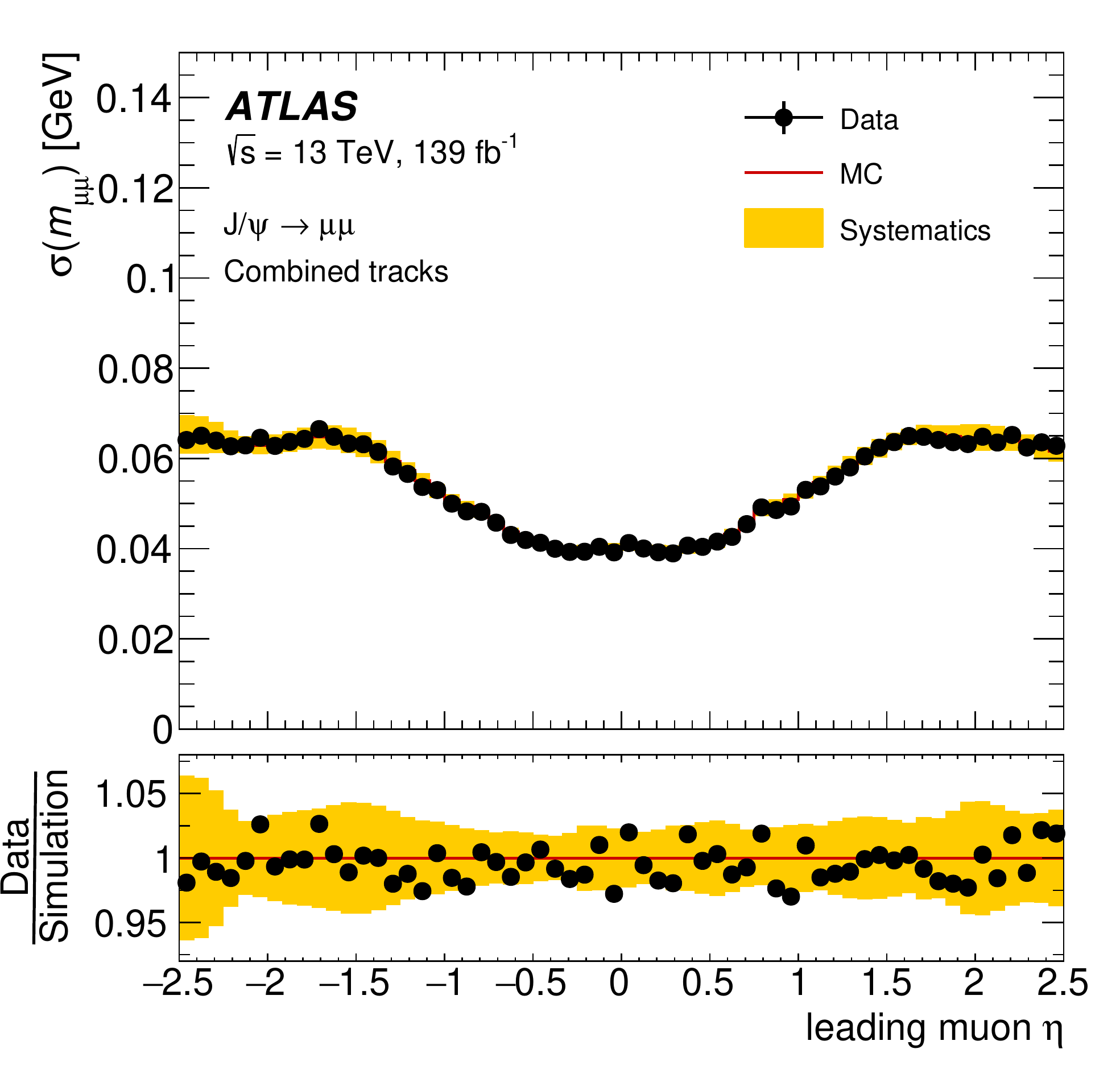}}
\subfloat[ ]{\includegraphics[width=0.49\linewidth]{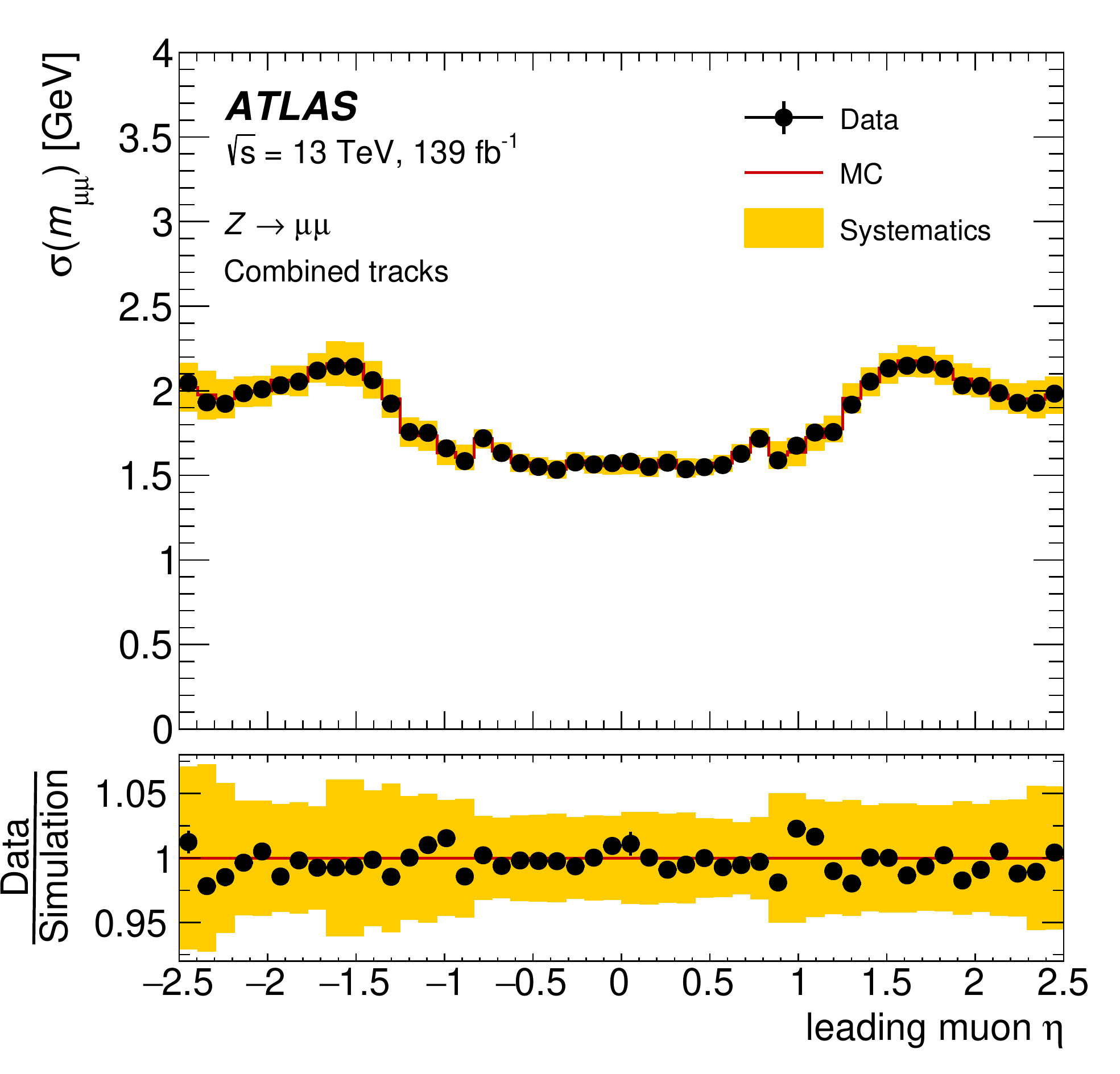}}
\caption{
Dimuon invariant mass resolution for CB muons   for  (a) $\Jmm$ and (b)  $\Zmm$ events for
data and  corrected simulation as a function of the pseudorapidity of the highest-$\pt$ muon.
The upper panels show the fitted resolution value for data
and corrected simulation. The lower panels show the ratios of the data to the MC simulations.
The error bars represent the statistical uncertainty;
the shaded bands represent the systematic uncertainty in the correction and the systematic uncertainty
in the extraction method added in quadrature. For the $\Zmm$  the estimator shown in the plot is only sensitive to the detector resolution and not to the natural width of the $Z$ boson, as discussed in the text.
}
\label{fig:width_eta}
\end{figure*}
 
\refF{fig:width_eta} displays the dimuon mass resolution $\sigma(m_{\mu\mu})$ as a function of the leading-muon $\eta$ for the two resonances.
The dimuon mass resolution is about 1.3\% and 1.6\%
at small $\eta$ values for the \jpsi and $Z$ bosons, respectively,
and increases to 2.1\% and 2.4\% in the endcaps.
This corresponds to a relative muon \pt\ resolution of 1.8\% and 2.3\% in the centre of the detector and 3.0\% and 3.4\% in the endcaps for \jpsi and $Z$ boson decays, respectively.
Uncertainties in the dimuon mass resolution range between 2\% and 5\% for the \jpsi and between 3\% and about 6\%  for the $Z$ boson, depending on the detector region.
 
\begin{figure*}[!htb]
\begin{center}
\includegraphics[width=0.49\linewidth]{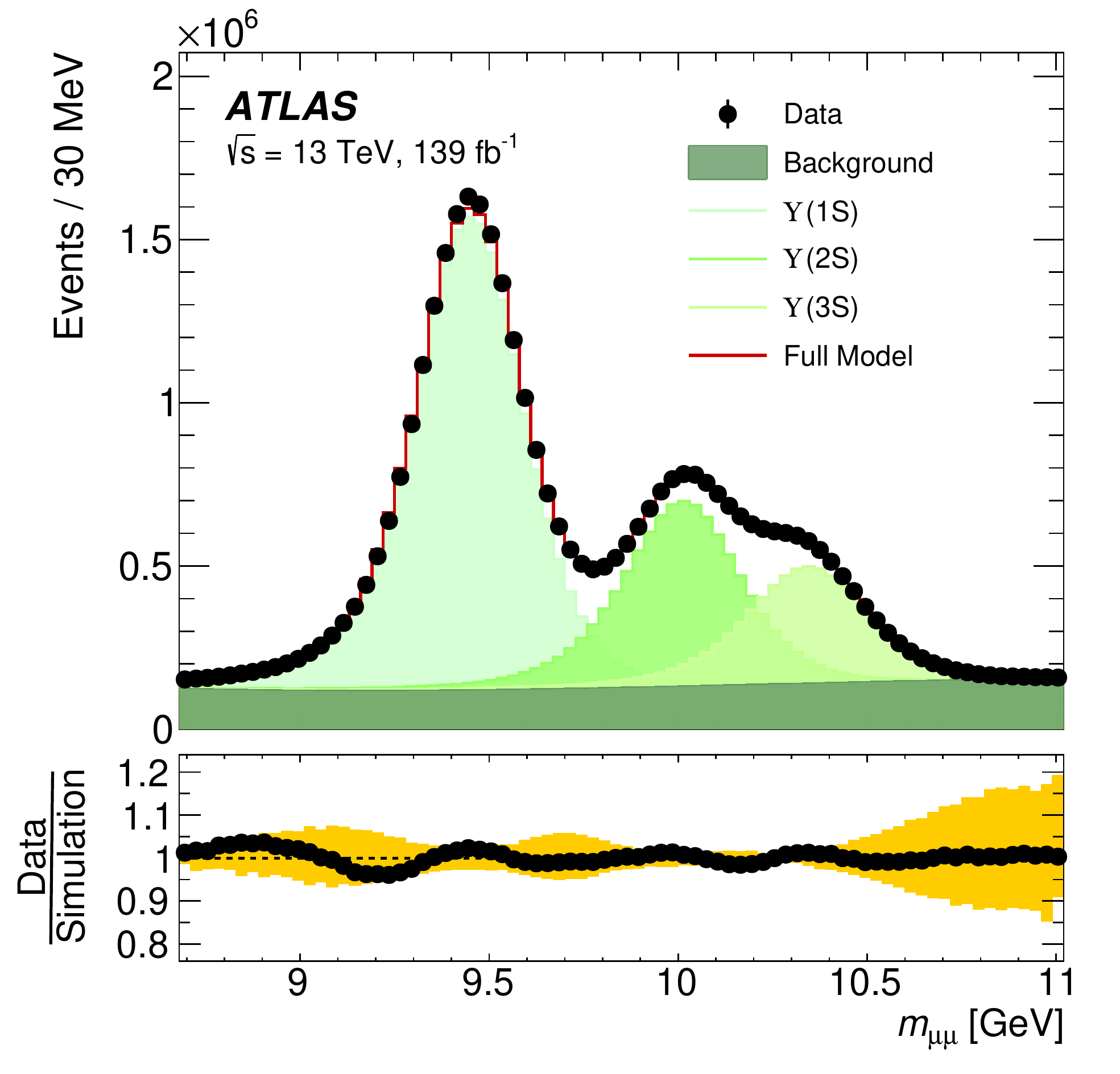}
\end{center}
\caption{Dimuon invariant mass distribution of  $\Upsmm$  candidate events reconstructed with CB muons.
The upper panels show the invariant mass distribution for data and for the signal simulation plus the background estimate.
The points show the data. The continuous line corresponds to the simulation with the MC momentum corrections applied. The band represents the effect of the systematic uncertainties in the MC momentum corrections.  The lower panel shows ratio of the data to the MC simulations.  The background is estimated from a fit to the data as described in the text. The sum of background and signal MC distributions is normalised to the data.}
\label{fig:UpsilonMass}
\end{figure*}
 
Using the same methodologies as for the $\Jmm$ decays, the measurement of the scale dependence and resolution is repeated on the fully independent set of data from $\Upsmm$ decays. The same approach as for the $\jpsi$ resonances models the $2S$ and $3S$ resonances of the $\Upsilon$. This entails constraining the resolution parameter of the $3S$ resonance to scale linearly with that of the $2S$ resonance and that of the $2S$ resonance to scale linearly with that of the $1S$ resonance. The mean parameters of the resolution models of the $2S$ and $3S$ are kept free independently in data and simulation, similarly to what is done for the $\jpsi$ resonance.
The background is, however,  modelled using a polynomial function derived from data. \refF{fig:UpsilonMass} shows the data and simulation agreement after applying the momentum corrections on simulation for the invariant mass distributions of the three resonances of the $\Upsmm$.
As done for the $\Jmm$ and $\Zmm$ decays, the momentum scale dependency and the resolution are extracted as a function of the pseudorapidity of the leading muon in the event. \refF{fig:YScaleAndResolution} compares the data and simulation after all corrections. The data agree with the simulation within the systematic uncertainties. The uncertainty in the momentum scale is up to 0.05\% in the central region and up to 0.1\% in the forward region. 
Similarly, the extracted resolution is in agreement within about 3\%.
Good agreement between the dimuon mass resolution measured in data and simulation is also observed when deriving the corrections independently for the ID and MS components of the combined tracks, as shown in \refA{app:results}.

\begin{figure*}[!htb]
\centering
\subfloat[ ]{\includegraphics[width=0.49\linewidth]{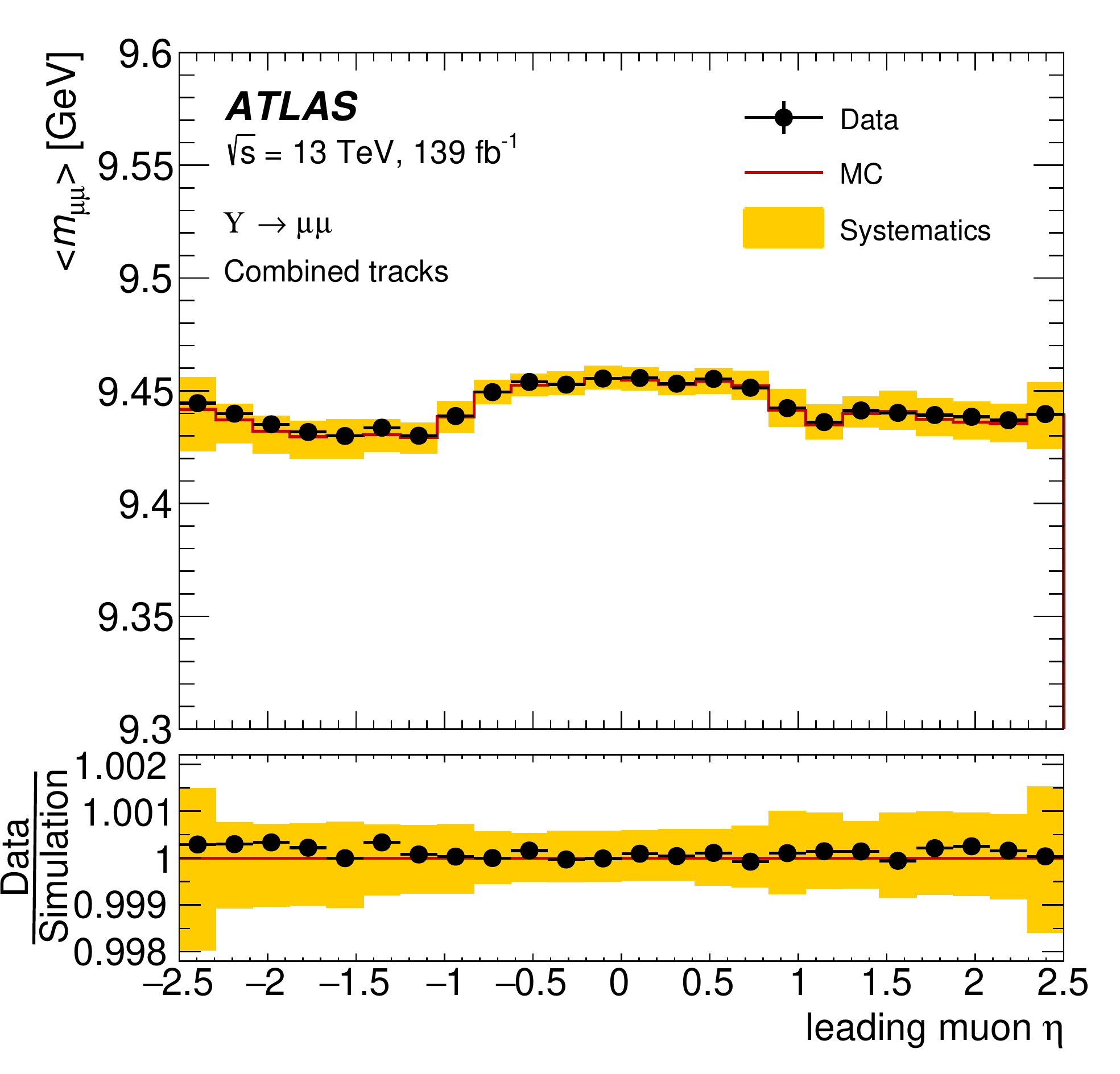}}
\subfloat[ ]{\includegraphics[width=0.49\linewidth]{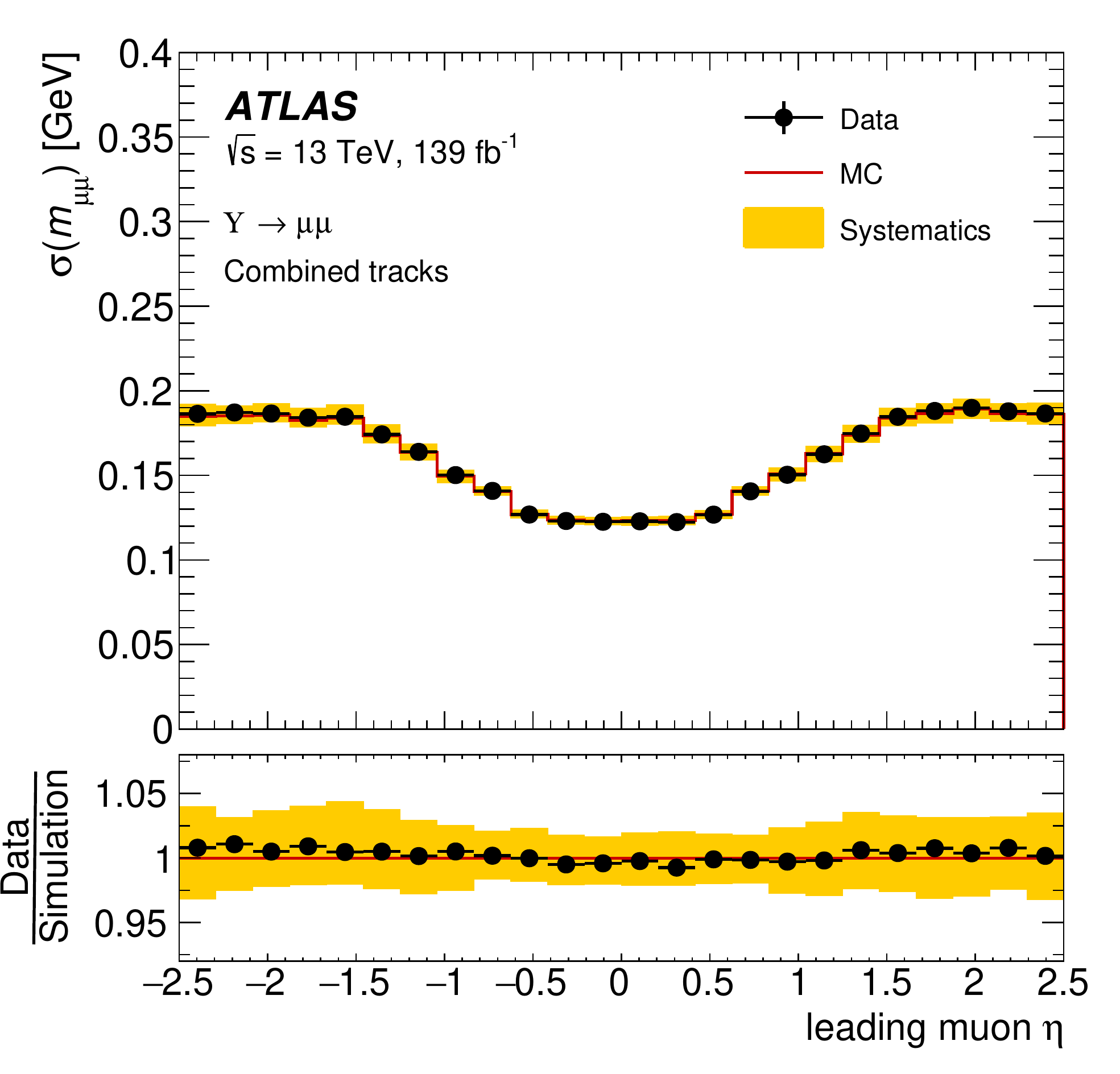}}
\caption{
(a) Fitted mean mass of the dimuon system for CB muons for $\Upsmm$ events
in data and corrected simulation as a function of the pseudorapidity of the highest-$\pt$ muon.  (b) Dimuon invariant mass resolution for CB muons for $\Upsmm$ events for data and  corrected simulation as a function of the pseudorapidity of the highest-$\pt$ muon. The upper panels show the results obtained on data and simulation after all corrections. The lower panels show the ratios of the data to the MC simulation. The error bars represent the statistical uncertainty;  the shaded bands represent the systematic uncertainty in the correction and the systematic uncertainty in the extraction method added in quadrature.}
\label{fig:YScaleAndResolution}
\end{figure*}

The relative dimuon mass resolution $\sigma_{\mu\mu}/m_{\mu\mu}$ is approximately proportional to the average momentum of the muons, as shown in \refE{eq:dmom}.
A direct comparison of the momentum resolution function determined with \jpsi{}, $\Upsilon$, and $Z$ boson decays can therefore be performed.
To remove the effect of the correlation between the measurement of the dimuon mass resolution and the $\pt$ of the muons, the following definition of transverse momentum is used:
\begin{equation}
\pt^{*} = \hat{m} \sqrt{ \frac{\sin{\theta_1}\sin{\theta_2}}{2 (1-\cos \alpha_{12})}},
\end{equation}
\noindent where $\hat{m}$ is a fixed value, corresponding to the best known value of the mass of the $Z$ boson, the $\jpsi$, and the $\Upsilon$, respectively. The variables  $\theta_1$ and $\theta_2$ are the polar angles of the two muons, and $\alpha_{12}$ is the opening angle of the muon pair.
The relative dimuon mass resolution from  $\Jmm$, $\Upsmm$, and $\Zmm$ events measured in data is compared with that obtained from calibrated simulation as a function of of $\pt^{*}$ in \refF{f:resol_vs_pt}.
The resolutions are in good agreement. Due to the larger number of events, the $\Jmm$ measurement extends to higher momenta than the $\Upsmm$ one.
 
\begin{figure*}[!htb]
\begin{center}
\includegraphics[width=0.49\linewidth]{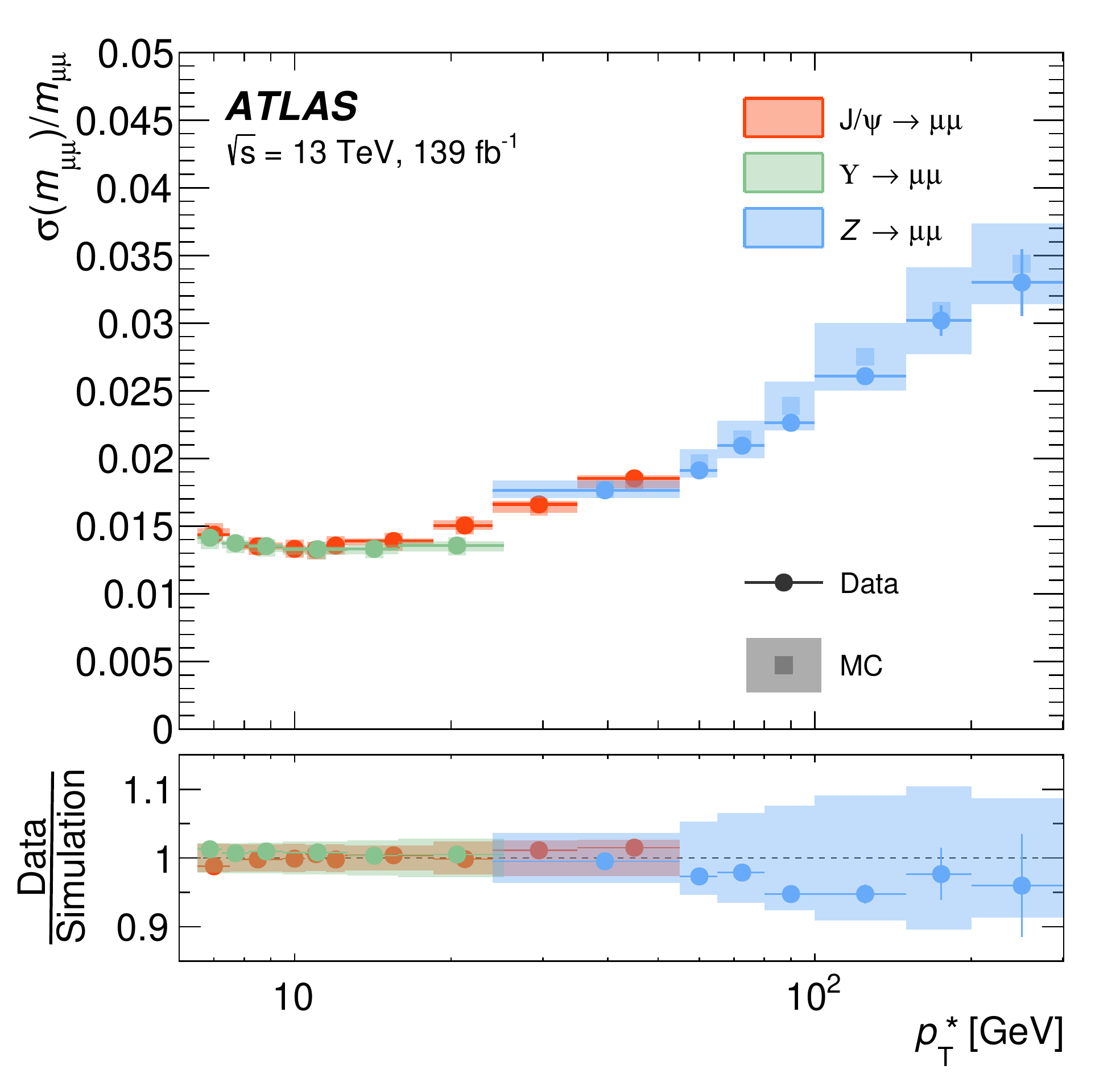}
\end{center}
\caption{
The upper panel shows the dimuon invariant mass resolution divided by the dimuon invariant mass for CB muons measured
from  $\Jmm$, $\Upsmm$ and  $\Zmm$ events as a  function of the $ \pt^{*}$ variable defined in the text. To account for the resolution differences induced by the different pseudorapidity distributions among muons from the three resonances, the data and simulations are re-weighted according to the pseudorapidity distribution of the muon with the highest momentum from $\Upsmm$ decays. The lower panel shows the ratio of the data to the MC simulation.
The error bars represent the statistical uncertainty, while the bands show the systematic uncertainties.
}
\label{f:resol_vs_pt}
\end{figure*}


\FloatBarrier
 
\section{Conclusion}
\label{sec:conclusion}

The momentum performance of the ATLAS muon reconstruction is measured using 139~\ifb{} of data
from  $pp$ collisions at $\sqrt{s} = 13$~\TeV\ recorded during the second run of LHC between 2015 and 2018.

Events from the $\Zmm$ process are used to correct the reconstructed data for a charge-dependent bias in the muon momentum measurement associated with detector alignment effects. This bias is reduced on average from up to 0.4~$\TeV{}^{-1}$ to $2\cdot10^{-4}~\TeV{}^{-1}$ for muons with $\pt<450~\GeV$ after the corrections are applied, with associated uncertainty at the level of 0.03~$\TeV{}^{-1}$ for muons with transverse momentum of about $45~\GeV$.
 
The scale and resolution of the muon momentum measurement is studied in detail using $\Jmm$ and $\Zmm$ decays.
These studies are used to correct the simulation, improving the agreement with the data and reducing the systematic uncertainties related to the muon calibration in physics analyses.
The improvements in the \pt\ correction methods described in this paper and the substantial number of $\Jmm$ and $\Zmm$ decays collected in the Run~2 data sample together improve the
precision of the momentum scale by up to a factor of two relative to the previous publication based on 3.2 \ifb{} of collected data~\cite{PERF-2015-10}. For $\Zmm$ decays, the uncertainty in the momentum scale varies from a minimum of $0.05\%$ for \mbox{$|\eta| <1$} to a maximum of \mbox{$0.15\%$} for \mbox{$|\eta|\sim 2.5$}.
 
The dimuon mass resolution is about 1.3\% (1.6\%)  at small values of pseudorapidity for \jpsi ($Z$) decays, and increases up
to 2.1\% (2.4\%) in the endcaps. This corresponds to a relative muon \pt\ resolution of 1.8\% (2.3\%) at small values of pseudorapidity and 3.0\% (3.4\%) in the endcaps for \jpsi ($Z$) decays.
After applying momentum corrections, the  \pt\ resolution in data and simulation agree within the quoted uncertainties, which are at the level of, or better than, 5\% (6\%) for \jpsi ($Z$) decays depending on the $\eta$ range.

Validation studies performed with $\Zmm$ and $\Jmm$ show that the corrections applied to the simulation bring the agreement with the data to the level of their estimated uncertainties. An additional, statistically independent validation performed using $\Upsmm$ decays confirms the correctness of the correction procedure.


\section*{Acknowledgements}


We thank CERN for the very successful operation of the LHC, as well as the
support staff from our institutions without whom ATLAS could not be
operated efficiently.
 
We acknowledge the support of
ANPCyT, Argentina;
YerPhI, Armenia;
ARC, Australia;
BMWFW and FWF, Austria;
ANAS, Azerbaijan;
CNPq and FAPESP, Brazil;
NSERC, NRC and CFI, Canada;
CERN;
ANID, Chile;
CAS, MOST and NSFC, China;
Minciencias, Colombia;
MEYS CR, Czech Republic;
DNRF and DNSRC, Denmark;
IN2P3-CNRS and CEA-DRF/IRFU, France;
SRNSFG, Georgia;
BMBF, HGF and MPG, Germany;
GSRI, Greece;
RGC and Hong Kong SAR, China;
ISF and Benoziyo Center, Israel;
INFN, Italy;
MEXT and JSPS, Japan;
CNRST, Morocco;
NWO, Netherlands;
RCN, Norway;
MEiN, Poland;
FCT, Portugal;
MNE/IFA, Romania;
MESTD, Serbia;
MSSR, Slovakia;
ARRS and MIZ\v{S}, Slovenia;
DSI/NRF, South Africa;
MICINN, Spain;
SRC and Wallenberg Foundation, Sweden;
SERI, SNSF and Cantons of Bern and Geneva, Switzerland;
MOST, Taiwan;
TENMAK, T\"urkiye;
STFC, United Kingdom;
DOE and NSF, United States of America.
In addition, individual groups and members have received support from
BCKDF, CANARIE, Compute Canada and CRC, Canada;
PRIMUS 21/SCI/017 and UNCE SCI/013, Czech Republic;
COST, ERC, ERDF, Horizon 2020 and Marie Sk{\l}odowska-Curie Actions, European Union;
Investissements d'Avenir Labex, Investissements d'Avenir Idex and ANR, France;
DFG and AvH Foundation, Germany;
Herakleitos, Thales and Aristeia programmes co-financed by EU-ESF and the Greek NSRF, Greece;
BSF-NSF and MINERVA, Israel;
Norwegian Financial Mechanism 2014-2021, Norway;
NCN and NAWA, Poland;
La Caixa Banking Foundation, CERCA Programme Generalitat de Catalunya and PROMETEO and GenT Programmes Generalitat Valenciana, Spain;
G\"{o}ran Gustafssons Stiftelse, Sweden;
The Royal Society and Leverhulme Trust, United Kingdom.
 
The crucial computing support from all WLCG partners is acknowledged gratefully, in particular from CERN, the ATLAS Tier-1 facilities at TRIUMF (Canada), NDGF (Denmark, Norway, Sweden), CC-IN2P3 (France), KIT/GridKA (Germany), INFN-CNAF (Italy), NL-T1 (Netherlands), PIC (Spain), ASGC (Taiwan), RAL (UK) and BNL (USA), the Tier-2 facilities worldwide and large non-WLCG resource providers. Major contributors of computing resources are listed in Ref.~\cite{ATL-SOFT-PUB-2021-003}.


\clearpage
\appendix
\part*{Appendix}
\addcontentsline{toc}{part}{Appendix}

This appendix complements the results presented in the paper with some additional material. \refA{app:sagitta} provides the measured charge-dependent biases before and after application of the correction procedure for the alternative method discussed in \refS{Sec:SagittaBias}. \refA{app:results} documents the measured performance when applying the corrections for CB tracks as obtained by the combination of corrections to ID and MS measurements.
 
\section{Charge-dependent momentum scale calibration in data}
\label{app:sagitta}
 
\begin{figure*}[htb]
\centering
\subfloat[ ]{\includegraphics[width=0.45\linewidth]{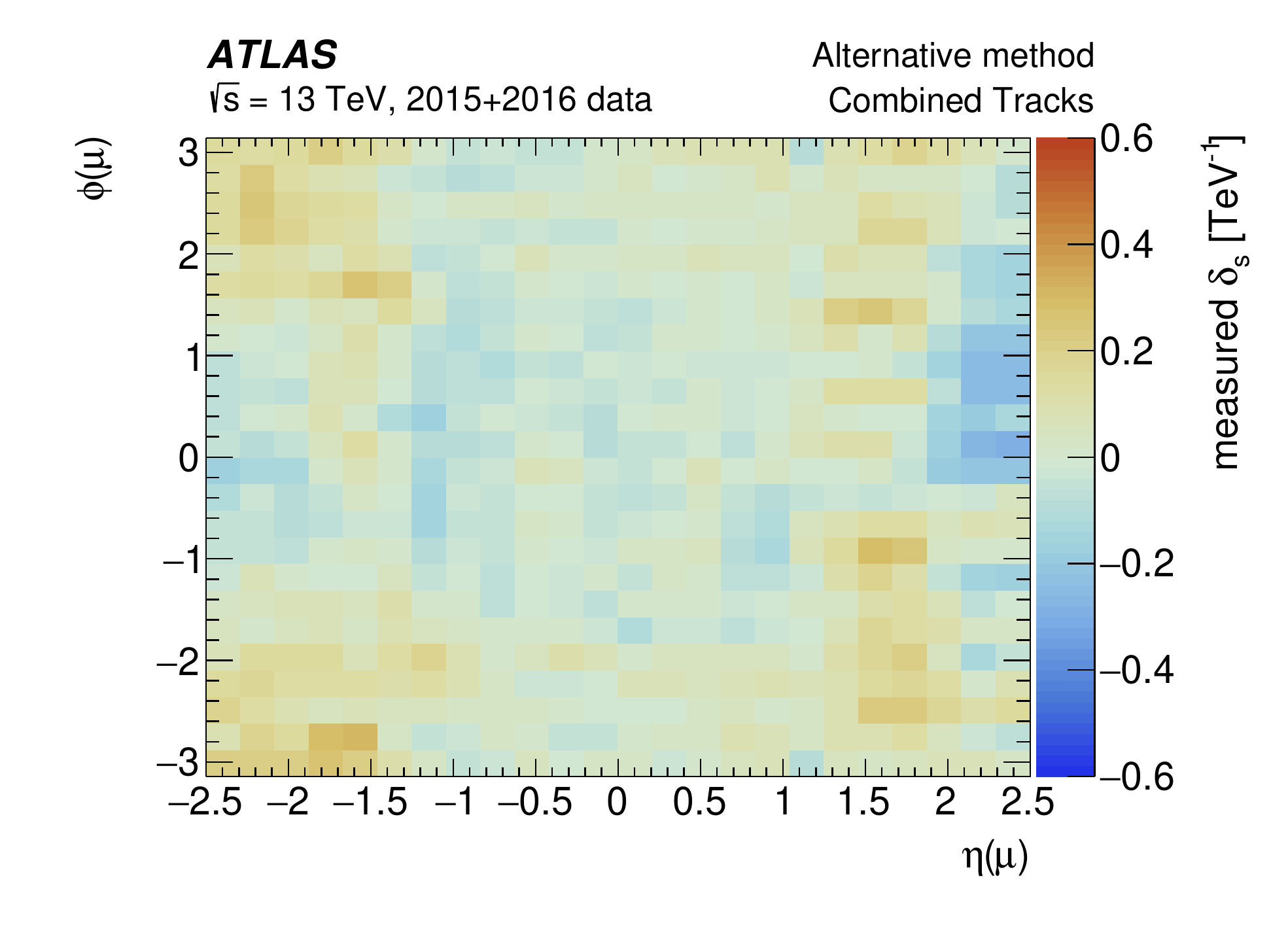}}
\subfloat[ ]{\includegraphics[width=0.45\linewidth]{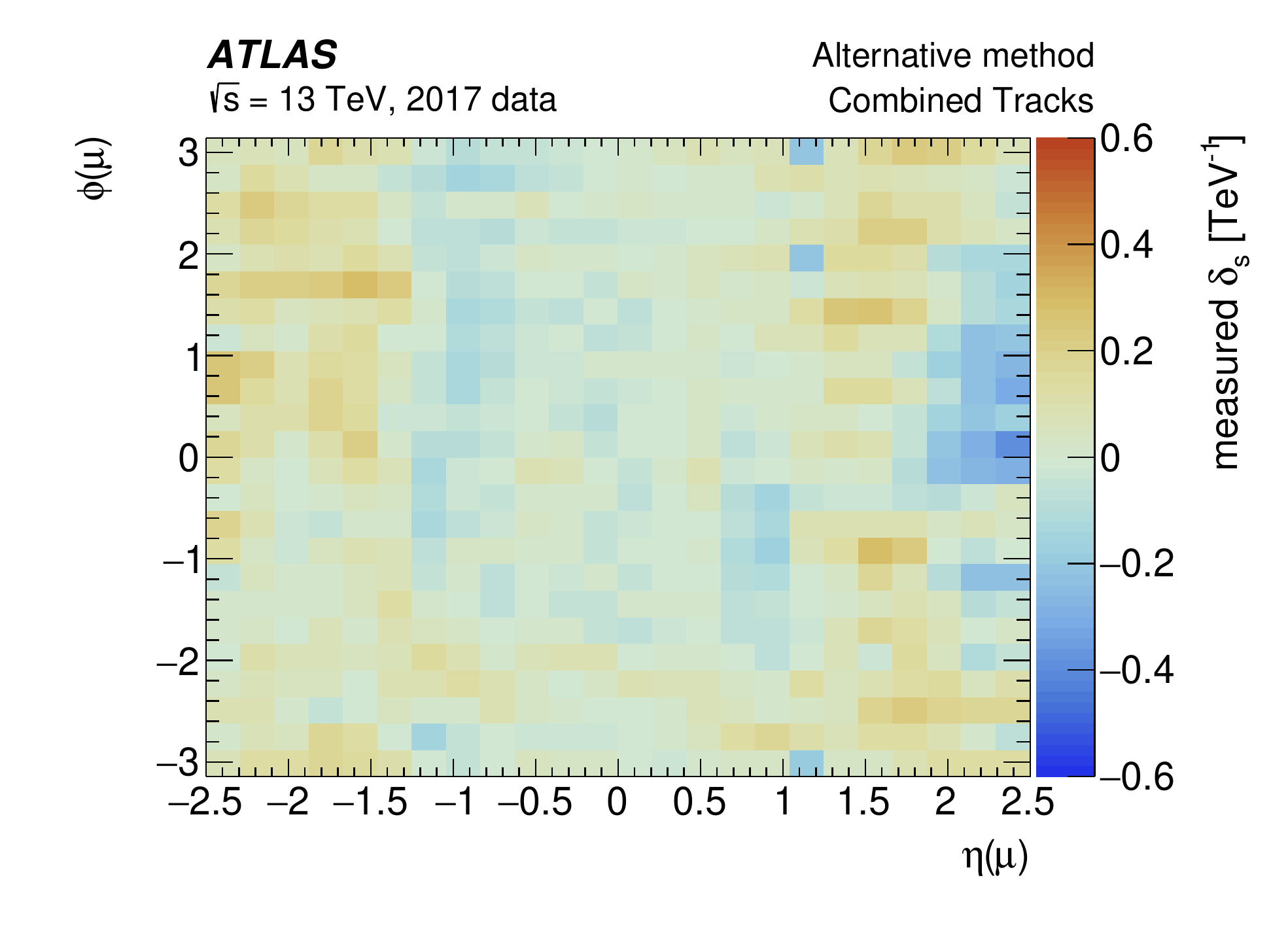}} \\
\subfloat[ ]{\includegraphics[width=0.45\linewidth]{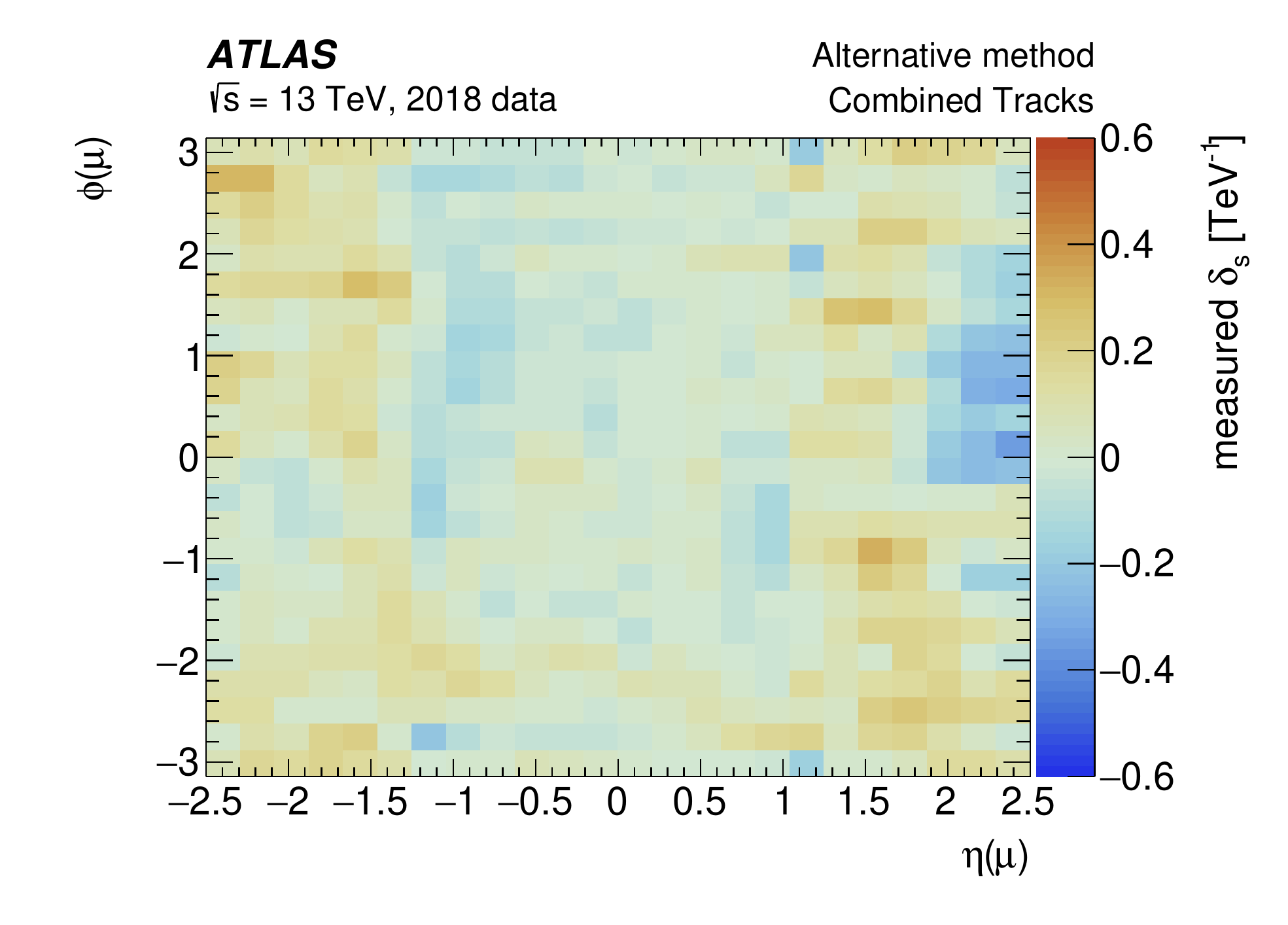}}
\caption{Charge-dependent biases on the muon \pt\ as evaluated on data for CB momenta after alignment and before applying the dedicated corrections using the alternative method described in \refS{Sec:SagittaBias}. The biases are shown separately for the three data-taking periods of (a) years 2015 and 2016, (b) 2017 and (c) 2018. The bias is defined in Eq.~(\ref{eq:bias_definition}).}
\label{fig:app_sgBefore}
\end{figure*}
 
As described in the main body of the paper, a complementary method to estimate the charge-dependent bias was also developed.
The details of the method are explained in \refS{Sec:SagittaBias}.
In this Appendix the results obtained with this method are presented.
Similarly to \refF{fig:sgBefore}, \refF{fig:app_sgBefore} shows the measured sagitta corrections as evaluated on CB momenta.
It can be seen that the biases also extend to about 0.4 $\TeV{}^{-1}$.
After using the correction procedure with the iterative method explained in \refS{Sec:SagittaBias}, using 25 iterations, the residual biases are re-evaluated with the same methodology.
The resulting values, as shown in \refF{fig:app_sgAfterCorr}, indicate the same reduction of the measured biases to values of less than $10^{-3}\TeV{}^{-1}$.
 
\begin{figure*}[!htb]
\centering
\subfloat[ ]{\includegraphics[width=0.45\linewidth]{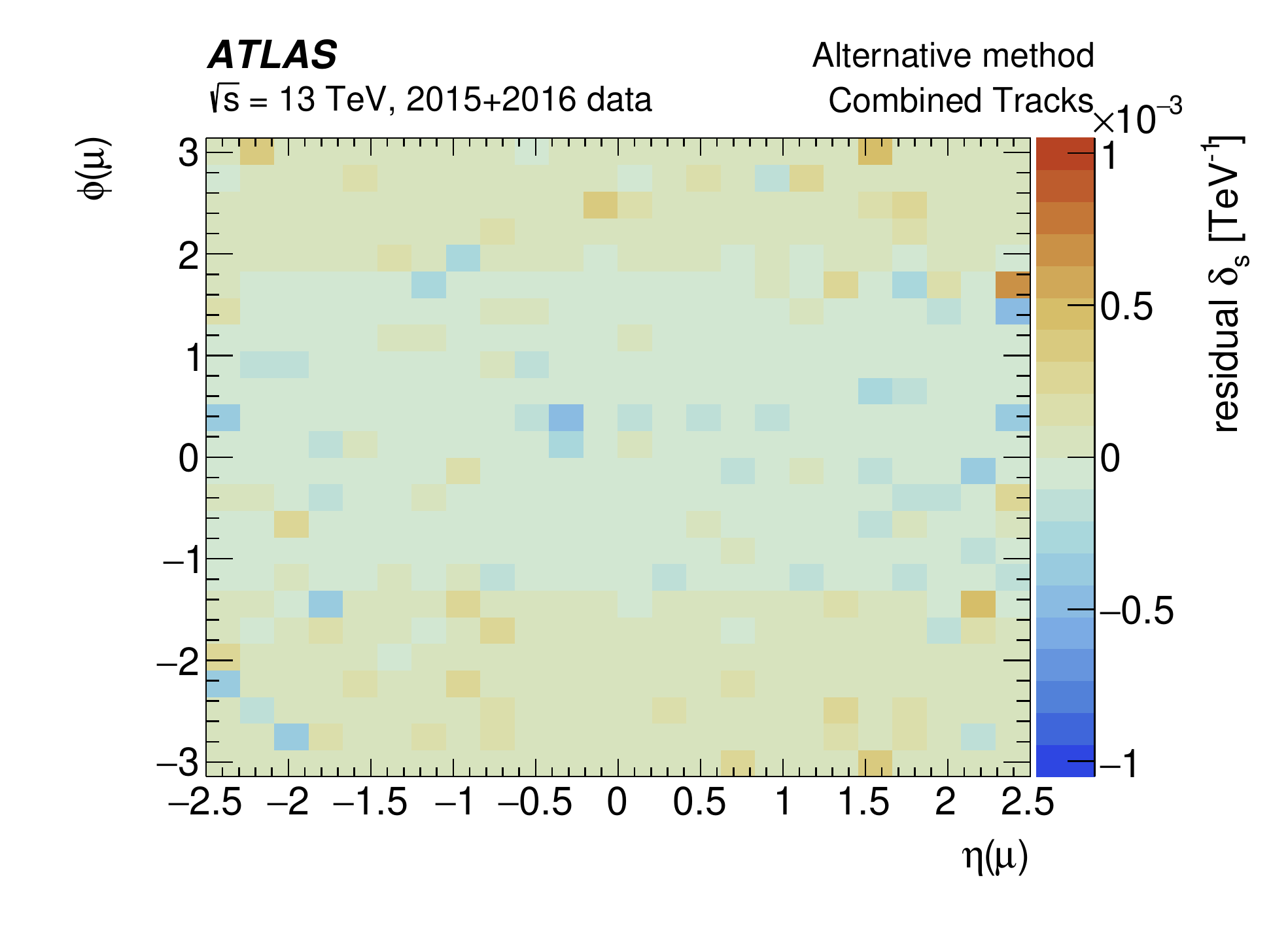}}
\subfloat[ ]{\includegraphics[width=0.45\linewidth]{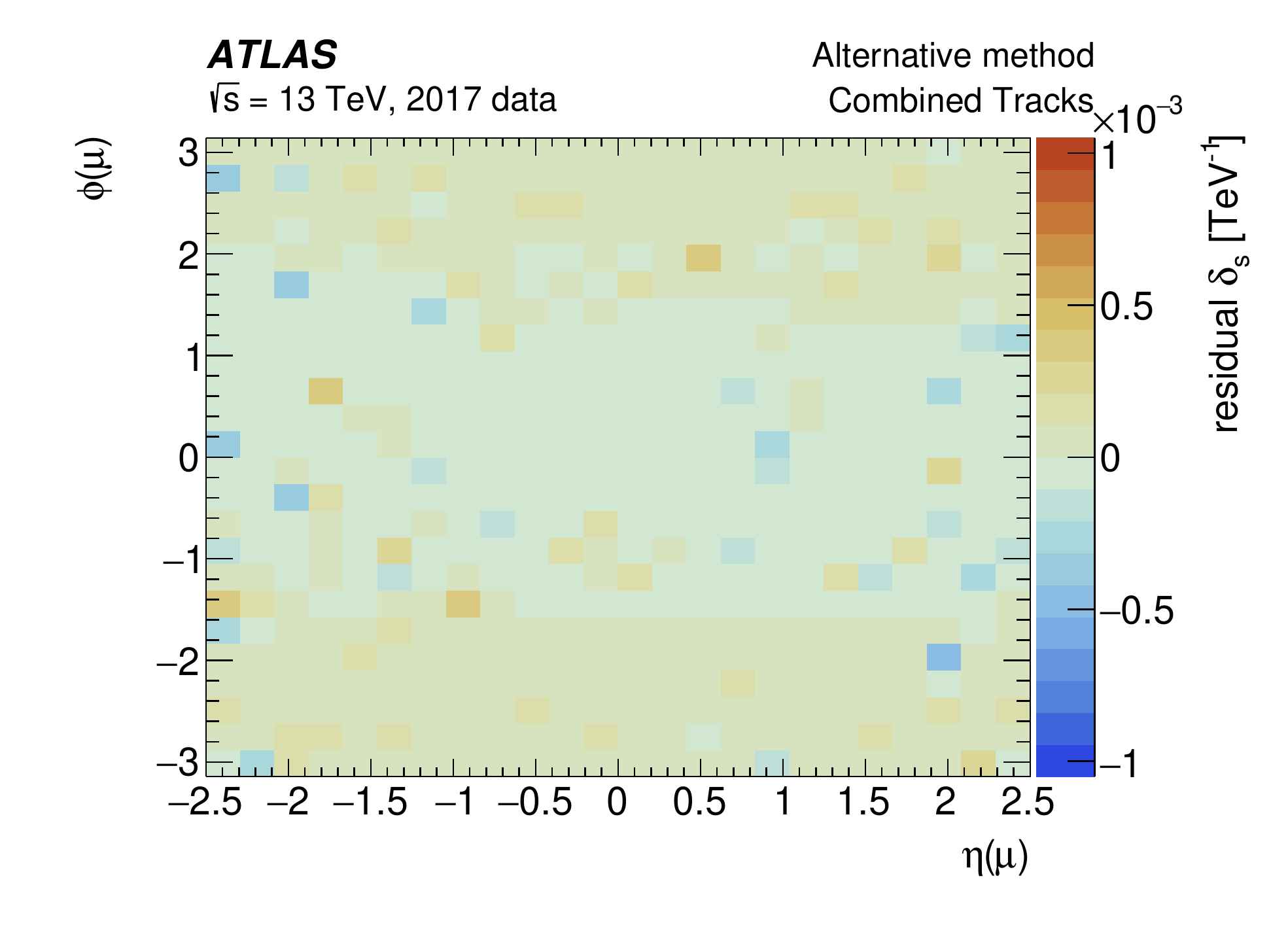}} \\
\subfloat[ ]{\includegraphics[width=0.45\linewidth]{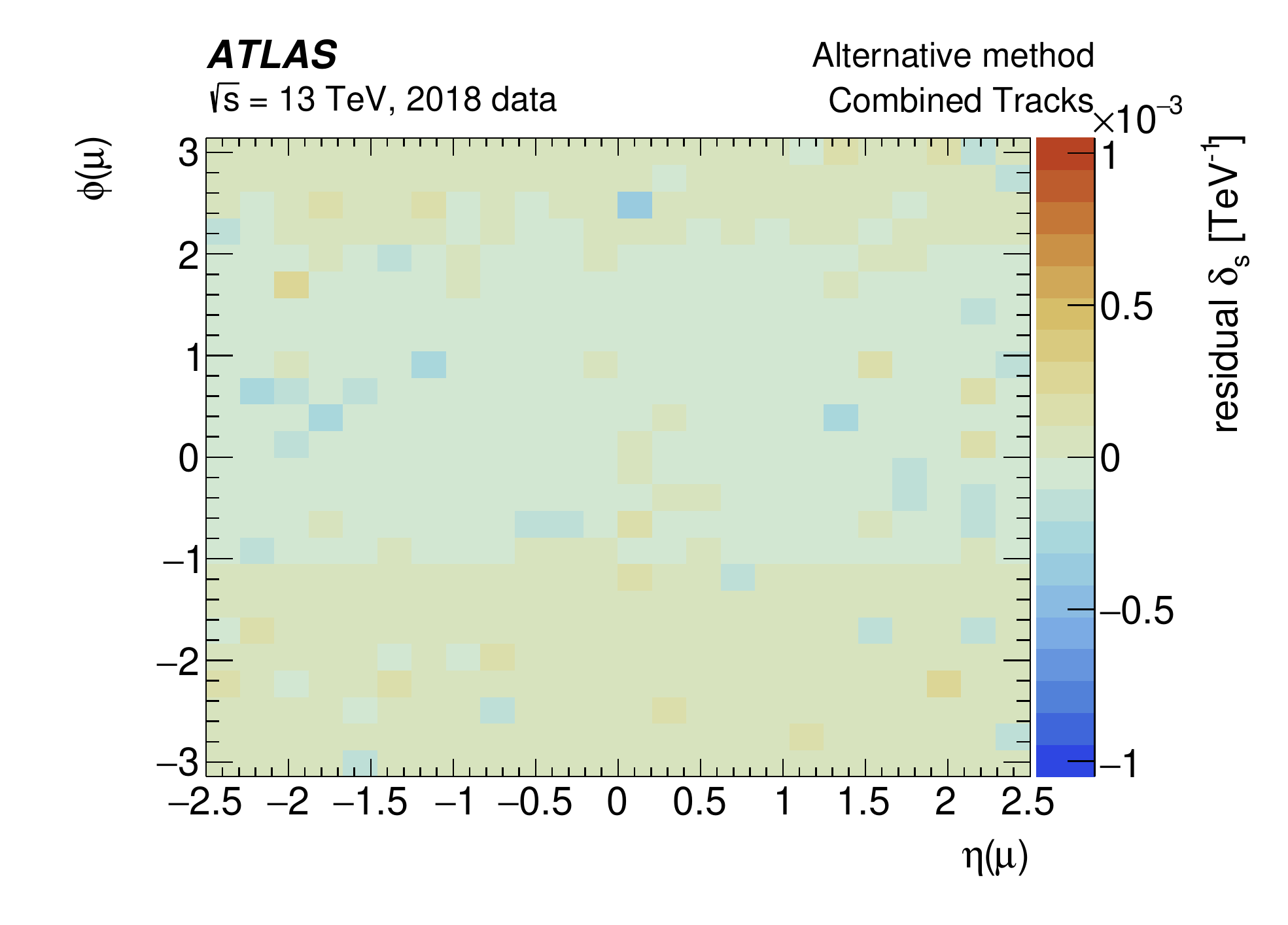}}
\caption{Residual charge-dependent biases on the muon \pt\ as evaluated on data for CB momenta after alignment and after application of the alternative correction procedure. The residual biases are shown separately for the three data-taking periods of (a) years 2015 and 2016, (b) 2017 and (c) 2018. The bias is defined in Eq.~(\ref{eq:bias_definition}).}
\label{fig:app_sgAfterCorr}
\end{figure*}

\section{Additional performance studies}
\label{app:results}
 
A second step, used to cross-check the validity of the corrections obtained directly for CB tracks, is introduced in \refS{Sec:CorrectionExtraction}. Because of the different parameterisation, a one-to-one comparison of the results obtained from the ID+MS method, as shown in Tables~\ref{tab:id_cor} and~\ref{tab:ms_cor}, and that of the CB method, as shown in \refT{tab:cb_cor}, is not possible. Therefore, the methods used to estimate the performance in \refS{sec::validation} are extended to the $\pt^{\textrm{Cor,ID+MS}}$ momenta as obtained from the ID+MS correction procedure. Using the same resonant decays from $\Jmm$, $\Upsmm$, and $\Zmm$ events in simulation, the dimuon momentum mass scale and the dimuon mass resolution are compared with the measurements obtained from data. The results are obtained with the same fitting methods detailed in \refS{sec::validation}. For the $\Jmm$ decays the results are presented in \refF{fig:app_JScaleAndResolution}, and they show a good agreement between data and simulation for both the mass scale and the mass resolution. Similarly, \refF{fig:app_YScaleAndResolution} and \refF{fig:app_ZScaleAndResolution} show the results obtained for $\Upsmm$ and $\Zmm$ decays, respectively. As expected from the systematic uncertainty study detailed in \refS{Sec:CorrectionExtraction}, the ID+MS methodology results in a reduced smearing uncertainty to a higher scale uncertainty than the results of the CB methodology. All results show an agreement within quoted uncertainties of the measured values between data and simulation.
 
\begin{figure*}[!htb]
\centering
\subfloat[ ]{\includegraphics[width=0.49\linewidth]{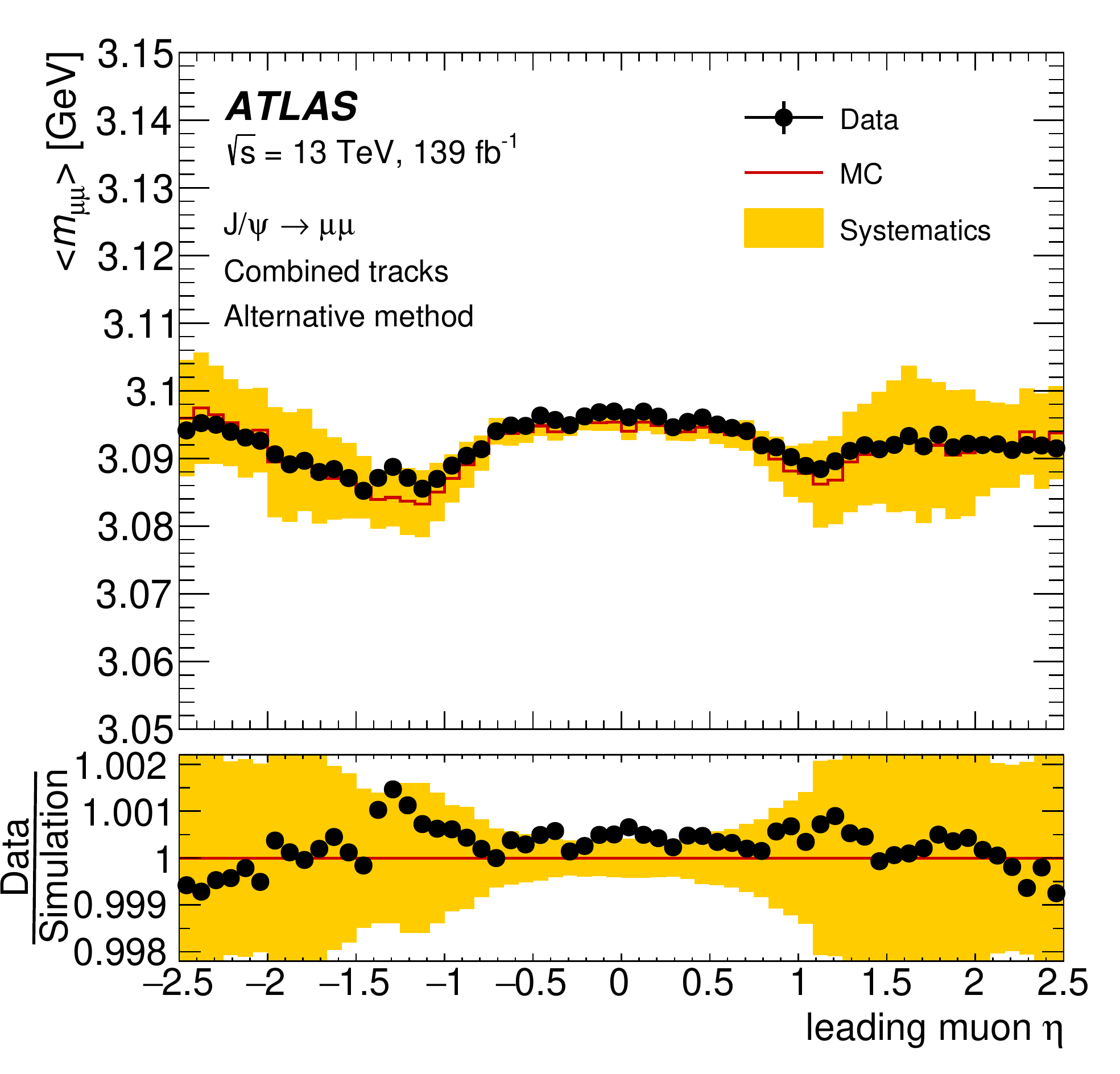}}
\subfloat[ ]{\includegraphics[width=0.49\linewidth]{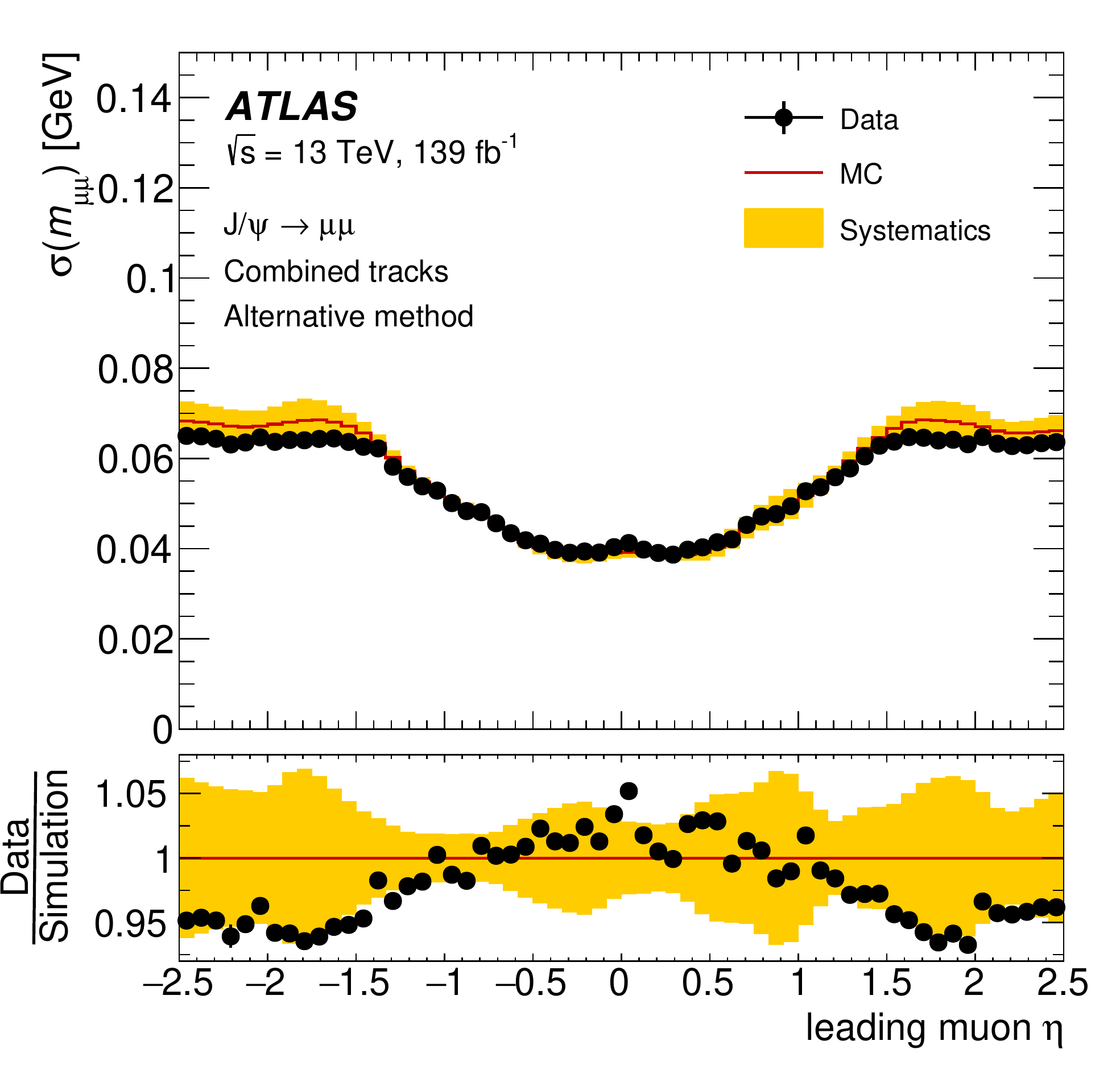}}
\caption{
(a)  Fitted mean mass of the dimuon system for CB muons for $\Jmm$ events data and corrected simulation as a function of the pseudorapidity of the highest-$\pt$ muon. (b) Dimuon invariant mass resolution for CB muons for $\Jmm$ events for data and  corrected simulation as a function of the pseudorapidity of the highest-$\pt$ muon. The upper panels show the results obtained on data and simulation after all corrections using the alternative ID+MS correction method for CB tracks. The lower panels show the ratios of the data to the MC simulation. The error bars represent the statistical uncertainty;  the shaded bands represent the systematic uncertainty in the correction and the systematic uncertainty in the extraction method added in quadrature.}
\label{fig:app_JScaleAndResolution}
\end{figure*}

\begin{figure*}[!htb]
\centering
\subfloat[ ]{\includegraphics[width=0.49\linewidth]{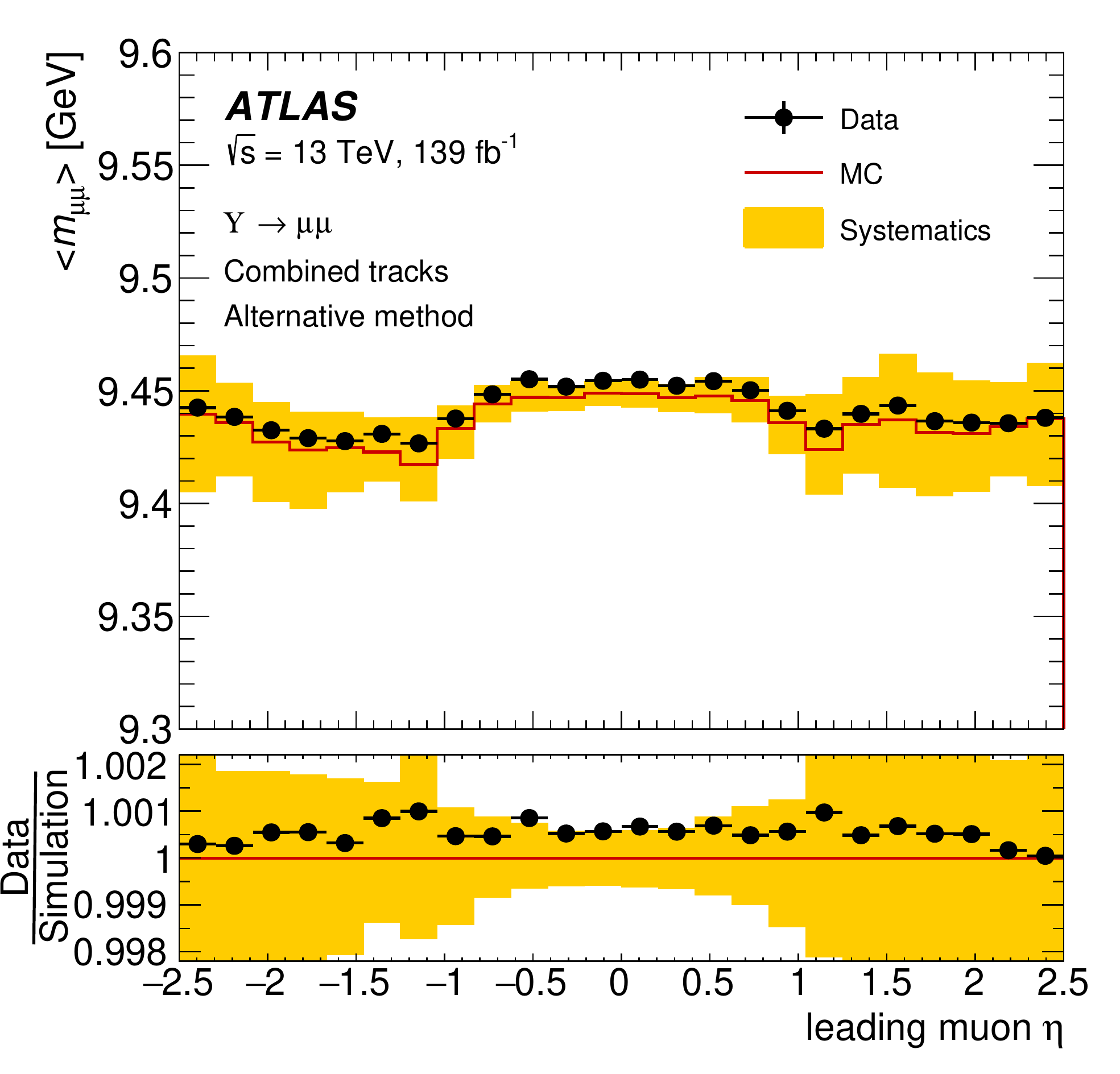}}
\subfloat[ ]{\includegraphics[width=0.49\linewidth]{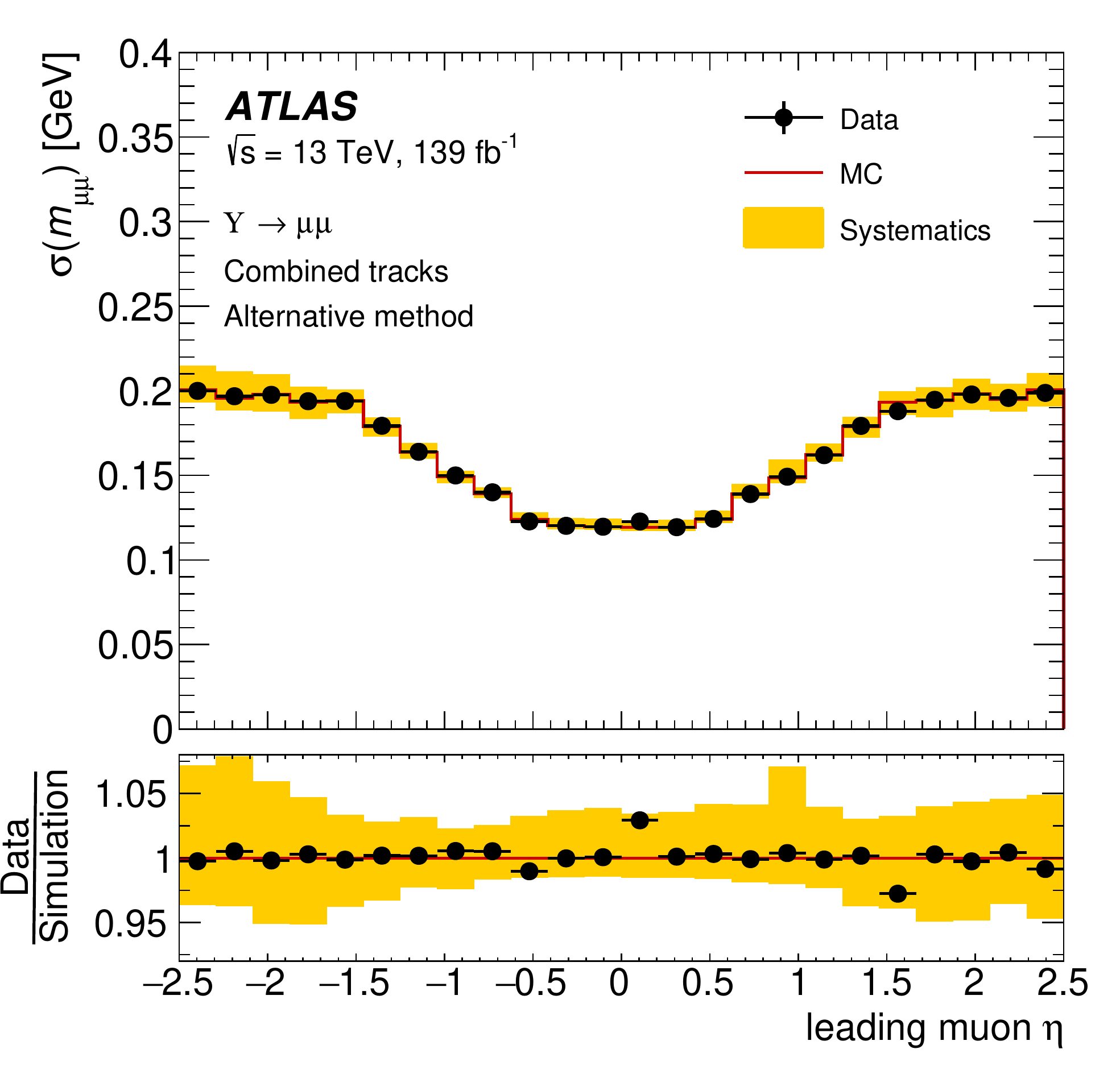}}
\caption{
(a)  Fitted mean mass of the dimuon system for CB muons for $\Upsmm$ events data and corrected simulation as a function of the pseudorapidity of the highest-$\pt$ muon. (b) Dimuon invariant mass resolution for CB muons for $\Upsmm$ events for data and  corrected simulation as a function of the pseudorapidity of the highest-$\pt$ muon. The upper panels show the results obtained on data and simulation after all corrections using the alternative ID+MS correction method for CB tracks. The lower panels show ratios of the data to the MC simulation. The error bars represent the statistical uncertainty;  the shaded bands represent the systematic uncertainty in the correction and the systematic uncertainty in the extraction method added in quadrature.}
\label{fig:app_YScaleAndResolution}
\end{figure*}

\begin{figure*}[!htb]
\centering
\subfloat[ ]{\includegraphics[width=0.49\linewidth]{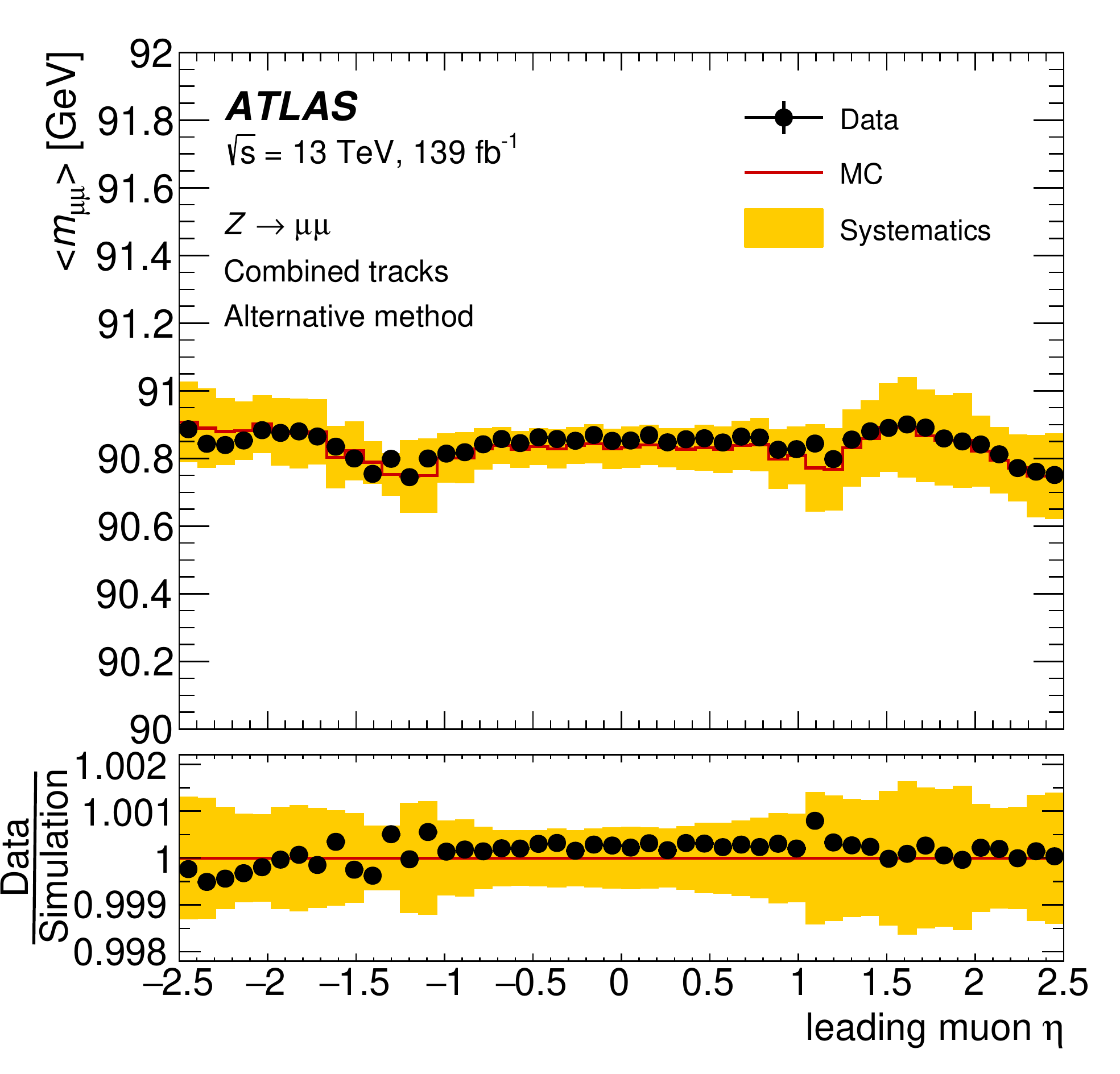}}
\subfloat[ ]{\includegraphics[width=0.49\linewidth]{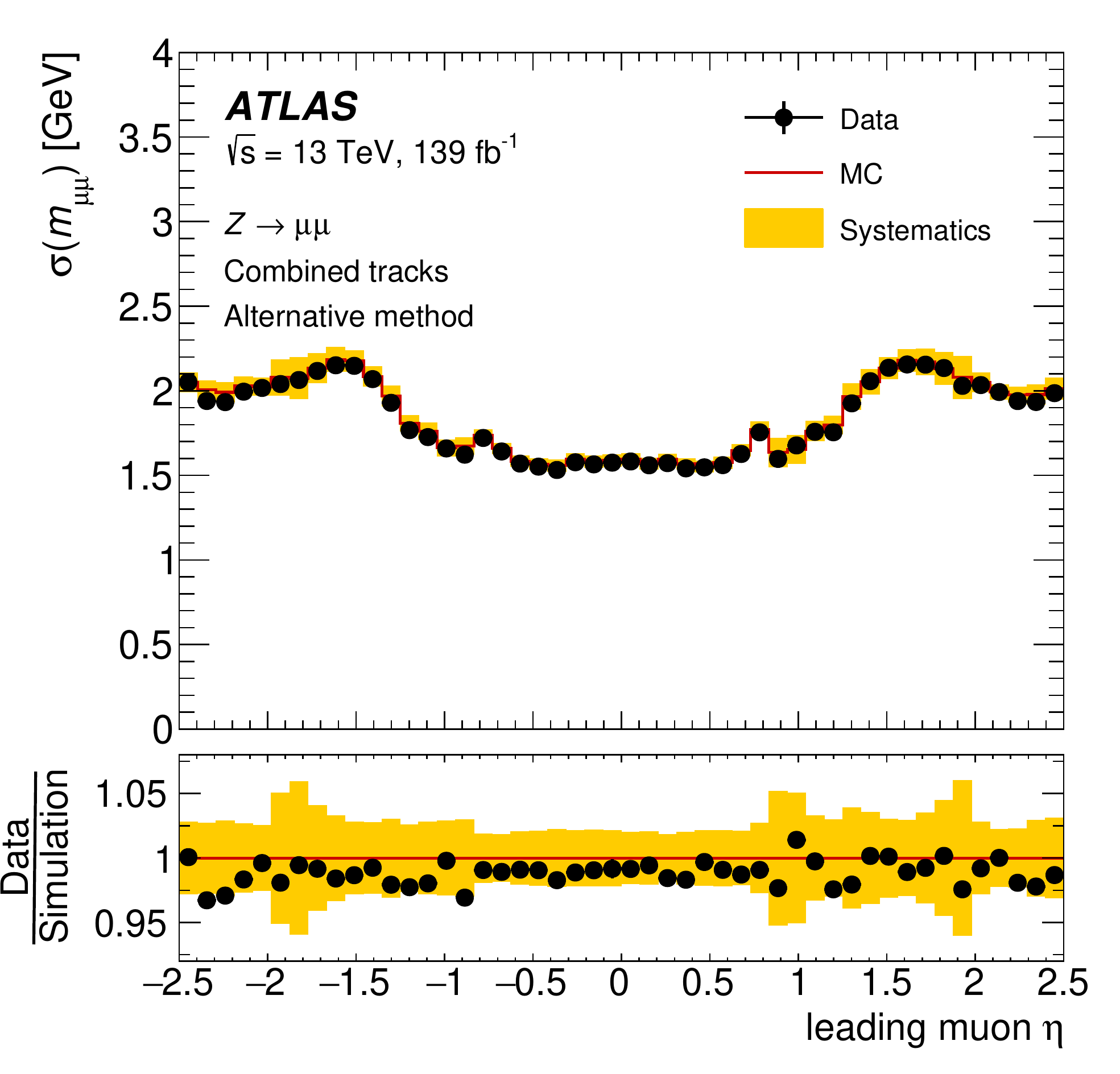}}
\caption{
(a)  Fitted mean mass of the dimuon system for CB muons for $\Zmm$ events data and corrected simulation as a function of the pseudorapidity of the highest-$\pt$ muon. (b) Dimuon invariant mass resolution for CB muons for $\Zmm$ events for data and  corrected simulation as a function of the pseudorapidity of the highest-$\pt$ muon. The upper panels show the results obtained on data and simulation after all corrections using the alternative ID+MS correction method for CB tracks. The lower panels show the ratios of the data to the MC simulations. The error bars represent the statistical uncertainty;  the shaded bands represent the systematic uncertainty in the correction and the systematic uncertainty in the extraction method added in quadrature.}
\label{fig:app_ZScaleAndResolution}
\end{figure*}
 
\clearpage


\printbibliography
 
\clearpage
\input{atlas_authlist}

\pagebreak

\end{document}

%% file: atlas_authlist.tex
 
\begin{flushleft}
\hypersetup{urlcolor=black}
{\Large The ATLAS Collaboration}

\bigskip

\AtlasOrcid[0000-0002-6665-4934]{G.~Aad}$^\textrm{\scriptsize 102}$,
\AtlasOrcid[0000-0002-5888-2734]{B.~Abbott}$^\textrm{\scriptsize 120}$,
\AtlasOrcid[0000-0002-1002-1652]{K.~Abeling}$^\textrm{\scriptsize 55}$,
\AtlasOrcid[0000-0002-8496-9294]{S.H.~Abidi}$^\textrm{\scriptsize 29}$,
\AtlasOrcid[0000-0002-9987-2292]{A.~Aboulhorma}$^\textrm{\scriptsize 35e}$,
\AtlasOrcid[0000-0001-5329-6640]{H.~Abramowicz}$^\textrm{\scriptsize 151}$,
\AtlasOrcid[0000-0002-1599-2896]{H.~Abreu}$^\textrm{\scriptsize 150}$,
\AtlasOrcid[0000-0003-0403-3697]{Y.~Abulaiti}$^\textrm{\scriptsize 117}$,
\AtlasOrcid[0000-0003-0762-7204]{A.C.~Abusleme~Hoffman}$^\textrm{\scriptsize 137a}$,
\AtlasOrcid[0000-0002-8588-9157]{B.S.~Acharya}$^\textrm{\scriptsize 69a,69b,p}$
% ...
\bigskip
\\

% ...
$^{29}$Physics Department, Brookhaven National Laboratory, Upton NY; United States of America.\\
% ...
$^{35}$$^{(a)}$Facult\'e des Sciences Ain Chock, R\'eseau Universitaire de Physique des Hautes Energies - Universit\'e Hassan II, Casablanca;$^{(b)}$Facult\'{e} des Sciences, Universit\'{e} Ibn-Tofail, K\'{e}nitra;$^{(c)}$Facult\'e des Sciences Semlalia, Universit\'e Cadi Ayyad, LPHEA-Marrakech;$^{(d)}$LPMR, Facult\'e des Sciences, Universit\'e Mohamed Premier, Oujda;$^{(e)}$Facult\'e des sciences, Universit\'e Mohammed V, Rabat;$^{(f)}$Institute of Applied Physics, Mohammed VI Polytechnic University, Ben Guerir; Morocco.\\
% ...
$^{55}$II. Physikalisches Institut, Georg-August-Universit\"{a}t G\"ottingen, G\"ottingen; Germany.\\
% ...
$^{69}$$^{(a)}$INFN Gruppo Collegato di Udine, Sezione di Trieste, Udine;$^{(b)}$ICTP, Trieste;$^{(c)}$Dipartimento Politecnico di Ingegneria e Architettura, Universit\`a di Udine, Udine; Italy.\\
% ...
$^{102}$CPPM, Aix-Marseille Universit\'e, CNRS/IN2P3, Marseille; France.\\
% ...
$^{117}$Department of Physics, New York University, New York NY; United States of America.\\
% ...
$^{120}$Homer L. Dodge Department of Physics and Astronomy, University of Oklahoma, Norman OK; United States of America.\\
% ...
$^{137}$$^{(a)}$Departamento de F\'isica, Pontificia Universidad Cat\'olica de Chile, Santiago;$^{(b)}$Millennium Institute for Subatomic physics at high energy frontier (SAPHIR), Santiago;$^{(c)}$Instituto de Investigaci\'on Multidisciplinario en Ciencia y Tecnolog\'ia, y Departamento de F\'isica, Universidad de La Serena;$^{(d)}$Universidad Andres Bello, Department of Physics, Santiago;$^{(e)}$Instituto de Alta Investigaci\'on, Universidad de Tarapac\'a, Arica;$^{(f)}$Departamento de F\'isica, Universidad T\'ecnica Federico Santa Mar\'ia, Valpara\'iso; Chile.\\
% ...
$^{150}$Department of Physics, Technion, Israel Institute of Technology, Haifa; Israel.\\
$^{151}$Raymond and Beverly Sackler School of Physics and Astronomy, Tel Aviv University, Tel Aviv; Israel.\\
% ...
$^{a}$ Also Affiliated with an institute covered by a cooperation agreement with CERN.\\
$^{b}$ Also at An-Najah National University, Nablus; Palestine.\\
$^{c}$ Also at Borough of Manhattan Community College, City University of New York, New York NY; United States of America.\\
$^{d}$ Also at Bruno Kessler Foundation, Trento; Italy.\\
$^{e}$ Also at Center for High Energy Physics, Peking University; China.\\
$^{f}$ Also at Center for Interdisciplinary Research and Innovation (CIRI-AUTH), Thessaloniki ; Greece.\\
% ...
$^{p}$ Also at Department of Physics, King's College London, London; United Kingdom.\\
% ...

\end{flushleft}
